\documentclass[11pt, a4paper]{article}
\usepackage[a4paper]{geometry}

\usepackage[longnamesfirst, round]{natbib}
\usepackage{url}

\usepackage[T1]{fontenc}              
\usepackage[utf8]{inputenc}         
\usepackage[english]{babel}

\usepackage{amsthm,amsmath,amsfonts,amssymb} 
\usepackage{amsthm}                   
\usepackage{graphicx}                 
\usepackage{subfig}                   
\usepackage{lscape}
\usepackage{booktabs}                          

\usepackage{listings}                 
\usepackage{tikz}

\usepackage{algorithm2e}
\usepackage{mathtools}
\usepackage{enumerate}

\usepackage[strict]{changepage}
\usepackage{relsize}
\usepackage{authblk}
\usepackage{epstopdf}
\usepackage{bbm}
\usepackage{placeins}
          
\usepackage{standalone}

\theoremstyle{plain}
\newtheorem{thm}{Theorem}[section]

\theoremstyle{definition}
\newtheorem{defn}{Definition}[section]

\newtheorem{exmp}{Example}[section]

\theoremstyle{remark}

\begin{document}
        \title{Constructing Priors that Penalize the Complexity of Gaussian Random Fields}
        \author[1]{Geir-Arne Fuglstad}
        \author[2]{Daniel Simpson}
        \author[3]{Finn Lindgren}
        \author[4]{Håvard Rue}
        \affil[1]{Department of Mathematical Sciences, NTNU, Norway}
        \affil[2]{Department of Statistical Sciences, University of Toronto, Canada}
        \affil[3]{School of Mathematics, University of Edinburgh, United Kingdom}
        \affil[4]{CEMSE Division, King Abdullah University of Science and Technology, Saudi Arabia}
        \maketitle

\begin{abstract}
Priors are important for achieving proper posteriors with physically
meaningful covariance structures for Gaussian random fields (GRFs) since the
likelihood typically
only provides limited information about the covariance structure
under in-fill asymptotics. 
We extend the recent Penalised Complexity
prior framework and develop a principled joint prior for the range 
and the marginal variance of one-dimensional, two-dimensional and three-dimensional Mat\'ern GRFs 
with fixed smoothness. 
The prior is weakly informative and penalises complexity by
shrinking the range towards infinity and the
 marginal variance towards zero. We propose guidelines for selecting the hyperparameters, and a simulation study shows that the new prior 
provides a principled alternative to reference priors that can leverage prior
knowledge to achieve shorter credible
intervals while maintaining good coverage.

We extend the prior to a non-stationary GRF parametrized
through local ranges and marginal standard deviations, and introduce
a scheme for selecting the hyperparameters based on the coverage of the parameters
when fitting simulated stationary data. The approach is
applied to a dataset of annual precipitation in southern Norway and
the scheme for selecting the hyperparameters leads to concervative
estimates of non-stationarity and improved 
predictive performance over the stationary model.
\end{abstract}

\noindent%
\vfill

\section{Introduction}
Gaussian random fields (GRFs) provide a simple and powerful tool for introducing
spatial or temporal dependence in Bayesian hierarchical models and are fundamental building blocks in spatial statistics
and non-parametric modelling, but even for stationary GRFs controlled only by range and marginal 
variance, the choice of
prior distribution remains a challenge. The prior 
is difficult to choose: a well-chosen prior will
stabilise the inference and improve the predictive performance, whereas a
poorly chosen prior can be next to catastrophic. 
The main focus in this paper is one-dimensional, two-dimensional and
three-dimensional GRFs with Mat\'ern covariance functions with fixed 
smoothness, but we also discuss how to extend the prior to 
non-stationary covariance structures.

The Mat\'ern covariance function leads a ridge in the likelihood
for the range and the marginal variance
\citep{warnes1987problems},
and there is no consistent estimator under in-fill asymptotics for these
parameters when the base space of the GRF is of dimension three or lower 
\citep{stein1999interpolation,Zhang2004}.
For these GRFs only a limited amount of information can be learned about the parameters from a bounded domain and
the prior affects the behaviour of the posterior of the parameters 
even under in-fill asymptotics.
For example, for a one-dimensional GRF with an exponential covariance function
observed on the interval \([0, 1]\), it is only the ratio of the range and the marginal variance that
can be estimated consistently, and not the range or the marginal variance 
separately~\citep{Ying1991}.

\begin{figure}
        \centering
        \subfloat[\(\rho = \sigma^2 = 1\)]{
                \includegraphics[width=4.3cm]{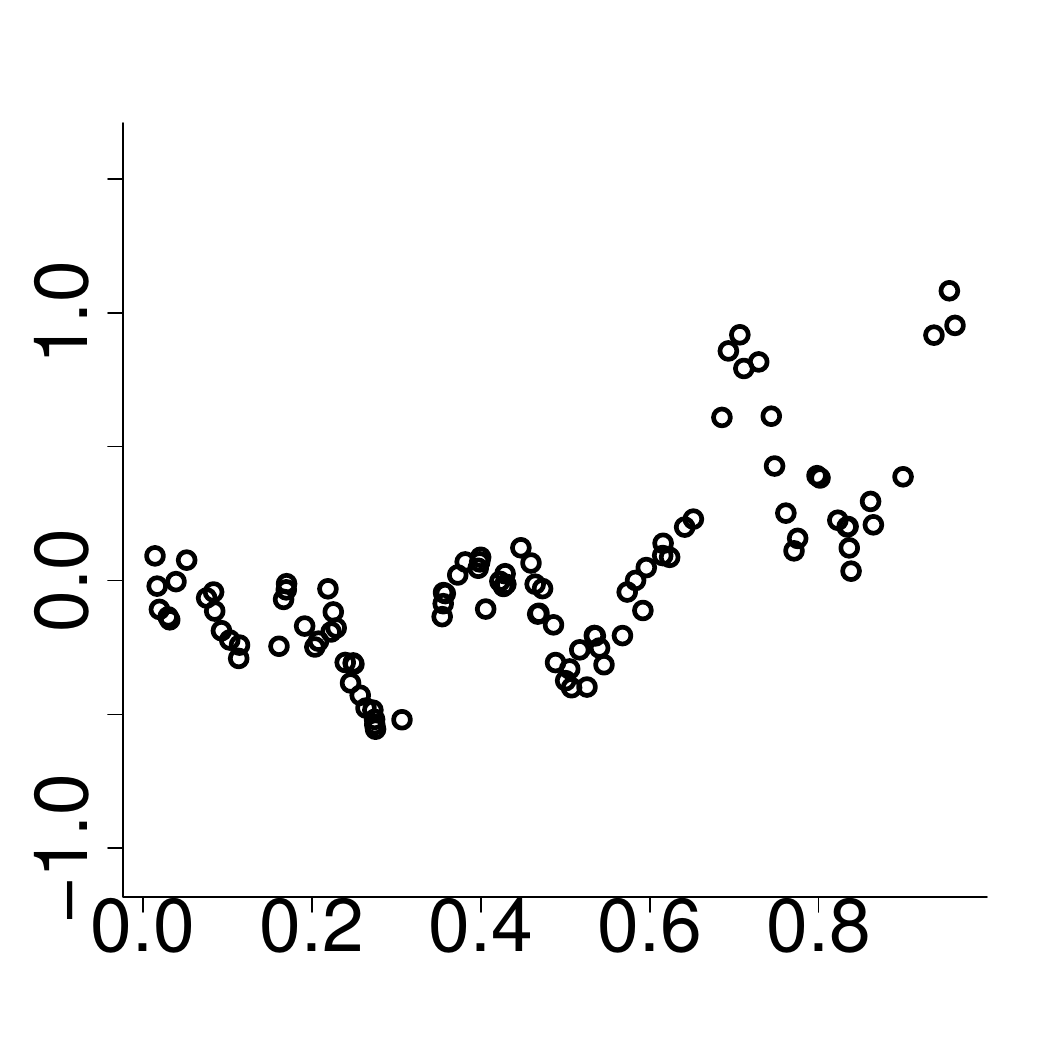}
                \label{fig:intrinsic:1}
        }
        \subfloat[\(\rho = \sigma^2 = 1000\)]{
                \includegraphics[width=4.3cm]{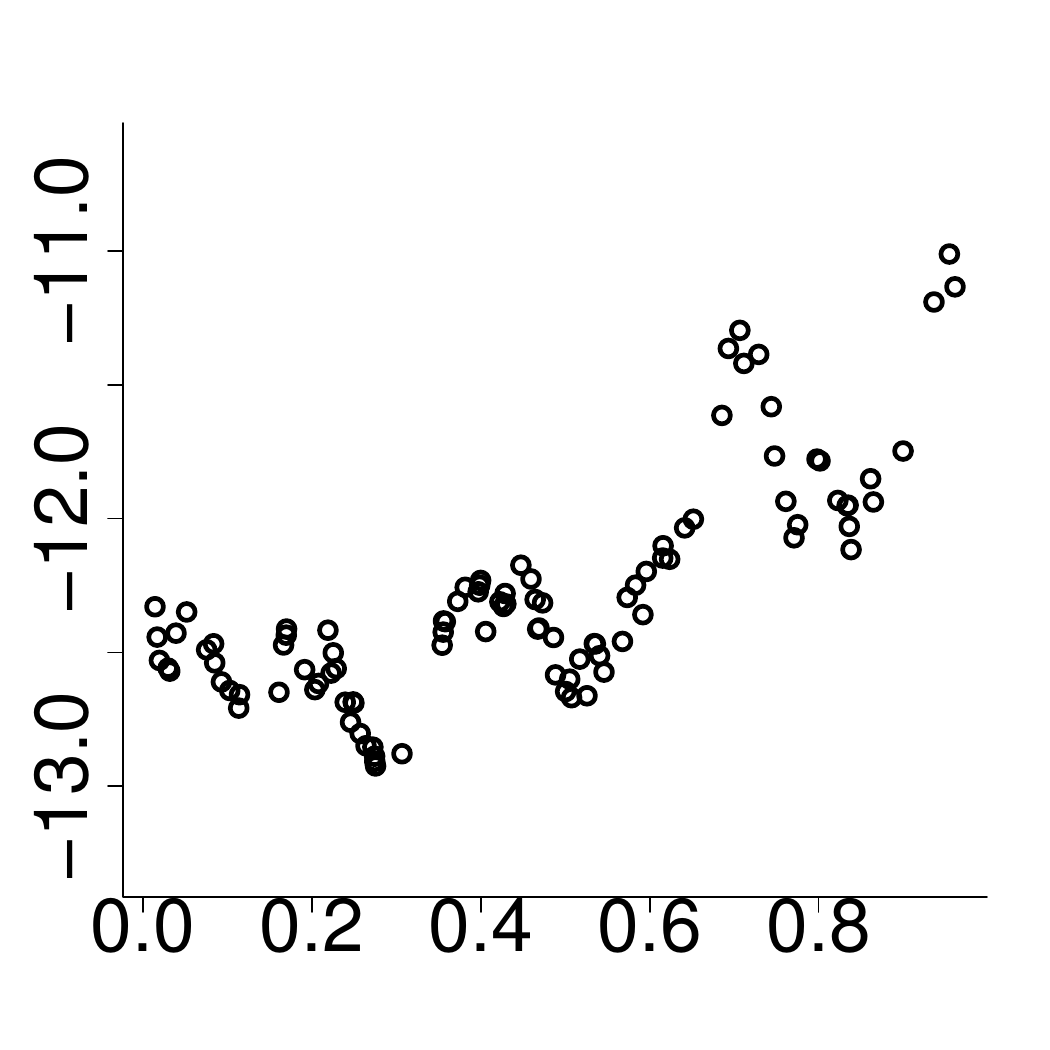}
                \label{fig:intrinsic:1000}
        }
        \subfloat[\(\rho = \sigma^2 = 1000000\)]{
                \includegraphics[width=4.3cm]{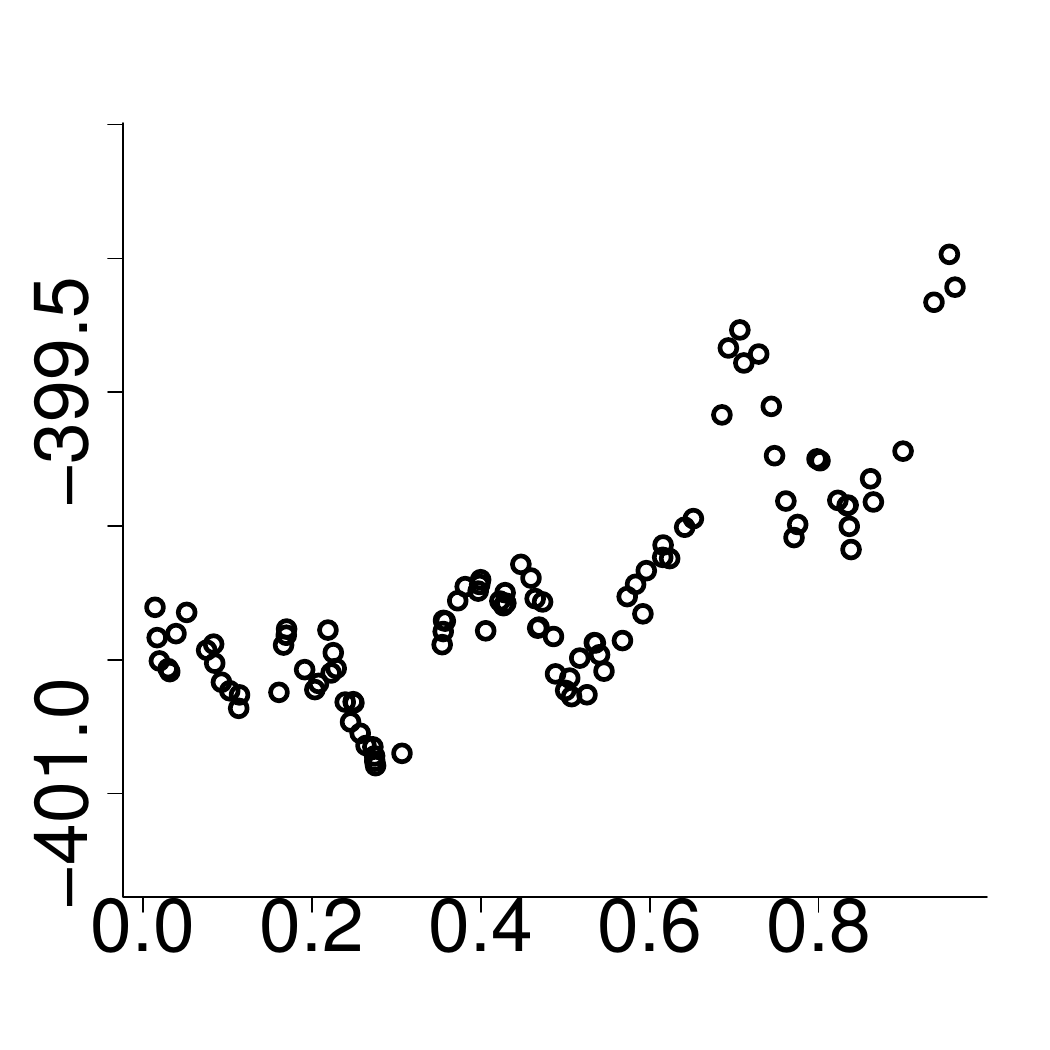}
                \label{fig:intrinsic:1000000}
        }
        \caption{Simulations with the exponential covariance function 
                         \(c(d) = \sigma^2 \mathrm{e}^{-d/\rho}\) for 
                         different values of \(\rho = \sigma^2\) using the same underlying 
                         realization of independent standard Gaussian random variables. 
                         The patterns of the values are almost the same, but
                         the levels differ.}
        \label{fig:intrinsic}
\end{figure}

This ratio also determines the asymptotic properties
of predictions  under in-fill
asymptotics with the exponential covariance function \citep{stein1999interpolation}, but
predictive distributions are not the only target for inference. 
Figure \ref{fig:intrinsic} shows that moves along the ridge in
the likelihood when using the exponential covariance function, changes
the level of the simulated observations, but that the pattern of the 
values around the level remains stable. 
These choices of parameters lead to similar predictive distributions
conditional on the observed data, but simulating unconditionally from 
GRFs with these parameters lead to highly different realizations.
In a real application where the values
in Figure~\ref{fig:intrinsic:1} were observed, the practitioner will
likely know that the ranges
and marginal variances that generate Figures~\ref{fig:intrinsic:1000} 
and~\ref{fig:intrinsic:1000000} are not physically meaningful
even if the spreads of values are consistent with the 
observed pattern. Therefore, we believe the practitioner
should be provided with a principled 
prior that allows him/her to include expert knowledge, 
in an interpretable way, about the range of parameters that 
are physically meaningful.

But to our knowledge, the only principled approach to prior selection for GRFs was 
introduced by~\citet{Berger2001}, who derived reference priors for a GRF
partially observed with no noise. Their work has been extended by several 
authors~\citep{Paulo2005, Kazianka2012, Kazianka2013} and,
critically, \citet{DeOliveira2007} allowed for Gaussian observation noise.
In the more restricted case of a GRF with a
Gaussian covariance function \citet{vandervaart2009} showed that the
inference asymptotically 
behaves well with an inverse gamma distribution on range, but they 
provide no guidance on which hyperparameters should be selected for the prior.

However, reference priors aim to be objective and 
are built on the fundamental principle of being
the least informative priors, in an information-theoretic sense, for Bayesian
inference \citep{berger2009}, and GRFs are often embedded in 
Bayesian hierarchical models that are too complex for deriving the
reference priors. Therefore, we propose a different construction that 
leads to a weakly informative prior that can leverage prior knowledge and is
appropriate for hierarchical models where models components are combined
linearly in the latent part of the model. In this setting,
the model construction tends to be modular
and priors should be constructed separately for each model
component. The GRFs are used to achieve 
the desired second-order structure
while the first-order structure of the model is handled by
separate model components, and we must construct a joint prior
for the range and marginal variance of a zero-mean Mat\'ern GRF.

This setting is similar to structured additive regression models where
\citet{klein2016} has shown that the Penalised Complexity (PC) prior 
framework developed by~\citet{Martins2014} behaves well when used for the
components of the models. This motivates the desire to use the PC prior
framework to construct a joint prior for the range and the marginal
variance of a Mat\'ern GRF, but there are three questions that must
be answered. Is the PC prior framework suitable for infinite-dimensional
model components? How can we deal with the fact that the KLD between
Mat\'ern GRFs in general is infinite? And how can we construct a multivariate
PC prior that properly accounts for the intrinsic link between range
and marginal variance due to the ridge in the likelihood?

In this paper we extend \citet{Martins2014} by answering these
questions, and we show that the principles
of the PC prior framework can be applied to construct a prior
for Mat\'ern GRFs that is independent of the observation process. 
This is technically more demanding than the direct approach, which
would be to construct the PC prior based on the finite-dimensional
observation process, but the rewards are a
prior that can be applied for any spatial design
and any observation process, and is computationally
inexpensive and has a much 
simpler form than the reference priors for GRFs 
in published literature. The resulting prior is 
weakly informative and shrinks towards a \emph{base model} with 
infinite range and zero
marginal variance through hyperparameters that indicate how strongly the user wishes
to shrink towards the base model.  

The stationary Mat\'ern GRF can be extended to a non-stationary GRF by adding extra flexibility in
the covariance function, but since the covariance 
structure of a GRF is only observed indirectly,
the estimated covariance structure can be highly sensitive
to the type of flexibility allowed and the prior used on the flexibility. We 
show that the PC prior developed
for the stationary Mat\'ern GRF can be extended further to a prior for a
non-stationary GRF where the non-stationarity is controlled by
covariates. The prior is motivated by \(g\)-priors and shrinkage properties, and
we consider one scheme for selecting the hyperparameters that reduces the risk of overfitting
the non-stationary GRF.

The joint PC prior for the range and the marginal variance of a
Mat\'ern GRF with a fixed smoothness parameter is derived in Section~\ref{sec:PenalisedComplex}. Then
in Section \ref{sec:SimStudy} a small simulation study is performed to evaluate the 
frequentist properties of the credible intervals and the behaviour of the joint posterior,
and we demonstrate that the prior is applicable also for logistic spatial regression
where the observation process is highly non-Gaussian. In Section \ref{sec:NonStat}
we discuss how to
extend the PC prior to a conservative prior for a non-stationarity model
for annual precipitation in southern Norway.
The paper ends with discussion and conclusions in Section \ref{sec:Discussion}.
The appendices contain proofs of the theorems, computer code, technical details and further discussion of many
of the topics addressed.

\section{Penalised Complexity prior}
\label{sec:PenalisedComplex}

\subsection{Framework}
Including a GRF in a model may lead to overfitting by, for example, estimating
spurious spatial trends or spurious
temporal trends.
\citet{Martins2014} suggest to handle the issue of overfitting by viewing model
components, such as GRFs, as flexible extensions of simpler,
less flexible \emph{base models} and then developing priors that shrink the components towards
their base models. For example,
they view a random effect with non-zero variance as an extension of a random effect with zero variance,
and construct a prior that shrinks the variance of the random effect towards zero.

The first step of their approach is to derive a distance from the base model to
its flexible extension using the Kullback-Leibler divergence (KLD). The purpose of
the distance is to provide a better parametrization of the model component
 where the size of the change in the parameter corresponds to the size of the change in the difference
between the model component and its base model.
In the setting
of this paper, this can be done
by describing the base model for the GRF by the Gaussian measure
$P_0$ and the flexible model by the Gaussian measure $P$, and then defining the distance
by $\text{dist}(P\vert\vert P_0) = \sqrt{2\text{KL}(P \vert \vert P_0)}$, where
 $\text{KL}(P \vert \vert P_0)$ is the KLD from $P_0$ to $P$
and is defined as follows.

\begin{defn}[Kullback-Leibler divergence]
        Let $P_0$ and $P$ be measures over the set $\chi$, where $P$ is absolutely continuous with respect
        to $P_0$, then the Kullback-Leibler divergence from $P_0$ to $P$ is defined as
        \[
                \text{KL}(P\vert\vert P_0) = \int_\chi \log\frac{\mathrm{d} P}{\mathrm{d} P_0} \mathrm{d}P,
        \]
        where $\mathrm{d}P/\mathrm{d}P_0$ is the Radon-Nikodym derivative of $P$ with respect to $P_0$.
\end{defn}

The KLD is used by \citet{Martins2014} and has the benefits that it has 
an information-theoretical interpretation as the information lost when using the base model $P_0$
to approximate $P$ and that it is an asymmetric distance 
from the ``preferred'' base model to the flexible extension. The square root is used in the
definition of the distance to bring the distance to the correct scale \citep{Martins2014}.

The second step of the prior construction is to define the prior on the derived distance using
three principles: Occam's razor, constant-rate penalisation and user-defined
scaling. Occam's razor means that the prior penalises more and more
strongly the further one is
from the base model and can be achieved by using constant-rate penalisation, where
the prior on the distance, $t$, satisfies
\[
        \frac{\pi(t+\delta)}{\pi(t)} = r^\delta, \qquad t, \delta > 0,
\]
for a constant decay-rate $0 < r < 1$. The only continuous distribution
with this property is the exponential distribution $\pi(t) = \lambda \exp(-\lambda t)$, for
$t > 0$,  where
the relative change in the prior when the distance increases by $\delta$ does not depend on the current
distance $t$. The justification for using a simple prior on distance is that the parametrization
 corresponds directly to the size of the changes in the distribution of the model component.

The prior has a hyperparameter $\lambda$ that must be set by the user and the principle
of user-defined scaling is used to provide an interpretable way to set its value.
The distance itself is typically not directly interpretable by the user and must be 
transformed to an interpretable size $Q(t)$. The prior information can then be included through, for example, tail probabilities
$P(Q(d) > U) = \alpha$ or $P(Q(d) < L) = \alpha$, where $U$ or $L$ is an upper or lower limit,
respectively, and $\alpha$ is the upper or lower tail probability of the prior distribution. 
Through this
construction the
PC prior combines the geometry of the parameter space
with prior belief about an interpretable size.

\subsection{Derivation}
\label{sec:unboundPrior}
The Mat\'ern covariance function has been studied extensively \citep{stein1999interpolation}, and it is
isotropic and can be defined as a function
of the distance between locations.
\begin{defn}[Mat\'ern covariance function]
        \label{def:KLD}
        A Mat\'ern covariance function $c:[0, \infty)\rightarrow\mathbb{R}$ can be parametrized
        through a marginal standard deviation \(\sigma\), 
        a range parameter \(\rho\), and a smoothness parameter
        \(\nu\), and is given by
        \[
                c_\nu(r; \sigma, \rho) = \sigma^2 \frac{2^{1-\nu}}{\Gamma(\nu)}\left(\sqrt{8\nu}\frac{r}{\rho}\right)^\nu K_\nu\left(\sqrt{8\nu}\frac{r}{\rho}\right),
        \]
        where \(K_\nu\) is the modified Bessel function of the second kind, order \(\nu\).
\end{defn}

The choice of $\sqrt{8\nu}$ in the definition follows \citet{Lindgren2011} and makes
$\rho$ the distance at which the correlation is approximately $0.1$.
This parametrization of the Mat\'ern covariance function has convenient
interpretations for the parameters, but the parametrization is not convenient
for deriving a PC prior. Therefore, we introduce an alternative parametrization.

\begin{defn}[Alternative parametrization of the Mat\'ern covariance function]
        \label{def:AltPar}
        Assume that the base space is \(\mathbb{R}^d\) and introduce 
        \begin{equation}
                \kappa = \sqrt{8\nu}/\rho\quad \text{and}\quad \tau = \sigma\kappa^\nu \sqrt{\frac{\Gamma(\nu + d/2)(4\pi)^{d/2}}{\Gamma(\nu)}}.
                \label{eq:tauPar}
        \end{equation}
\end{defn}

This parametrization has the benefit
that it describes what can, $\tau$, and what cannot, $\kappa$, be consistently estimated under in-fill
asymptotics when the dimension of the base space $d \leq 3$. When $\kappa$
is changed, but $\tau$ is fixed, the resulting Gaussian measures are equivalent and the KLD between
the GRFs is finite, but if $\tau$ is changed, the resulting Gaussian measures are singular and
the KLD between the GRFs is infinite \citep{Zhang2004}. By assumption $\nu$ is fixed, and
the joint prior is derived in two steps: first $\pi(\tau | \kappa)$ and then $\pi(\kappa)$. The parameter
$\tau$ can be consistently estimated under in-fill asymptotics, so 
the derivation of the PC prior for $\tau | \kappa$
must be based on a finite-dimensional observation (but will not
depend on the spatial design).

\begin{thm}[PC prior for $\tau | \kappa$]
        \label{thm:PC_tauIkappa}
        Let \(u\) be a GRF defined on \(\mathcal{D}\subset\mathbb{R}^d\) with a Mat\'ern covariance function
        with parameters $\tau$, $\kappa$ and $\nu$.  If
        the GRF is observed on $\boldsymbol{s}_1, \boldsymbol{s}_2, \ldots, \boldsymbol{s}_n\in\mathcal{D}$,
        then conditionally on \(\kappa\) the PC prior for $\tau$ with base model $\tau = 0$ is
        \[
                \pi(\tau | \kappa) = \lambda \exp(-\lambda \tau), \quad \tau > 0,
        \]
        where \(\lambda > 0\) is a hyperparameter.
\end{thm}
\begin{proof}
        See Appendix A.1.
\end{proof}

Since the prior shrinks towards zero variance conditionally on $\kappa$, we suggest to select the 
hyperparameter $\lambda$ by limiting the upper tail probability $\alpha$
that the marginal standard deviation of the GRF will exceed $\sigma_0$. That is
by selecting $\sigma_0$ and $\alpha$ such that
\(
        \mathrm{P}(\sigma > \sigma_0 | \kappa) = \alpha,
\)
where \(\sigma\) is the marginal standard deviation corresponding to
\(\tau\) and \(\kappa\). Alternatively, one can set the hyperparameter by selecting
the tail probability that the GRF at an arbitrary location exceeds a chosen value, 
but this does not lead to a simple analytic expression.

\begin{thm}
        \label{thm:PCtau_P}
        The PC prior for $\tau|\kappa$ satisfies \(\mathrm{P}(\sigma > \sigma_0 | \kappa) = \alpha\) if 
        \[      
                \lambda(\kappa) = -\kappa^{-\nu}\sqrt{\frac{\Gamma(\nu)}{\Gamma(\nu + d/2)(4\pi)^{d/2}}} \frac{\log(\alpha)}{\sigma_0}.
        \]
\end{thm}
\begin{proof}
        See Appendix A.2.        
\end{proof}

The PC prior for $\kappa$ can also be based on the finite-dimensional distribution
corresponding to the observation locations, but this would lead to a computationally
expensive prior because calculating
KLDs between Gaussian distributions with dense covariance matrices
has a cubic complexity in the number of observation locations. 
We seek to overcome this challenge by constructing the 
PC prior for $\kappa$ using the infinite-dimensional GRF instead of the finite-dimensional
observations. This is possible because changes in $\kappa$ result in finite values for the
 KLD 
for the infinite-dimensional GRF when
$\tau$ is fixed. In the proofs it is assumed that 
the GRF itself exists on an arbitrarily large ambient domain. In the
next section we discuss how the prior could be derived under the assumption that the GRF only exists
on the area from which the observations were made.

\begin{thm}[PC prior for $\kappa$]
        \label{thm:PC_kap_unb}
        Let $u$ be a GRF defined on $\mathbb{R}^d$, where $d\leq 3$, with a Mat\'ern
        covariance function with parameters $\tau$, $\kappa$ and $\nu$. 
        The PC prior for $\kappa$ with base model $\kappa = 0$ is
        \[
                \pi(\kappa) = \frac{d}{2}\lambda\kappa^{d/2-1}\exp\left(-\lambda \kappa^{d/2}\right), \quad \kappa > 0,
        \]
        where $\lambda>0$ is a hyperparameter.
\end{thm}
\begin{proof}
        See Appendix A.3.
\end{proof}

The prior in Theorem \ref{thm:PC_kap_unb} is a Weibull distribution with
shape parameter $d/2$ and scale parameter $\lambda^{-d/2}$, and is a
heavy-tailed distribution.
Since the prior shrinks the range towards infinity ($\kappa = 0$), we suggest
to set the hyperparameter by controlling the tail probability
that the range is below a certain limit.

\begin{thm}
        \label{thm:kappa_lam}
        The prior for $\kappa$ satisfies
        $
                \mathrm{P}(\rho < \rho_0) = \alpha
        $
        if 
        \[
                \lambda = -\left(\frac{\rho_0}{\sqrt{8\nu}}\right)^{d/2}\log(\alpha)
        \]
\end{thm}
\begin{proof}
	See Appendix A.4.
\end{proof}

Combining the priors for $\tau | \kappa$ and $\kappa$ provides
the main results of this paper, which are the joint PC prior for $(\kappa, \tau)$ and the 
joint PC prior for $(\rho, \sigma)$.

\begin{thm}[PC prior for the Mat\'ern ($\kappa$, $\tau$)]
        \label{thm:PC}
        Let $u$ be a GRF defined on $\mathbb{R}^d$, where $d\leq 3$, with a Mat\'ern
        covariance function with parameters $\tau$, $\kappa$ and $\nu$.
        The joint PC prior based on the base models $\tau = 0$ and $\kappa = 0$
        is
        \[
                \pi(\kappa, \tau) = \frac{d}{2}\lambda_1\lambda_2(\kappa) \kappa^{d/2-1}\exp(-\lambda_1 \kappa^{d/2} - \lambda_2(\kappa) \tau), \quad \kappa >0, \tau>0,
        \]
        where \(\mathrm{P}(\rho < \rho_0) = \alpha_1
        \) and
        \(\mathrm{P}(\sigma > \sigma_0 \vert \kappa) = \alpha_2\)
        are achieved by
        \[
                \lambda_1 = -\left(\frac{\rho_0}{\sqrt{8\nu}}\right)^{d/2}\log(\alpha_1) \quad \text{and} \quad \lambda_2(\kappa) = -\kappa^{-\nu}\sqrt{\frac{\Gamma(\nu)}{\Gamma(\nu + d/2)(4\pi)^{d/2}}} \frac{\log(\alpha_2)}{\sigma_0}.
        \]

\end{thm}
\begin{proof}
        See Appendix A.5.
\end{proof}

\begin{thm}[PC prior for the Mat\'ern ($\rho$, $\sigma$)]
        \label{cor:PC}
        Let $u$ be a GRF defined on $\mathbb{R}^d$, where $d\leq 3$, with a Mat\'ern
        covariance function with parameters $\sigma$, $\rho$ and $\nu$. Then the joint PC prior 
        corresponding to a base model with infinite range and zero variance is
        \[
                \pi(\sigma, \rho) = \frac{d}{2}\tilde{\lambda}_1 \tilde{\lambda}_2 \rho^{-d/2-1} \exp(-\tilde{\lambda}_1 \rho^{-d/2} - \tilde{\lambda}_2 \sigma), \quad \sigma >0, \rho > 0,
        \]
        where  \(\mathrm{P}(\rho < \rho_0) = \alpha_1
        \) and \(\mathrm{P}(\sigma > \sigma_0) = \alpha_2\) are achieved by
        \[
                \tilde{\lambda}_1 = -\log(\alpha_1) \rho_0^{d/2} \quad \text{and} \quad \tilde{\lambda}_2 = -\frac{\log(\alpha_2)}{\sigma_0}.
        \]
\end{thm}
\begin{proof}
        See Appendix A.6.
\end{proof}

The prior is easy and fast to compute regardless of the number of observations
 and $d = 2$ provides the two-dimensional spatial
case that is used in Sections \ref{sec:SimStudy} and \ref{sec:NonStat}.

\subsection{Restrictions and extensions}
The results derived in the previous section do not hold when 
$d > 4$ since in this case both the range and the marginal variance are consistently estimable under
in-fill asymptotics and it is not possible to make moves in the parameter space 
for which the KLD is finite. It is unknown whether the results hold for $d = 4$ since
it is an open question whether the parameters can be consistently estimated 
for that case \citep{anderes2010}.
This means that the assumption on the dimension, $d \leq 3$, used to derive the joint prior
is important and cannot be removed.

Most of the technical difficulties in the previous section is caused by the desire
to work with continuously indexed GRFs instead of discretely indexed
observation processes. The benefit 
is that the prior is not dependent on the spatial design, which is a good property
because the GRF also exists on other locations than on those it was observed. 
 In particular, the prior does not need to be changed 
if data is made available at new observation locations and 
the prior is meaningful when predictions are made at a higher resolution than the observed
data or for a larger observation area. In the former case there is
more difference between small ranges than a prior based on the observed locations would
indicate and in the latter case there is a larger difference between large ranges than
a prior based on the observed locations would indicate.

Similarly, if the GRF were assumed to exists only on the area on which the observations
were made, the upper tail behaviour of the prior for the range would be wrong if the posterior
is used to make predictions on a larger domain. A longer discussion is provided
 in Appendix B, but the short story is: when the
range changes, the properties
of the GRF change even if those changes cannot be detected on the arbitrary observation
locations or observation domain, and the construction of the prior should account for these changes.

In most applications the covariance function is chosen from the 
Mat\'ern family of covariance functions 
and a prior applicable only for this family is of great interest. However,
the approach in the paper could be extended to other isotropic families of covariance
functions that are defined through a marginal variance and a spatial scale. If
the spatial scale is consistently estimable, the techniques in
the paper are not applicable. If the spatial scale is not consistently estimable, 
the main challenge is to know which combination of the parameters that is consistently estimable. 
When this information is known, one can let $\kappa$ be the spatial
scale and let $\tau$ be the consistently estimable parameter, and one can
likely use a similar proof as in this paper. However, it is, in general, not known which parameters are consistently
estimable for different families of covariance functions
and it is outside the scope of this paper to go
investigate further.

\section{Simulation study}
\label{sec:SimStudy}
The series of papers on reference priors for GRFs starting with \citet{Berger2001} 
evaluated the priors by studying frequentist properties of the resulting 
Bayesian inference. A prior intended for use as a default prior should lead to
good frequentist properties such as 
frequentist coverage of the equal-tailed \(100(1-\alpha)\%\) Bayesian
credible intervals that is close to the nominal \(100(1-\alpha)\%\).  In this
paper, the study is replicated with one key difference: no covariates are included.
This choice is made because the PC prior is derived for a zero-mean GRF, and
if a mean were desired, it would be handled by extending the hierarchical model
with another latent component that had its own, separate prior.
Without covariates the reference prior approach results in the Jeffreys' rule
prior as there are no 
nuisance parameters to integrate out when constructing the spatial reference prior.
Furthermore, we compute the $100(1-\alpha)\%$ highest posterior density (HPD) 
intervals \citep{HPDinterval}
to investigate whether skewness of the posteriors result in substantially
different conclusions for HPD credible intervals compared to quantile-based credible intervals.

We start by selecting 25 locations, 
$\boldsymbol{s}_1, \boldsymbol{s}_2, \ldots, \boldsymbol{s}_{25}$, at random
in $[0,1]^2$ and generate realizations,
$\boldsymbol{u} = (u(\boldsymbol{s}_1), u(\boldsymbol{s_2}), \ldots, u(\boldsymbol{s}_{25}))$, 
using a GRF with an exponential covariance
function \(c(r) = \exp(-2r/R_0)\) for true ranges $R_0 = 0.1$ and 
$R_0 = 1$. The data is then fitted using a GRF with the exponential covariance function
\(c(r) = \sigma^2\exp(-2r/\rho)\), where the unknown parameters are marginal variance $\sigma^2$ and
range $\rho$. Four priors are considered:
the PC prior (PriorPC), the Jeffreys' rule prior (PriorJe), and the 
Jeffreys prior for variance combined with a bounded uniform prior on range (PriorUn1) and
a bounded uniform prior on the logarithm of range (PriorUn2). The most important and interesting
results are presented in this section, while the full details of the simulation study
are provided in Appendix D.

We begin with a general discussion on the differences in results observed between 
quantile-based credible intervals and HPD credible intervals, and then
proceed with discussion about specific results.
In general, the marginal posteriors are highly skew and the HPD credible intervals
are substantially shorter than the equal-tailed credible intervals,
but comparisions of average lengths remain consistent between the two approaches
because the relative differences are similar. Further, the coverage was further
away from the nominal level for the HPD credible intervals than the quantile-based credible
intervals for PriorJe and PriorPC, and the coverage of the credible intervals
was more sensitive to the hyperparameters of PriorPC for HPD credible intervals than for
quantile-based credible intervals. The coverage of the 
HPD credible intervals was closer to the nominal level than the quantile-based credible 
intervals for PriorUn1 and PriorUn2,
but since our main focus are PriorPC and PriorJe we use the equal-tailed 95\% credible
intervals in what follows.

\begin{figure}
        \centering
        \includegraphics[width=7cm]{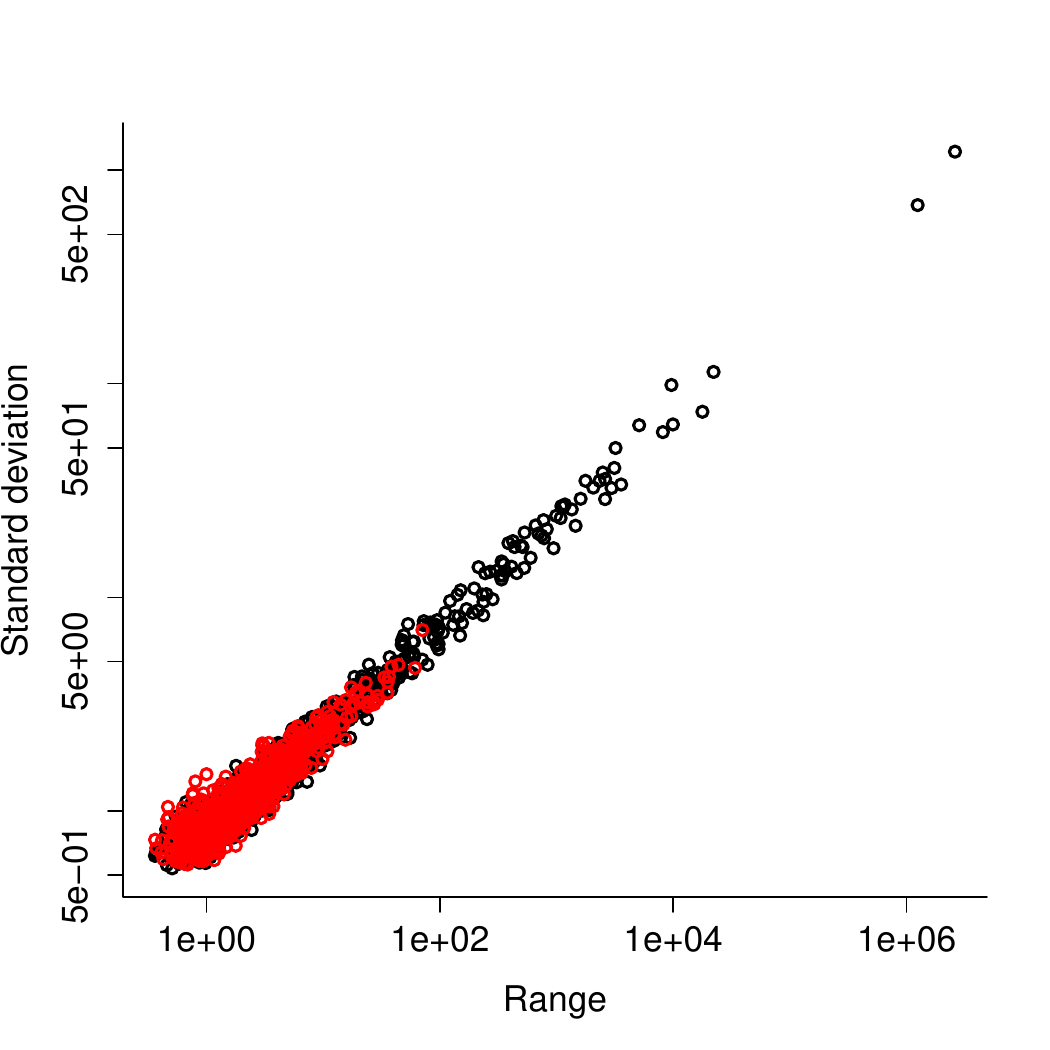}
        \caption{Samples from the joint posterior of range and marginal 
                 standard deviation. The red circles are samples using the 
                 PC-prior and the black circles are samples using the 
                 Jeffreys' rule prior.}
        \label{fig:jointPostMain}
\end{figure}

First, one observation with true range equal to $1.0$ is selected and the model
is fitted with PriorJe, and with PriorPC with hyperparameters 
selected such that $\mathrm{P}(\rho < 0.1) = 0.05$ and
$\mathrm{P}(\sigma > 10) = 0.05$. The latter corresponds to a probability
of $0.025$ that the value of the GRF at an arbitrary location will exceed
$10$. The resulting samples from the posterior are shown
in Figure \ref{fig:jointPostMain} and the figure shows that when PriorJe is used, the
MCMC sampler explores areas far out in the tail, whereas when
PriorPC is used, the prior restricts the movement away from the upper tail. This
means that when prior knowledge is available, PriorPC can be used to
achieve credible intervals that are more reasonable.

Second, we study the sensitivity of the coverage and the lengths
of the credible intervals to the choice of the hyperparameters in PriorPC and
look for general guidelines for selecting the hyperparameters.  
We choose to set the hyperparameters in PriorPC through 
$\mathrm{P}(\rho < \rho_0) = 0.05$ and $\mathrm{P}(\sigma > \sigma_0) = 0.05$.
The results show that choosing $\sigma_0$ lower than the true standard
deviation or $\rho_0$ higher than the true range results in too low 
coverage for both the marginal variance and the range.
Selecting $\sigma_0$ to be $2.5$, $10$ or $40$ times the true standard
deviation and $\rho_0$ to be $1/10$ or $1/2.5$ times the true range 
results in good coverage both for the marginal variance and the
range for both values of the true range. Selecting $\rho_0$ to be $1/40$ times the true range degrades
the coverage for the range when the true range is $0.1$, but leads
to good coverage when the true range is $1.0$, while the coverage for
the marginal variance is good for both values of the true
range. Thus the study indicates that 
good coverage properties are achieved
when $\sigma_0$ is selected between $2.5$ to $40$ times the
true standard deviation and $\rho_0$ is set to between $1/10$ and $1/2.5$
times the true range. Further, shorter credible intervals are achieved
for smaller values of $\sigma_0$ and 
smaller values of $\rho_0$, and for the values tested the best balance between
good coverage and shortest lengths of the credible intervals is achieved
for $\sigma_0$ equal to $2.5$ times the true standard deviation and $\rho_0$ equal to
$1/10$ times the true range.

Third, we compare the properties when using PriorPC, PriorJe, PriorUn1 and
PriorUn2. PriorJe results in 98.3\% coverage with 
average length of the credible intervals of 0.78 for range and 96.7\% coverage and average length of the credible
intervals of 2.6 for marginal variance for true range \(R_0= 0.1\), and 95.6\% coverage with average
length of the credible intervals of 376 for range and 95.6\% coverage with average length of
the credible intervals of 295 for variance for \(R_0 = 1\). The lengths of the credible
intervals are shorter when using PriorPC with the hyperparameters suggested
in the previous paragraph than when using PriorJe. The average lengths of the
credible intervals are around $1.4$ and $3.1$ for marginal variance for
true ranges $0.1$ and $1.0$, respectively. Note that the use of HPD intervals significantly
reduces the average length of the credible intervals for range and marginal variance for PriorJe
to 95 and 75, respectively, but they are still long and there are no hyperparameters that
can be used to reduce them.
For PriorUn1 the coverage and average lengths of the credible intervals are
sensitive to the upper limit on
range, and for PriorUn2 the coverage is good and has little sensitivity to the lower
and upper limit on range, while the average lengths of the credible intervals are
sensitive to the upper limit.

Fourth, we investigate whether the behaviour found for PriorPC changes
when the observation process is changed to a less informative observation
process. For each realization with true range $R_0 = 0.1$ probabilities
are calculated through a probit link, $\mathrm{probit}(p_i) = u(\boldsymbol{s}_i)$,
and binomial data is simulated using $y_i | p_i \sim \mathrm{Binomial}(20, p_i)$.
The data is then fitted using the true logistic spatial regression model and
the coverage and average lengths of the credible intervals are estimated
for marginal variance and range. The results show that
the properties found using direct observations of the spatial field also
holds for the spatial logistic regression, and the only significant difference
is that the average lengths of the credible intervals are larger.

Overall, the simulation study shows that 
with respect to computation time and ease of use versus coverage and lengths of
the credible intervals PriorUn2 and PriorPC appear to be the best choices. If coverage
is the only concern, PriorUn2 performs the best, but if one also wants to control
the length of the credible intervals by disallowing unreasonably high variances,
PriorPC offers the most interpretable alternative. Furthermore, choosing the optimal values
for $\sigma_0$ and $\rho_0$ or missing the optimal
values by less than one order provides good coverage
and lengths of the credible intervals.

\section{Example: Extending to non-stationarity}
\label{sec:NonStat}
Neither stationary nor non-stationary GRFs provide true representations of
reality, but the extra flexibility in the covariance structure of a 
non-stationary GRF may provide a better fit to the data 
than a stationary GRF. Therefore,
we consider how to extend the prior for the stationary model to a prior
for a non-stationary model, with the goal of improving predictions,
using a dataset of annual precipitation. The
details are technical and can be found in Appendix G, but
this section provides a condensed version.

\begin{figure}
        \centering
        \subfloat[Observation]{
          \includegraphics[height = 7cm]{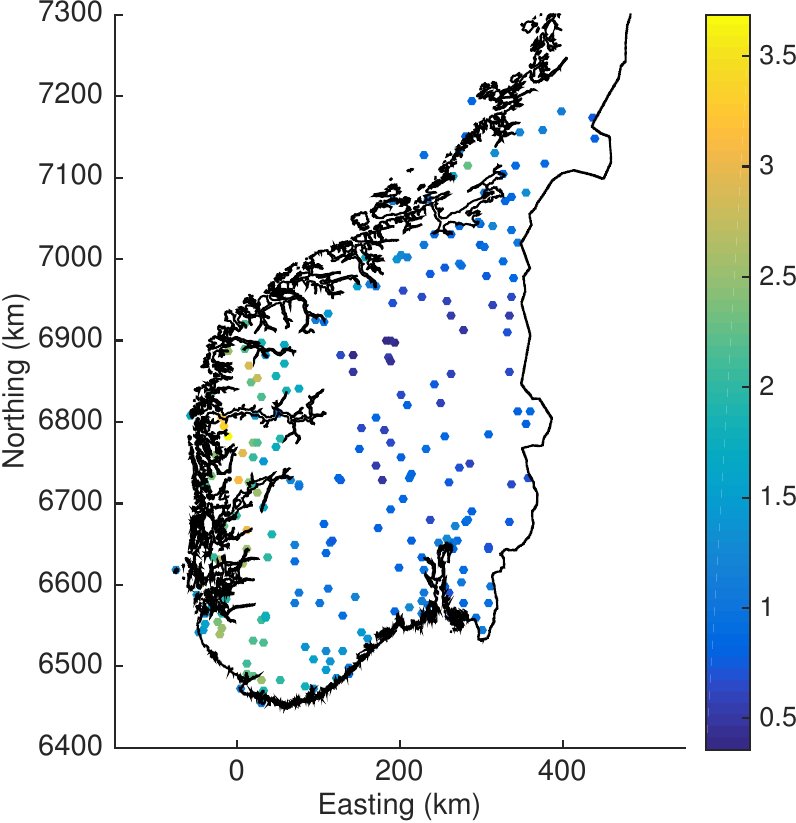}
        }
        \subfloat[Prediction with non-stationary model]{
          \includegraphics[height = 7.5cm]{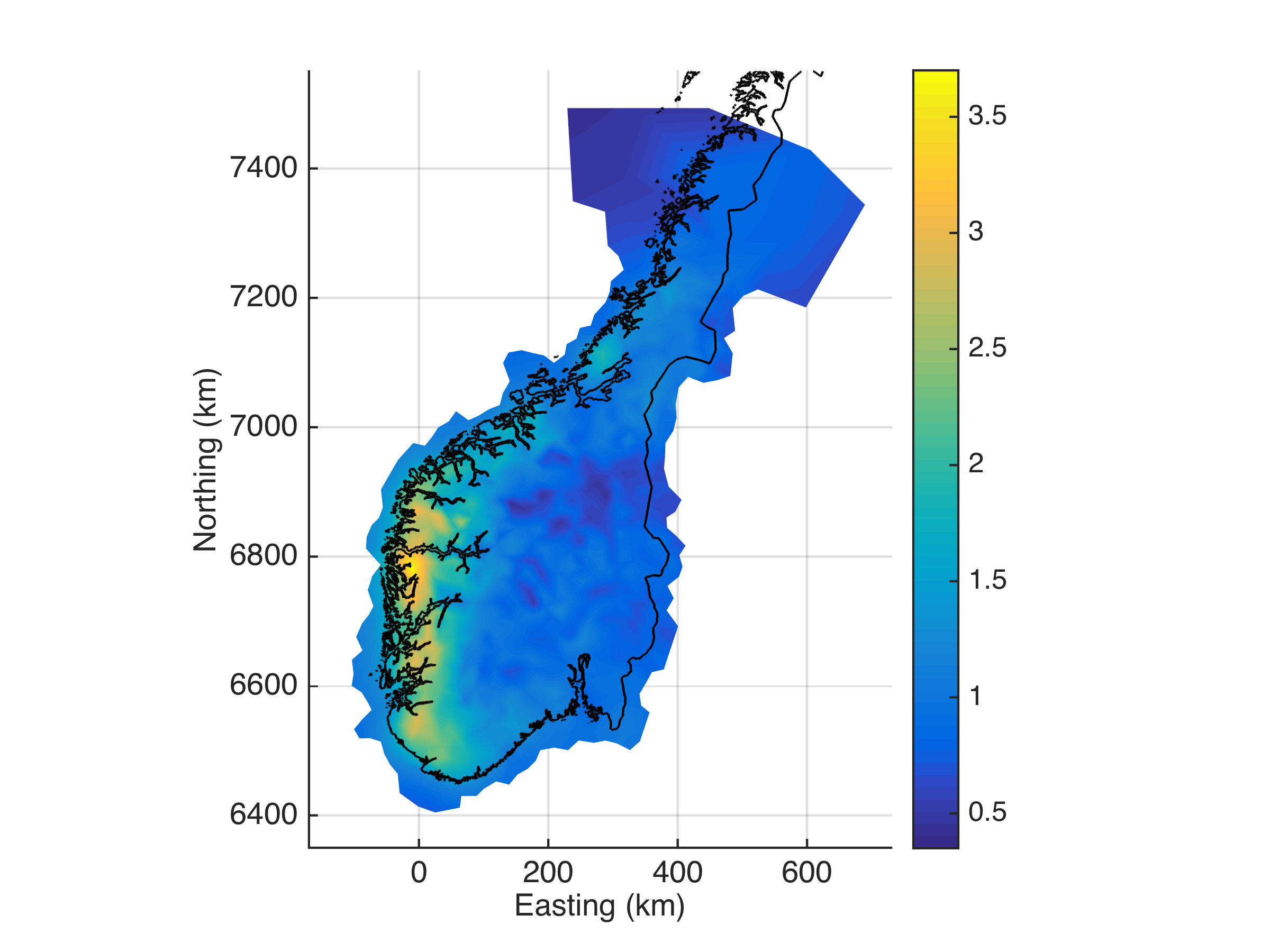}
        }
        \caption{Total precipitation for the one year period September
                         1, 2008, to August 31, 2009, for 233 measurement 
                         stations in southern Norway measured in meters in a) and predictions
                         from the non-stationary model in b). Coordinate system is UTM33.}
        \label{fig:NorwayObs}
\end{figure}

We use a dataset consisting of total annual precipitation for the one year period
September 1, 2008, to August 31, 2009, for the 233 measurement stations in southern
Norway shown in Figure~\ref{fig:NorwayObs}. The dataset has previously been used by
\citet{ingebrigtsen2014spatial,ingebrigtsen2014estimation} to study
the use of elevation as a covariate 
in the covariance structure and associated priors. They used an intercept
and a linear effect of the elevations of the stations in the first-order structure and used
the elevation as a covariate in the second-order structure. We will follow their choice of
covariates in the first-order structure, but use two covariates in the second-order structure:
elevation and the  magnitude of the gradient of the 
elevation.

\begin{figure}
        \centering
        \subfloat[Elevation (km)]{
                \includegraphics[width = 7.7cm]{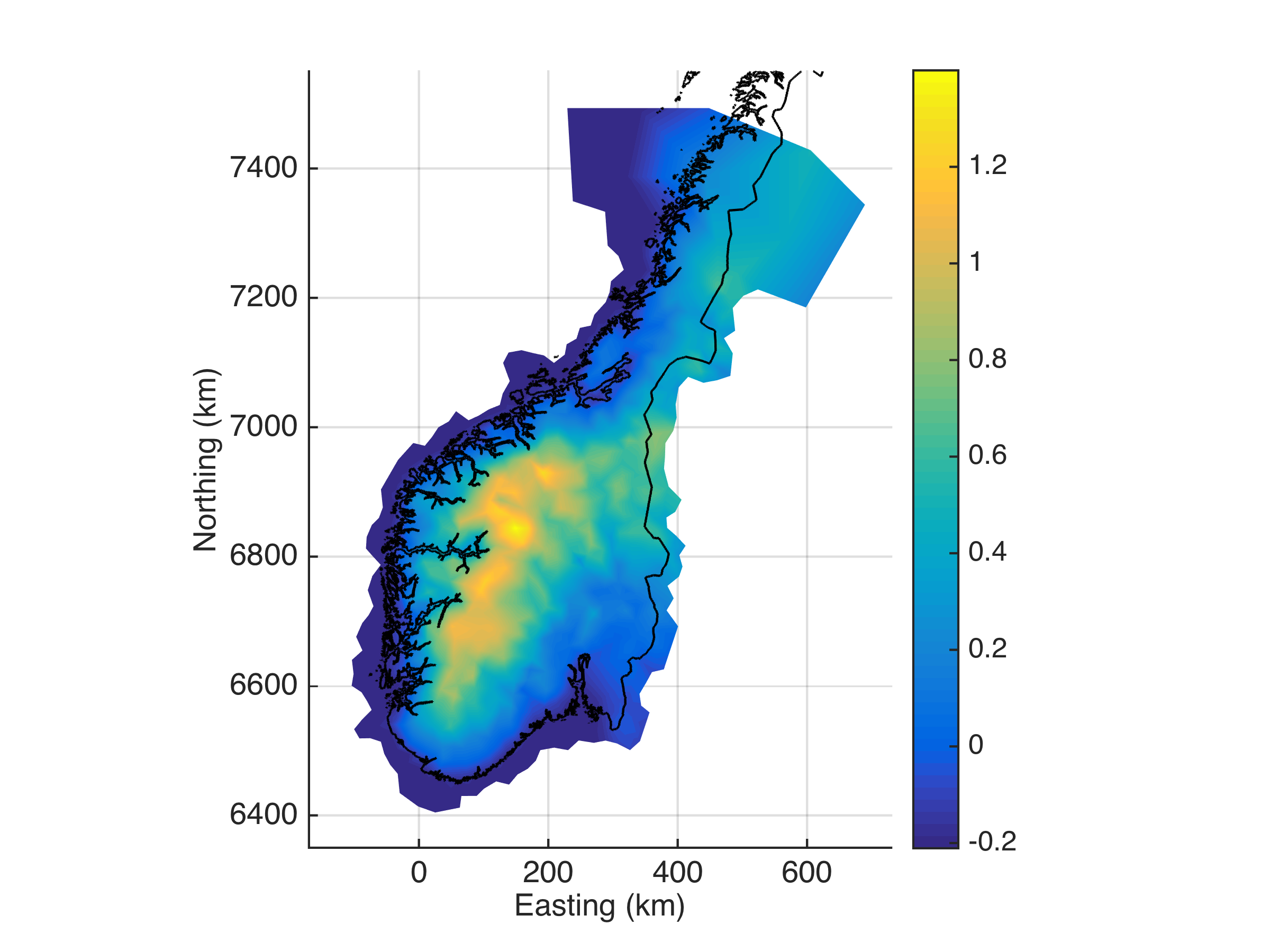}
                \label{fig:covCov:elevMain}
        }
        \subfloat[Magnitude of gradient (100m/km)]{
                \includegraphics[width = 7.7cm]{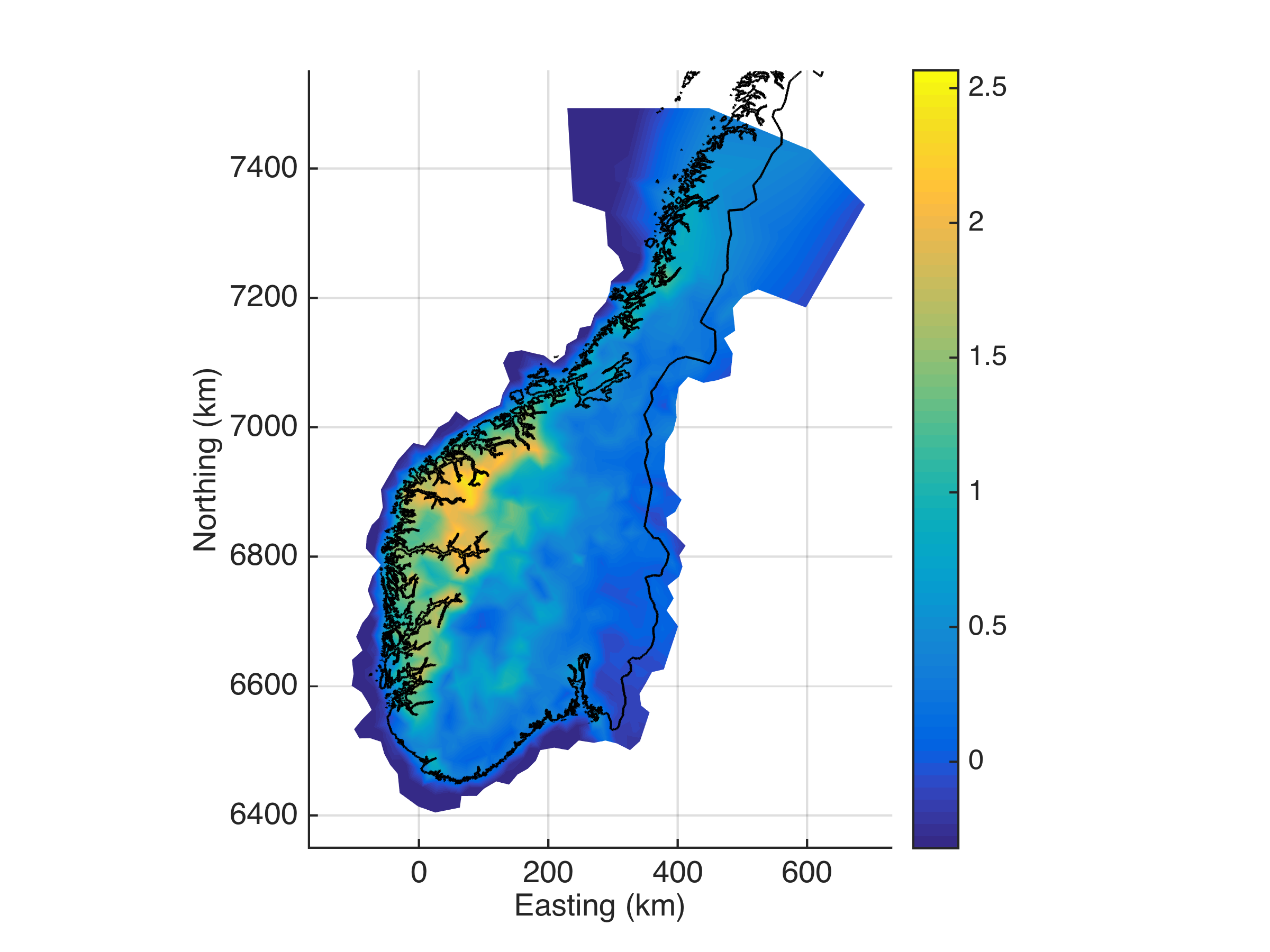}
                \label{fig:covCov:gradMain}
        }
        \caption{The covariates \protect\subref{fig:covCov:elevMain} elevation and
                 \protect\subref{fig:covCov:gradMain} magnitude of the gradient used 
                 for the covariance structure.}
        \label{fig:covCovMain}
\end{figure}

We use the simple geostatistical model
\begin{equation}
        y_i = \beta_0 + x_i\beta_1 + u(\boldsymbol{s}_i) + \epsilon_i, \qquad i = 1, 2, \ldots, 233,
        \label{eq:example:model}
\end{equation}
where for station \(i\), \(y_i\) is the observation made at location \(\boldsymbol{s}_i\), 
\(x_i\) is the elevation of the station, \((\beta_0, \beta_1)\) are the coefficients of the
fixed effects, \(u(\cdot)\) is the spatial effect, and \(\epsilon_i\) is the nugget effect.
The nuggets are i.i.d. \(\epsilon_i\sim\mathcal{N}(0, \sigma_\mathrm{N}^2)\), and
the spatial effect is constructed with the SPDE approach of \citet{Lindgren2011} and the 
stationary version has two parameters: spatial range \(\rho\) and marginal variance of the
spatial field \(\sigma^2\). The non-stationary version is constructed as shown in Appendix G
 and uses the two covariates shown in   
Figure~\ref{fig:covCovMain} in the second-order structure. The spatial field is
orthogonalized against the intercept and the two covariates in the second-order structure
to avoid confounding between the first-order structure and the second-order structure.

The stationary model uses the PC prior developed in this paper for the spatial field and
the PC prior for precision parameter from \citet{Martins2014} for the precision of
the nugget effect. The hyperparameters are selected to satisfy
$\mathrm{P}(\rho < 10) = 0.05$,
$\mathrm{P}(\sigma > 3) = 0.05$
and $\mathrm{P}(\sigma_\mathrm{n} > 3) = 0.05$,
and the model is fitted to the data with INLA \citep{rue2009approximate}. 
With this prior we consider a standard deviation greater than 3 large for both
the GRF and the nugget effect, and a range less than 10 km
unlikely based on the spatial
scale that we are working on. The MAP estimates are \(\hat{\sigma}_\mathrm{N} = 0.13\), 
\(\hat{\rho} = 219\) and \(\hat{\sigma} = 0.72\), and will be used
in our scheme for setting the 
hyperparameters in the prior for the non-stationarity.

The non-stationarity is described by a function $R(\cdot)$ that
describes how the local range varies and a function $S(\cdot)$ that
describes how the marginal variance varies. The two covariates in
the second-order linear structure enters linearly in $\log(R(\cdot))$
and $\log(S(\cdot))$, and 
the coefficients, \(\boldsymbol{\theta}_1\), of the two linear covariates in 
\(\log(R(\cdot))\) are given the prior
\begin{align*}
        \boldsymbol{\theta}_1 | \tau_1 &\sim \mathcal{N}(\boldsymbol{0}, S_1/\sqrt{\tau_1}) \\
        \tau_1 &\sim \frac{\lambda_1}{2}\tau_1^{-3/2}\mathrm{e}^{-\lambda_1/\sqrt{\tau_1}} 
        \end{align*}
and the coefficients, \(\boldsymbol{\theta}_2\), of the two linear covariates in 
\(\log(S(\cdot))\) are given a similar prior, but with hyperparameter
\(\lambda_2\). Further details are found in Appendix G. 

The hyperparameters $\lambda_1$ and $\lambda_2$ are
selected based on the frequentist coverages of the non-stationarity parameters when
fitting the non-stationary model to stationary data.
Specifically, we use the MAP estimates of the stationary model to simulate 100
datasets from the stationary model with \(\beta_0 = \beta_1 = 0\), set values
for the hyperparameters \(\lambda_1\) and \(\lambda_2\), fit 
a non-stationary model with \(\beta_0 = \beta_1 = 0\) to each of datasets, 
and calculate the frequentist coverage of the the \(95\%\) credible intervals
of the non-stationarity parameters. It is overly expensive to run the
model 100 times and we use a cheaper approximation in INLA that is 
conservative. We tried several values for the hyperparameters
\(\lambda_1\) and \(\lambda_2\) and found that \(\lambda_1 = \lambda_2 = 20 \)
provides coverage that is close to the nominal \(95\%\) for \(\boldsymbol{\theta}_1\)
and \(\boldsymbol{\theta}_2\). 

The non-stationary model was then fitted using an MCMC sampler and 
the resulting posterior means of the range and the standard deviation are
shown in Appendix G, and they are not included here
since the focus is on improving predictions. The figures show that the non-stationary model moves away from the
stationary model even under the conservative prior.

The leave-one-out log-score is estimated
from the samples of the MCMC sampler, and we find the score 0.13 for the
stationary model and 0.22 for the non-stationary model. 
The leave-one-out estimates for
the continuous rank probability score (CRPS) are
0.092 for the stationary model and 0.083 for the non-stationary model.
Experimentation with the strictness of the prior showed that further improvements 
were possible by making the prior weaker, but that making the prior too weak leads
to worse scores. The prior and the procedure for selecting the hyperparameters appears
to introduce a reasonable level of conservativeness for this dataset. 

If we run the model with the same hyperparameters and
remove the non-stationarity in the local range, the CRPS is 0.086, and if we remove
the non-stationarity in the marginal standard deviations,
the CRPS is 0.081. This shows that the covariates in the local range appear to be contributing
more to the improved predictions than the covariates in the standard deviation, and that
using all four covariates has degraded the performance slightly compared to using only
a non-stationary local range. This demonstrates that guaranteeing improvements when
including more covariates in the second-order structure is difficult. So
a procedure for constructing conservative priors are critically important for non-stationary
models, but the prediction scores of the models must be compared to ensure that the
non-stationary model improves the predictions.

\section{Discussion} 
\label{sec:Discussion}
The main challenge for constructing multivariate PC priors based on
a measure of distance from a base model is that a joint prior for the parameters
cannot be uniquely determined from a prior for
the distance. \citet{Martins2014} present a general
approach where conditional on the value of the distance, $D$, the probability density is uniformly 
distributed on the set of parameters that specify models that are a distance
$D$ from the base model, but the approach is not parametrization-invariant
and it is not clear for which parametrization of range and marginal variance
that this approach would be appropriate. However, in this paper we have shown
that the key for properly extending the PC prior framework to a joint prior
for range and marginal variance in Mat\'ern GRFs 
with fixed smoothness is to 
use knowledge about the parameter space to split the construction of the 
multivariate PC prior into a sequential construction of univariate 
PC priors. This demonstrates
that the principles of the PC prior framework are applicable for model components
with complex parameter spaces that contain intrinsically linked parameters, but 
that the simple idea of distance
from a base model must be combined with
careful consideration of the parameter space.

The construction of the joint prior based on the 
infinite-dimensional distribution of the GRF instead of the finite-dimensional
 distribution of an observation from the GRF is technically more challenging that the 
finite-dimensional examples in \citet{Martins2014}. But the calculation of the KLD can be handled using the spectrum
of the GRF and the fact that the KLD
is infinite for general changes in the parameters can be
overcome by careful reparametrization and a sequential construction of the prior.
The benefits gained from the extra difficulty are that the PC prior for
 Mat\'ern GRFs with fixed smoothness and the extension to the non-stationary GRF are computationally inexpensive
since they have simple forms, are appropriate for hierarchical models since they work with
any observation process, and can be applied for sequential analysis of data since they 
do not depend on the design of the experiment. 

Setting the hyperparameters for the stationary part of the model can be done
based on statements about what constitutes a large standard deviation or
a large deviation from zero for the spatial field, and
what constitutes a small range. This allows the users to choose to limit
the preference for intrinsic models 
 and thus provide
more sensible posterior inference for the problem at hand. In the simulation study
we observe
good coverage of the equal-tailed 95\% credible intervals
when the prior satisfies $\mathrm{P}(\sigma > \sigma_0) = 0.05$
and $\mathrm{P}(\rho < \rho_0) = 0.05$, where $\sigma_0$ is between $2.5$
to $40$ times the true marginal standard deviation and $\rho_0$ is between
$1/10$ and $1/2.5$ of the true range. The lengths of the credible intervals
depend on the values chosen for $\sigma_0$ and $\rho_0$, but are shorter
than for the reference prior and consistent with the information put 
into the prior. The recommendations are based on the quantile-based
credible intervals because the coverage of the 95\% HPD credible intervals is
further away from the nominal level and more sensitive to hyperparameters
than the equal-tailed 95\% credible intervals when the PC prior is used.

It is difficult to elicit expert knowledge about the hyperparameters for 
a non-stationary GRF since the second-order structure is not observed directly, and 
we discuss an alternative way to set the hyperparameters based on the frequentist coverage
of the credible intervals. Using the new prior and the associated scheme for selecting the 
hyperparameters, we find a better fit for the non-stationary GRF than with the stationary GRF
when applied to the dataset of annual precipitation in 
southern Norway measured both with leave-one-out CRPS and log scores.

The paper shows that the PC prior framework provides a useful tool for deriving
a principled joint prior for the range and the marginal variance of
a Mat\'ern GRF with fixed smoothness, and that the ideas of
the framework are useful for constructing priors that limit flexibility
also for non-stationary GRFs where exact derivation is not possible.

\appendix
\section{Proofs}
    \subsection{Theorem 2.1}
        \label{app:PC_tauIkappa}
        \begin{proof}
            For a fixed value of  \(\kappa\), the covariance
            matrix of \(\boldsymbol{u} = (u(\boldsymbol{s}_1, u(\boldsymbol{s}_2), \ldots, u(\boldsymbol{s}_n))\) is
            \(\Sigma(\tau) = \tau^2 \Sigma_0\), where $\Sigma_0$ depends
            on the values of \(\kappa\) and $\nu$ and the locations at which the process is observed. This means that
            \(
                    \boldsymbol{u} | \tau, \kappa \sim \mathcal{N}_n(\boldsymbol{0}, \tau^2 \Sigma_0),
            \)
            and we need to derive the PC prior for the scale parameter of a multivariate Gaussian distribution 
            with
            the base model \(\tau = 0\). 

            This can be formulated as constructing the PC prior for a precision parameter, which
            was done in \citet[Appendix A.2]{Martins2014}, and a transformation to a scale parameter results in
            \(\pi(\tau | \kappa) = \lambda \exp(-\lambda \tau)\), for \(\tau > 0\),
            where $\lambda > 0$.
        \end{proof}

    \subsection{Theorem 2.2}
    \label{app:PCtau_P}
    \begin{proof}
        The marginal standard deviation is given by $\sigma = \tau \kappa^{-\nu}C^{-1}$, where 
        \[
            C = \sqrt{\frac{\Gamma(\nu + d/2)(4\pi)^{d/2}}{\Gamma(\nu)}}.
        \]
        The probability
        $\mathrm{P}(\sigma > \sigma_0|\kappa) = \alpha$ is equivalent to
        $\mathrm{P}(\tau > \sigma_0\kappa^{\nu}C|\kappa) = \alpha$ and under the
        prior distribution this leads to
        \begin{align*}
                \exp(-\lambda \sigma_0\kappa^\nu C) &= \alpha \\
                \lambda &= -\kappa^{-\nu}\sqrt{\frac{\Gamma(\nu)}{\Gamma(\nu + d/2)(4\pi)^{d/2}}} \frac{\log(\alpha)}{\sigma_0}.
        \end{align*}
    \end{proof}

    \subsection{Theorem 2.3}
        \label{app:proof_kap_unb}
        \begin{proof}
                Restrict u to the subset $[0, L]^d \subset \mathbb{R}^d$ and let
                $\kappa =  \kappa_0 > 0$ denote the base model. Let $1/L = o(\kappa_0)$, then
                the covariance function on $[0,L]^d$ is, for small $\kappa_0$, well approximated by
                \[
                        c(\boldsymbol{s}, \boldsymbol{t}) = \sum_{\boldsymbol{w}\in \frac{2\pi}{L}\mathbb{Z}^d} f_\nu(\boldsymbol{w};\kappa, \tau) \exp\left(-\imath\langle \boldsymbol{w}, \boldsymbol{s}-\boldsymbol{t}\rangle\right) \left(\frac{2\pi}{L}\right)^d,
                \]
                where 
                $f_\nu(\boldsymbol{w}; \kappa, \tau) = ({2\pi})^{-d} \tau^2(\kappa^2 +\vert\vert\boldsymbol{w}\vert\vert^2)^{-(\nu+d/2)}$ is
                the spectral density of a Mat\'ern GRF on $\mathbb{R}^d$ with parameters
                $\tau$, $\kappa$ and $\nu$ (See \cite{Lindgren2011}). 
                 Further, the KLD for
                the periodic approximation from 
                \(\kappa_0\) to \(\kappa\), for a fixed \(\tau\), is (based on~\citet[Thm. 6.4.6]{bogachev1998gaussian})
                \begin{align*}
                        \mathrm{KL}(\kappa, \kappa_0) &= \frac{1}{2}\sum_{\boldsymbol{w} \in \frac{2\pi}{L}\mathbb{Z}^d} \left[\frac{f_\nu(\boldsymbol{w}; \kappa, \tau)}{f_\nu(\boldsymbol{w}; \kappa_0, \tau)} - 1 - \log \frac{f_\nu(\boldsymbol{w}; \kappa, \tau)}{f_\nu(\boldsymbol{w}; \kappa_0, \tau)}\right] \\
                        &= \frac{1}{2}\sum_{\boldsymbol{w} \in \frac{2\pi}{L}\mathbb{Z}^d} \left[\frac{(\kappa_0^2 + ||\boldsymbol{w}||^2)^\alpha}{(\kappa^2 +||\boldsymbol{w}||^2)^\alpha} - 1 - \log \frac{(\kappa_0^2 + ||\boldsymbol{w}||^2)^\alpha}{(\kappa^2 + ||\boldsymbol{w}||^2)^\alpha}\right],
                \end{align*}
                where $\alpha = \nu + d/2$.

                The sum can be divided in two parts: the zero frequency $E_0$ and the other
                frequencies $E_1$. The zero frequency term is
                \[
                        E_0 = \frac{1}{2}\left[\left(\frac{\kappa_0^2}{\kappa^2}\right)^\alpha - 1 - \log\left(\frac{\kappa_0^2}{\kappa^2}\right)^\alpha\right]
                \]
                and the remaining terms are
                \begin{align*}
                        E_1 &= \frac{1}{2}\left(\frac{L\kappa}{2\pi}\right)^d \sum_{\boldsymbol{w} \in \frac{2\pi}{L\kappa}\mathbb{Z}^d, \boldsymbol{w}\neq 0} \left[\frac{((\kappa_0/\kappa)^2 + ||\boldsymbol{w}||^2)^\alpha}{(1 + ||\boldsymbol{w}||^2)^\alpha} - 1 - \log \frac{((\kappa_0/\kappa)^2 + ||\boldsymbol{w}||^2)^\alpha}{(1 + ||\boldsymbol{w}||^2)^\alpha}\right]\left(\frac{2\pi}{L\kappa}\right)^d \\
                        &=  \frac{1}{2}\left(\frac{L\kappa}{2\pi}\right)^d \left\{\int_{\mathbb{R}^d} \left[\frac{((\kappa_0/\kappa)^2 + ||\boldsymbol{w}||^2)^\alpha}{(1 + ||\boldsymbol{w}||^2)^\alpha} - 1 - \log \frac{((\kappa_0/\kappa)^2 + ||\boldsymbol{w}||^2)^\alpha}{(1 + ||\boldsymbol{w}||^2)^\alpha}\right] \mathrm{d}\boldsymbol{w} + o(1)\right\}
                \end{align*}
                as $\kappa_0 \rightarrow 0$. For any fixed value of $L$, we may include $(L/(2\pi))^d$ in the hyperparameter $\lambda$ in the
                PC prior. Thus we consider the rescaled terms
                \[
                        \tilde{E}_0 = \frac{1}{2} \left(\frac{2\pi}{L}\right)^d\left[\left(\frac{\kappa_0^2}{\kappa}\right)^\alpha - 1 - \log\left(\frac{\kappa_0^2}{\kappa}\right)^\alpha\right]
                \]
                and
                \[
                        \tilde{E}_1 =  \frac{1}{2} \kappa^d \left\{\int_{\mathbb{R}^d} \left[\frac{((\kappa_0/\kappa)^2 + ||\boldsymbol{w}||^2)^\alpha}{(1 + ||\boldsymbol{w}||^2)^\alpha} - 1 - \log \frac{((\kappa_0/\kappa)^2 + ||\boldsymbol{w}||^2)^\alpha}{(1 + ||\boldsymbol{w}||^2)^\alpha}\right] \mathrm{d}\boldsymbol{w} + o(1)\right\}.
                \]

                Let $\kappa_0\rightarrow 0$ with $1/L = o(\kappa_0)$, then $\tilde{E}_0 \rightarrow 0$ and, for $d \leq 3$,
                \[
                        \tilde{E}_1 \rightarrow  \frac{1}{2}\kappa^d \int_{\mathbb{R}^d} \left[\left(\frac{||\boldsymbol{w}||^2}{(1 + ||\boldsymbol{w}||^2)}\right)^\alpha - 1 - \log \left(\frac{||\boldsymbol{w}||^2}{(1 + ||\boldsymbol{w}||^2)}\right)^\alpha\right] \mathrm{d}\boldsymbol{w} = \frac{1}{2} C_\alpha \kappa^d,
                \]
                where the finiteness of the integral is shown to hold in Appendix H, and 
                $C_\alpha$ is a constant that depends on $\alpha$ and $d$.

                The distance from the base model is $\tilde{\text{dist}}(\kappa) = \sqrt{2\cdot \frac{1}{2}C_\alpha\kappa^d}$,
                where we can absorb the constants in the hyperparameter of the 
                PC prior, so we choose the distance
                \(
                        \text{dist}(\kappa) = \kappa^{d/2}.
                \)
                The PC prior using an exponential distribution on the distance
                is then
                \begin{align*}
                        \pi(\kappa) &= \lambda \exp(-\lambda \kappa^{d/2}) \frac{\mathrm{d}}{\mathrm{d}\kappa} \kappa^{d/2} = \frac{d \lambda}{2}\kappa^{d/2-1} \exp(-\lambda \kappa^{d/2}), \quad \kappa > 0.
                \end{align*}

        \end{proof}

    \subsection{Theorem 2.4}
    \label{app:kappa_lam}
    \begin{proof}
        The probability $\mathrm{P}(\rho < \rho_0) = \alpha$ is equivalent to
        $\mathrm{P}(\kappa > \sqrt{8\nu}/\rho_0) = \alpha$ and under the prior distribution
        we find
        \begin{align*}
                \exp(-\lambda (\sqrt{8\nu}/\rho_0)^{d/2}) &= \log(\alpha) \\
                \lambda &= -\left(\frac{\rho_0}{\sqrt{8\nu}}\right)^{d/2}\log(\alpha).
        \end{align*}
    \end{proof}

        \subsection{Theorem 2.5}
        \label{app:thmPC}
        \begin{proof}
                Using Theorems 2.1 and 2.3, we find 
                the joint prior
                \begin{align*}
                        \pi(\kappa, \tau) &= \pi(\kappa) \pi(\tau | \kappa) \\
                                                          &= \frac{d}{2}\lambda_1 \kappa^{d/2-1}\exp(-\lambda_1 \kappa^{d/2}) \lambda_2 \exp(-\lambda_2 \tau) \\
                                                          &= \frac{d}{2}\lambda_1\lambda_2\kappa^{d/2-1}\exp(-\lambda_1\kappa^{d/2}-\lambda_2\tau), \quad \tau > 0, \kappa > 0.
                \end{align*}
                And Theorems 2.2 and 2.4 gives
                \[
                        \lambda_1 = -\left(\frac{\rho_0}{\sqrt{8\nu}}\right)^{d/2}\log(\alpha_1) \quad \text{and} \quad \lambda_2 = -\kappa^{-\nu}\sqrt{\frac{\Gamma(\nu)}{\Gamma(\nu + d/2)(4\pi)^{d/2}}} \frac{\log(\alpha_2)}{\sigma_0}.
                \]
        \end{proof}

        \subsection{Theorem 2.6}
        \label{app:PCcor}
        \begin{proof}
                Since there is no dependence of $\rho$ on $\tau$, the change of variables from $(\kappa, \tau)$ to 
                $(\rho, \sigma)$ can be divided in two steps. First, $\rho = \sqrt{8\nu}/\kappa$ so we find
                \begin{align*}
                        \pi(\rho) & = \pi(\kappa = \sqrt{8\nu}/\rho) \left\vert \frac{\mathrm{d}}{\mathrm{d}\rho} \sqrt{8\nu}\rho^{-1}\right\vert \\
                                          & = \frac{d}{2}\lambda_1 (\sqrt{8\nu})^{d/2-1} \rho^{-d/2+1}\exp(-\lambda_1 (\sqrt{8\nu})^{d/2} \rho^{-d/2}) \sqrt{8\nu}\rho^{-2} \\
                                          & = \frac{d}{2} \tilde{\lambda}_1 \rho^{-d/2-1}\exp(-\tilde{\lambda}_1 \rho^{-d/2}), \quad \rho > 0,
                \end{align*}
                where
                \(
                        \tilde{\lambda}_1 = -\log(\alpha_1)\rho_0^{d/2}.
                \)

                Second, $\sigma = \tau \kappa^{-\nu} C^{-1}$, where
                \[
                    C = \sqrt{\frac{\Gamma(\nu + d/2)(4\pi)^{d/2}}{\Gamma(\nu)}}.
                \] 
                Note that conditioning on $\kappa$ is equivalent to conditioning on $\rho$. So
                the density $\pi(\sigma | \rho)$ can be found by
                \begin{align*}
                        \pi(\sigma | \rho) &= \pi(\sigma | \kappa) = \pi(\tau = \sigma\kappa^\nu C | \kappa) \left\vert \frac{\partial}{\partial \sigma} \sigma \kappa^{\nu} C \right\vert \\
                                        &= \lambda_2 \exp(-\lambda_2\sigma\kappa^\nu C) \kappa^\nu C \\
                                        &= \tilde{\lambda}_2 \exp(-\tilde{\lambda}_2 \sigma), \quad \sigma > 0,
                \end{align*}
                where \(\tilde{\lambda}_2 = -\frac{\log(\alpha_2)}{\sigma_0}.
                \)
                So the joint density is
                \[
                        \pi(\rho, \sigma) = \pi(\rho)\pi(\sigma | \rho) = \frac{d}{2}\tilde{\lambda}_1\tilde{\lambda}_2\rho^{-d/2-1}\exp(-\tilde{\lambda}_1\rho^{-d/2}-\tilde{\lambda}_2\sigma), \quad \rho>0, \sigma>0.
                \]
        \end{proof}

\section{Detailed discussion for bounded domains}
\label{sec:boundPrior}

The derivations in the main paper use the assumption that
the size of the domain is large compared to the range of the base model. 
This is a reasonable assumption if the underlying 
GRF exists on a larger domain than the area on which observations have been made
since the prior should be based on the distribution of the GRF
on the domain where it is defined.  However, if 
it is known that the GRF only exists on a bounded domain, it would be reasonable to 
base the derivation instead on the bounded domain and not a larger
ambient domain.

When the parameter $\kappa$ increases, the variance of the process 
increases, but the spread of the observations relative to each other
does not change. Since there is no larger ambient space in which this
effect could be distinguished from adding an intercept to the model, it
is more meaningful to have a base model with finite range.

When the priors are derived based on bounded domains, there will typically not
be any analytic expressions available. One
exception is the exponential covariance function on the one-dimensional domain
$[0, L]$ where the exact expression for the distance between the models
specified by $(\kappa, \tau)$ and $(\kappa_0, \tau)$ can be derived (see Section \ref{sec:derivation}) and is
given by
\begin{equation}
        \text{dist}_\mathrm{1D,exp}(\kappa || \kappa_0) = \sqrt{\frac{\kappa_0}{\kappa} - 1 - \log\left(\frac{\kappa_0}{\kappa}\right) + L\left(\frac{\kappa_0^2}{2\kappa} - \kappa_0 + \frac{\kappa}{2}\right)}.
        \label{eq:exact1D}
\end{equation}
In this case it is clear that the term $\log(\kappa)$ dominates
when $\kappa$ is small if  
$\sqrt{8\nu}/\kappa_0$ is of the same order as $L$ or larger.
In the following example one can see how the prior for $\kappa$
calculated based on this distance differs from the one derived in the previous section

\begin{exmp}[One-dimensional exponential covariance function]
        \label{exmp:1DexpExact}
        Let a GRF with an exponential covariance function be observed on $[0, 1]$. 
        There is a one-to-one correspondence between $\kappa$ and $\rho$, so the 
        distance given in Equation \eqref{eq:exact1D} can be expressed in range by
        \[
                \mathrm{dist}_{\rho_0}(\rho) = \sqrt{\frac{\rho}{\rho_0} - 1 - \log\left(\frac{\rho}{\rho_0}\right) + \sqrt{8\nu}\left(\frac{\rho}{2\rho_0^2} - \frac{1}{\rho_0} + \frac{1}{2\rho}\right)},
        \]
        where $\rho_0$ is the range of the base model. This distance
        results in the prior
        \[
                \pi_1(\rho) = \lambda_1\exp(-\lambda_2 \mathrm{dist}_{\rho_0}(\rho)) \frac{1}{2\mathrm{dist}_{\rho_0}(\rho)}\left(\frac{1}{\rho}-\frac{1}{\rho_0} + \frac{\sqrt{8\nu}}{2}\left[\frac{1}{\rho^2}-\frac{1}{\rho_0^2}\right]\right), \quad 0 < \rho < \rho_0,
        \]
        where $\lambda_1 = -\log(\alpha)/\mathrm{dist}_{\rho_0}(R_0)$ ensures that $\mathrm{P}(\rho < R_0) = \alpha$.

        For comparison, the prior for $\rho$ based on an unbounded domain is 
        \[
                \pi_2(\rho) = \frac{\lambda_2}{2} \rho^{-3/2} \exp(-\lambda_2 \rho^{-1/2}), \quad \rho > 0,
        \]
        where $\lambda_2 = -\log(\alpha)R_0^{1/2}$ ensures that $\mathrm{P}(\rho < R_0) = \alpha$.

        The parameter values $R_0 = 0.05$ and $\alpha = 0.05$ are chosen, and the prior based on the unbounded domain
        and the priors based on the bounded domain for $\rho_0 = 2$ and $\rho_0 = 4$ are shown in Figure \ref{fig:1DExpExact}.
        The figure shows that the two prior constructions are similar for $\rho < 0.25$ for
        $\rho = 2$ and for $\rho < 0.5$ for $\rho_0 = 4$. For higher values of range, $\pi_2$ distributes the
        probability mass over the entire positive line and has a faster decay than $\pi_1$. The priors will
        not correspond to each other in the case that $\rho_0 \rightarrow \infty$. In that case $\pi_1$ will
        have a decay of approximately $1/\rho$ whereas $\pi_2$ has a decay of approximately $\rho^{-3/2}$.

        \begin{figure}
                \centering
                \subfloat[Base model $\rho_0 = 2$]{
                        \includegraphics[width = 7cm]{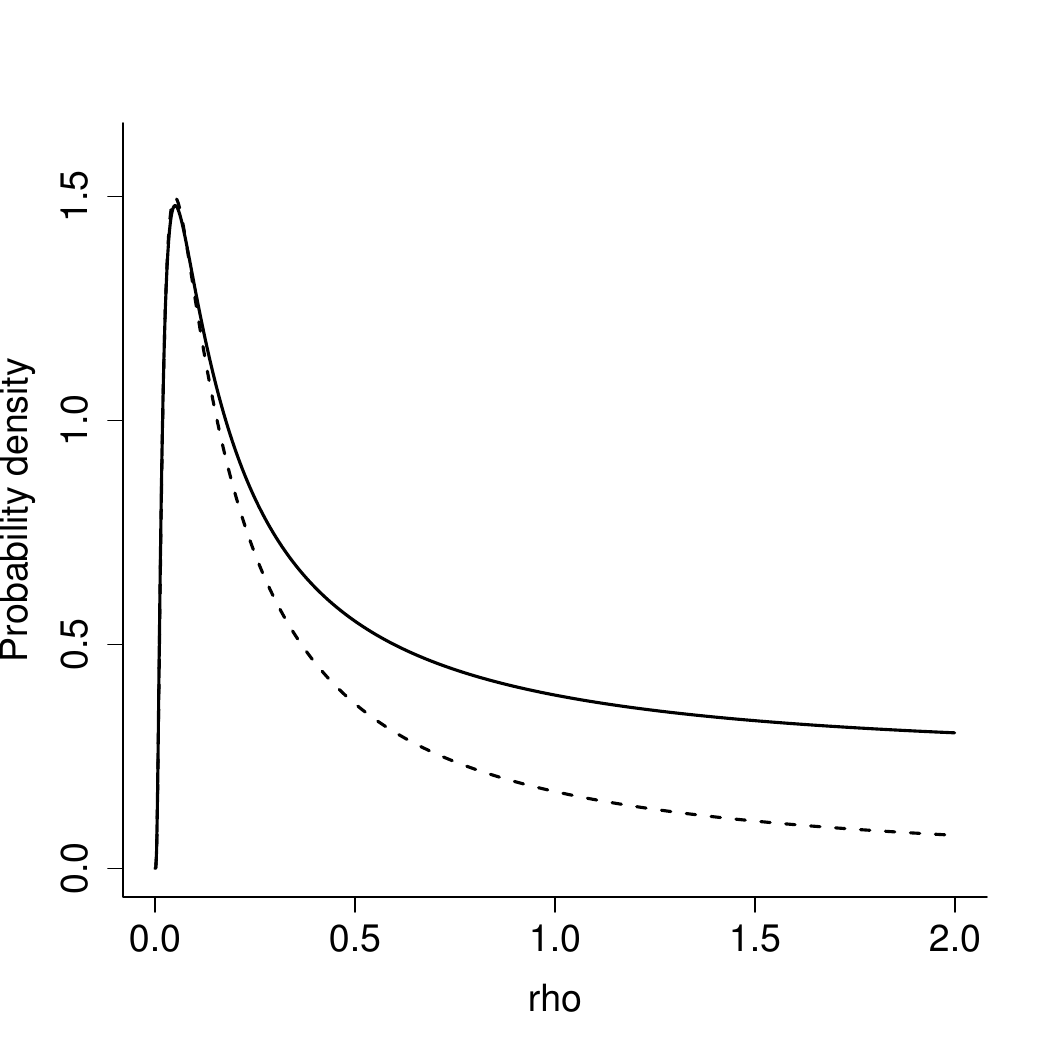}
                        \label{fig:1DExpExact:rho0_2}
                }
                \subfloat[Base model $\rho_0 = 4$]{
                        \includegraphics[width = 7cm]{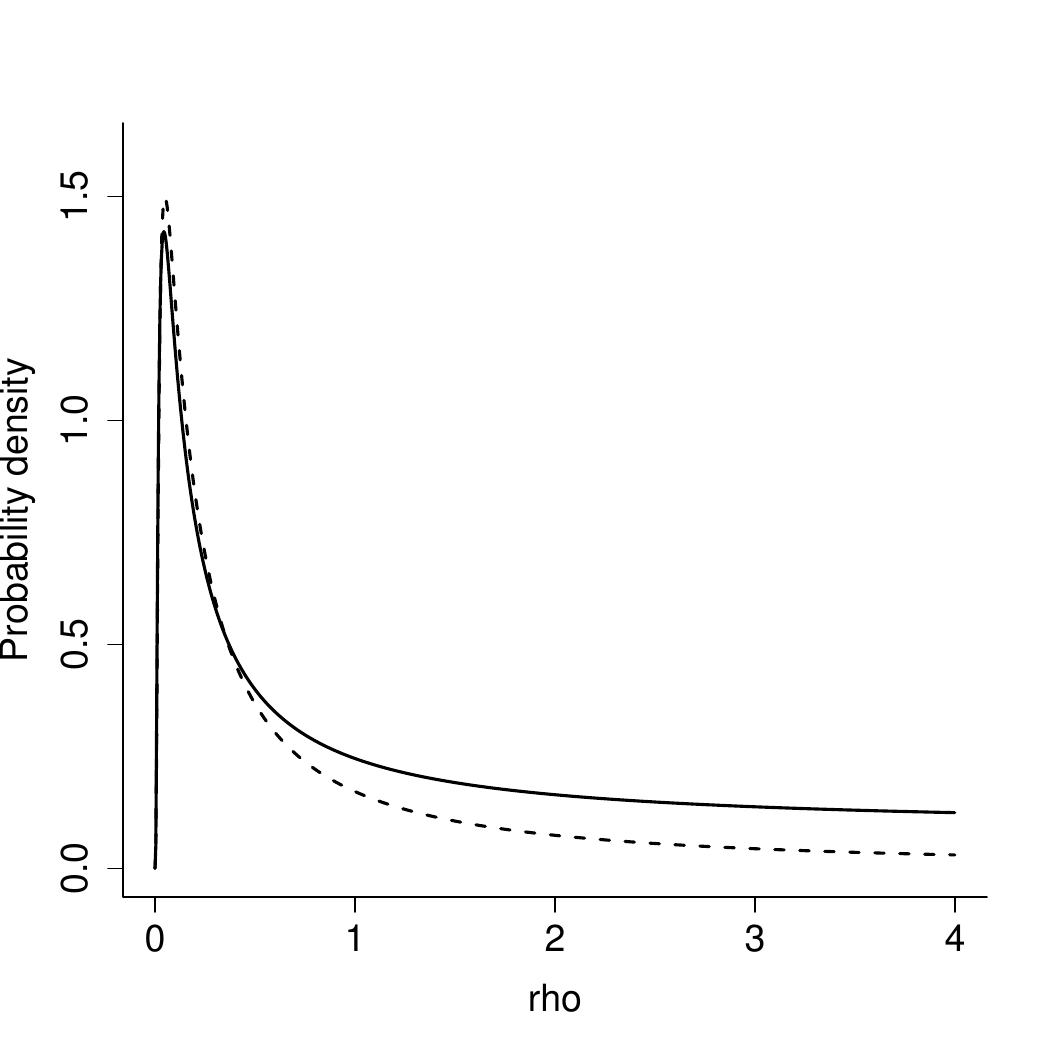}
                        \label{fig:1DExpExact:rho0_4}
                }
                \caption{The PC prior based on the unbounded domain is shown as dashed lines in each
                                 subfigure and the prior based on the bounded domain is shown as solid lines
                                 for base model \protect\subref{fig:1DExpExact:rho0_2} $\rho_0 = 2$ \protect\subref{fig:1DExpExact:rho0_4}
                                 $\rho_0 = 4$. The prior based on the unbounded domain continues to $\rho = \infty$.}
                \label{fig:1DExpExact}
        \end{figure}
\end{exmp}

For other covariance functions and for bounded domains of dimension 2 and 3, the analytic
expressions for the distances are not known, but the priors can be approximated numerically.
The values of the prior on an interval $\kappa \in [A, B]$ can be computed by selecting a
grid of sufficiently dense locations in the domain and then calculating the KLD based
on this finite set of locations. It would be possible to use a fully design-dependent prior 
where the KLD is calculated based on the observation locations in the dataset of interest, but
this provides, potentially, undesired behaviour. Even for a bounded domain one is interested in 
doing predictions outside the observed locations on a much higher resolution and in that case 
using a prior that ignores all properties of the higher resolution process would not be advisable.
If one wants to be able to do predictions at arbitrarily high resolution, 
the prior should be constructed based on the
infinite-dimensional GRF defined on the full bounded domain. The following example demonstrates
how calculations may be done in the two-dimensional case and the consequence of using
a too low resolution in the calculation of the prior.

\begin{exmp}[Two-dimensional Mat\'ern covariance function with $\nu = 3/2$]
        Let a GRF with a Mat\'ern covariance function with smoothness $\nu = 1.5$ be observed on $[0, 1]^2$, and
        let the base model be given by $\rho_0 = 4$. We calculate priors $\pi_1$, $\pi_2$ and $\pi_3$ for
        the range based on a regular grid of $10\times 10$ points, $20\times 20$ points and $40 \times 40$
        points, respectively, and prior $\pi_4$, which is the prior calculated based on an unbounded domain.
        For each prior the hyperparameter is set such that $P(\rho < 0.05) = 0.05$.

        The calculated priors are shown in Figure \ref{fig:2DExact} and demonstrate that
        the lower tail behaviour varies strongly dependent on the number of locations
        used to calculate the PC prior. One can see that the lower the resolution is,
        the higher the values of the prior in the left-hand side tail. Intuitively, this
        is because the distance decreases more slowly as a function of range as range
        goes to zero the lower the resolution is, and this causes more probability
        mass to be placed further out in the tail. This is because most of the differences
        in the GRFs for low ranges cannot be detected when it is observed at lower resolution.
        However, since the properties of the prior are used when making predictions at higher resolutions,
        we suggest to use the prior based on the continuous process instead of the discrete
        observation process.

        As the resolution of the grid used to calculate the prior using the bounded domain
        goes to infinity, the prior will agree better and better with the prior based on
        the unbounded domain for low values of the range. In Example \ref{exmp:1DexpExact}
        the overlap for low values of range is clear since the analytic expression for
        infinite resolution is used.

        \begin{figure}
                \centering
                \subfloat[Original axes]{
                        \includegraphics[width = 6.5cm]{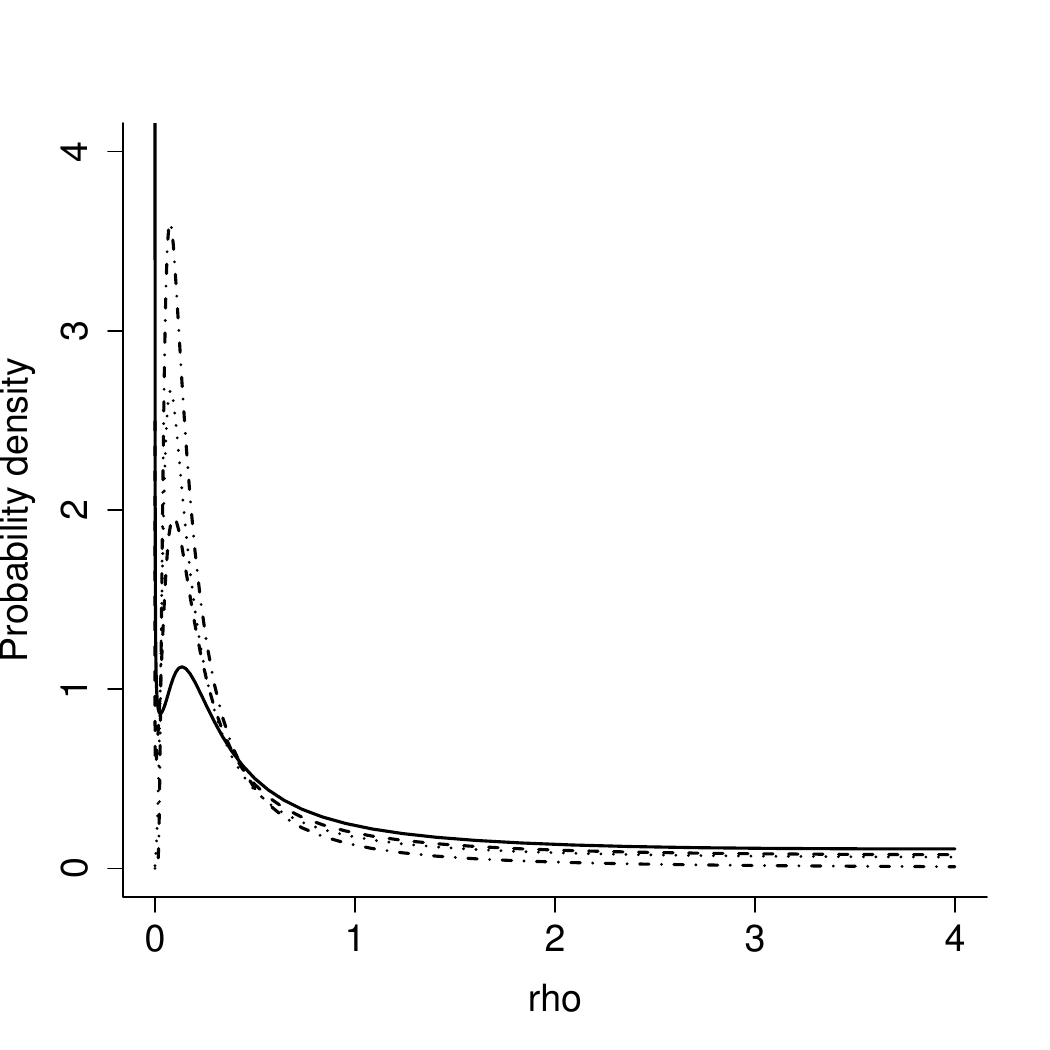}
                        \label{fig:2DExact:Orig}
                }
                \subfloat[Log-log axes]{
                        \includegraphics[width = 6.5cm]{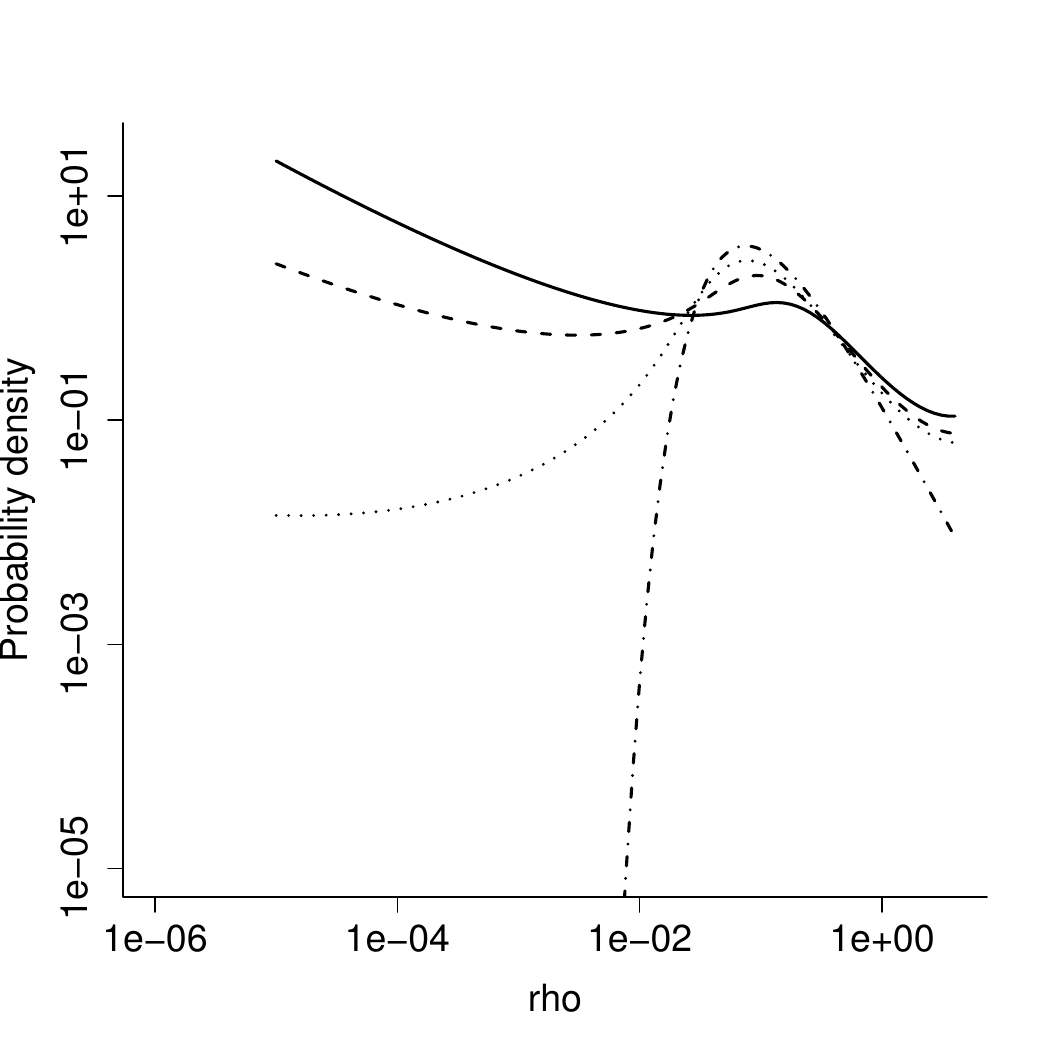}
                        \label{fig:2DExact:LogLog}
                }
                \caption{PC priors for range in a Mat\'ern GRF observed on $[0,1]^2$ with
                                 smoothness $\nu = 3/2$. The prior based on the unbounded domain is shown as
                                 dash-dot-dashed line, and the priors for the bounded domain 
                                 based on $10\times 10$ points, $20 \times 20$
                                 points and $40 \times 40$ are shown as a solid line, a dashed line and a dotted line,
                                 respectively. For each of the priors based on bounded domains the base
                                 model is $\rho_0 = 4$.}
                \label{fig:2DExact}
        \end{figure}
\end{exmp}

Even though there are cases in which the domain of the GRF is naturally
limited to the area on which the observations are made, the use of priors
based on an unbounded domain have many advantages. The priors
based on the bounded domain are more expensive to calculate and require
the additional choice of a range for the base model, but have the same 
lower tail behaviour as the prior based on the unbounded domain and only
the behaviour for higher ranges changes. Overall, we find that the unbounded
domain prior is most appealing.

\section{Distance for exponential covariance function on bounded one-dimensional domain}
\label{sec:derivation}
\subsection{Goal}
Let \(u_\kappa\) be a stationary GRF with the exponential covariance function,
\begin{equation}
        c(d) = \frac{1}{2\kappa} \mathrm{e}^{-\kappa d},
        \label{eq:ExpCovFun}
\end{equation}
where \(\kappa>0\). This way of writing the exponential covariance
function differs from the traditional parametrization using the range
and the marginal variance, and is chosen because the KLD between the 
distributions
described by different values \(\kappa>0\) is finite. The parametrization describes
how to move in the parameter space while keeping the KLD finite.
The goal of this appendix is to calculate the KLD between the distributions of 
\(u_{\kappa}\) and \(u_{\kappa_0}\) on the interval \([0,L]\)

\subsection{Discretization} 
The direct computations for the interval \([0,L]\) are difficult. So we
first consider the KLD for the distributions of \(u_\kappa\) and
\(u_{\kappa_0}\) at the observation points \(t_i = i\Delta t\), for 
\(i=0,1, \ldots, N\), where \(\Delta t = L/N\). The spatial field
\(u_\kappa\) can be described as a stationary solution of the stochastic 
differential equation
\[
        \mathrm{d}u_\kappa(t) = -\kappa u_\kappa(t)\mathrm{d}t + \mathrm{d} W(t),
\]
where \(W\) is a standard Wiener processes, and written explicitly as
\[
        u_\kappa(t) = \int_{-\infty}^{t}\! \mathrm{e}^{-\kappa(t-s)}\, \mathrm{d}W(s).
\]
This expression shows that 
\[
        u_\kappa(t_{i+1}) | u_\kappa(t_{i}) \sim \mathcal{N}(\mathrm{e}^{-\kappa\Delta t}u_\kappa(t_{i}), \sigma_\kappa^2),
\]
where  
\[
        \sigma_\kappa^2 = \mathrm{Var}[u_\kappa(t+\Delta t)|u_\kappa(t)] = \int_{t}^{t+\Delta t} \! \mathrm{e}^{-2\kappa(t+\Delta t-s)} \, \mathrm{d}s
                = \frac{1-\mathrm{e}^{-2\kappa \Delta t}}{2\kappa}.
\]

This is an AR(1) process with initial condition 
\(u_\kappa(t_0)\sim\mathcal{N}(0, (2\kappa)^{-1})\), which means that
\(\boldsymbol{u}_\kappa = (u_\kappa(t_0), \ldots, u_\kappa(t_N))\) 
has a multivariate Gaussian distribution with mean \(\boldsymbol{0}\) and precision matrix
\begin{equation}
        \mathbf{Q}_\kappa = \frac{1}{\sigma_\kappa^2}
\begin{bmatrix} 
        1    & -\mathrm{e}^{-\kappa\Delta t}      &        &        &     & \\
        -\mathrm{e}^{-\kappa\Delta t}   &  1+\mathrm{e}^{-2\kappa\Delta t}      & -\mathrm{e}^{-\kappa\Delta t}     &        &     & \\ 
             & \ddots  & \ddots & \ddots &     & \\
             &         &  -\mathrm{e}^{-\kappa\Delta t}    &   1+\mathrm{e}^{-2\kappa\Delta t}    &  -\mathrm{e}^{-\kappa\Delta t} & \\
             &         &        &   -\mathrm{e}^{-\kappa\Delta t}   &   1 & \\
\end{bmatrix}.
\label{eq:Qmatrix}
\end{equation}

\subsection{Kullback-Leibler divergence}
The vectors \(\boldsymbol{u}_{\kappa_0}\) and \(\boldsymbol{u}_{\kappa}\) have
multivariate Gaussian distributions and the KLD from the distribution described by \(\kappa_0\)
to the distribution described by \(\kappa\) is
\[
        \mathrm{KL}(\kappa, \kappa_0) = \frac{1}{2}\left[\mathrm{tr}(\mathbf{Q}_{\kappa_0}\mathbf{Q}_\kappa^{-1})-(N+1) - \log\left(\frac{|\mathbf{Q}_{\kappa_0}|}{|\mathbf{Q}_{\kappa}|}\right)\right].
\]
We are interested in taking the limit \(\Delta t \rightarrow 0\) to find the value
corresponding to the KLD from \(u_{\kappa_0}\) to \(u_{\kappa}\). This is done
in two steps: first we consider the trace and the \(N+1\) term, and then
we consider the log-determinant term.

\subsubsection{Step 1}
Let \(f_\kappa = 1/\sigma_\kappa^2\), then the trace term can be written as
\begin{align*}
        &\mathrm{tr}(\mathbf{Q}_{\kappa_0}\Sigma_\kappa) \\
        &\phantom{hei}= f_{\kappa_0}\left[2c_{\kappa}(0)+\sum_{i=1}^{N-1}(1+\mathrm{e}^{-2\kappa_0\Delta t})c_\kappa(0)-2\sum_{i=1}^N\mathrm{e}^{-\kappa_0\Delta t} c_\kappa(\Delta t)\right]\\
        &\phantom{hei}= f_{\kappa_0}\left[2c_{\kappa}(0)+(N-1)(1+\mathrm{e}^{-2\kappa_0\Delta t})c_\kappa(0)-2N\mathrm{e}^{-\kappa_0\Delta t} c_\kappa(\Delta t)\right].\\
\end{align*}
We extract the first summand and parts of the last summand, and combine
with \(2\) from the \(N+1\) term, to find the limit
\begin{align*}
        2 f_{\kappa_0}[c_\kappa(0)-\mathrm{e}^{-\kappa_0\Delta t}c\kappa(\Delta t)]-2
                    &= 2 f_{\kappa_0}\frac{1-\mathrm{e}^{-(\kappa+\kappa_0)\Delta t}}{2\kappa}-2 \\
                    &= \frac{\kappa+\kappa_0}{\kappa} \frac{f_{\kappa_0}/\Delta t}{f_{\kappa+\kappa_0}/\Delta t}-2 \\
                    &\rightarrow \frac{\kappa_0-\kappa}{\kappa}.
\end{align*}
For the remaining summands and the remaining \(N-1\) from the \(N+1\)
term, we can simplify the expression as
\begin{align*}
        &S_3(\Delta t) \\
        &\phantom{hei}=(N-1)f_{\kappa_0}\left[\left(1+\mathrm{e}^{-2\kappa_0\Delta t}\right)c_\kappa(0)-2\mathrm{e}^{-\kappa_0\Delta t}c_\kappa(\Delta t) \right] -(N-1)\\
          &\phantom{hei}=(N-1)f_{\kappa_0}\left[\left(1+\mathrm{e}^{-2\kappa_0\Delta t}\right)\frac{1}{2\kappa}-2\frac{\mathrm{e}^{-(\kappa_0+\kappa)\Delta t}}{2\kappa} \right]-(N-1) \\
          &\phantom{hei}=(N-1)f_{\kappa_0}\frac{1}{2\kappa}\bigg[1 + (1 - 2\kappa_0\Delta t + \frac{4\kappa_0^2 (\Delta t)^2}{2})\\
          &\phantom{hei=hhh}-2(1-(\kappa_0+\kappa)\Delta t + \frac{(\kappa_0+\kappa)^2(\Delta t)^2}{2})+o((\Delta t)^2)\bigg]-(N-1) \\
          &\phantom{hei}=(N-1)f_{\kappa_0}\frac{1}{2\kappa}\bigg[(-2\kappa_0+2(\kappa_0+\kappa)) \Delta t \\
          &\phantom{hei=hhh}+(2\kappa_0^2-(\kappa_0+\kappa)^2)(\Delta t)^2+o((\Delta t)^2)\bigg]-(N-1)\\
          &\phantom{hei}=(N-1)f_{\kappa_0}\left[\Delta t+\frac{2\kappa_0^2-(\kappa_0+\kappa)^2}{2\kappa}(\Delta t)^2+o((\Delta t)^2)\right]-(N-1)\\
          &\phantom{hei}=\left(\frac{L}{\Delta t}-1\right)\left(\frac{1}{\Delta t}+\kappa_0+o(1)\right)\bigg[\Delta t \\
          &\phantom{hei=hhh}+\frac{2\kappa_0^2-(\kappa_0+\kappa)^2}{2\kappa}(\Delta t)^2+o((\Delta t)^2)\bigg]-\left(\frac{L}{\Delta t}-1\right),
\end{align*}
and see that the products involving \(o(1)\) tend to zero
\begin{align*}
        S_3(\Delta t) &= L\left[\frac{1}{\Delta t} + \frac{2\kappa_0^2-(\kappa_0+\kappa)^2}{2\kappa}-\frac{1}{\Delta t}\right] +L\kappa_0- [1+o(1)]+1\\
                &= L\frac{4\kappa_0^2-(\kappa_0+\kappa)^2}{2\kappa}+L\kappa_0+o(1)\\
                &= L\left(\kappa_0+\frac{\kappa_0^2}{2\kappa}-\kappa_0-\frac{\kappa}{2}\right)+o(1).
\end{align*}
Thus the limit is
\[
        \mathrm{tr}(\mathbf{Q}_{\kappa_0}\Sigma_\kappa)-(N+1) \rightarrow \frac{\kappa_0}{\kappa}-1+L\left(\frac{\kappa_0^2}{2\kappa}-\frac{\kappa}{2}\right).
\]

\subsubsection{Step 2}
The determinant of the matrix in Equation~\eqref{eq:Qmatrix} 
can be found by summing rows upwards, and we see that
\[
        \vert \mathrm{Q} \vert = \sigma^{-2(N+1)}(1-\mathrm{e}^{-2\kappa\Delta t}) = 2\kappa \sigma^{-2N}.
\]
Note that in the limit \(\kappa\rightarrow 0\), \(f\rightarrow \Delta t\) so
the determinant behaves asymptotically as \(\kappa\). This means that
\begin{align*}
\log\left(\frac{|\mathbf{Q}_{\kappa_0}|}{|\mathbf{Q}_\kappa|}\right) &= \log\left(\frac{2\kappa_0 f_{\kappa_0}^N}{2\kappa f_\kappa^N}\right) \\
    &= \log\left(\frac{\kappa_0}{\kappa}\right) + N \log\left(\frac{f_{\kappa_0}}{f_\kappa}\right)
\end{align*}
and we need to find the limit of the second part,
\begin{align*}
        &N \log\left(\frac{f_{\kappa_0}}{f_\kappa}\right) \\
                &\phantom{hei}= \frac{L}{\Delta t}\left[\log\frac{1}{f_{\kappa}}-\log\frac{1}{f_{\kappa_0}}\right]\\
                &\phantom{hei}= \frac{L}{\Delta t}\left[\log\left(\frac{1}{2\kappa}\left(1-\mathrm{e}^{-2\kappa\Delta t}\right)\right)-\log\left(\frac{1}{2\kappa_0}\left(1-\mathrm{e}^{-2\kappa_0\Delta t}\right)\right)\right]\\
                &\phantom{hei}= \frac{L}{\Delta t}\left[\log\left(\Delta t - \kappa (\Delta t)^2 +o((\Delta t)^2)\right)-\log\left(\Delta t - \kappa_0(\Delta t)^2 + o((\Delta t)^2)\right)\right]\\
                &\phantom{hei}= \frac{L}{\Delta t}\left[\log\left(1 - \kappa \Delta t +o(\Delta t)\right)-\log\left(1 - \kappa_0\Delta t + o(\Delta t)\right)\right]\\
                &\phantom{hei}= \frac{L}{\Delta t}\left[-\kappa \Delta t + \kappa_0 \Delta t +o(\Delta t)\right]
\end{align*}

Thus the limit is
\[
\log\left(\frac{|\mathbf{Q}_{\kappa_0}|}{|\mathbf{Q}_\kappa|}\right) \rightarrow \log\left(\frac{\kappa_0}{\kappa}\right) +L(\kappa_0-\kappa)
\]

\subsection{Full KLD}
The combination of the limits from the two steps gives the full KLD,
\begin{align}
        \mathrm{KL}(\kappa, \kappa_0) &= \frac{1}{2}\left[\frac{\kappa_0}{\kappa}-1+L\left(\frac{\kappa_0^2}{2\kappa}-\frac{\kappa}{2}\right)-\log\left(\frac{\kappa_0}{\kappa}\right)-L(\kappa_0-\kappa)\right]\notag\\
                &= \frac{1}{2}\left[\frac{\kappa_0}{\kappa}-1-\log\left(\frac{\kappa_0}{\kappa}\right) + L\left(\frac{\kappa_0^2}{2\kappa}-\kappa_0+\frac{\kappa}{2}\right)\right]\label{eq:1dExpKLD}.
\end{align}


\section{Details for the simulation study}
\label{sec:SimStud}
In this section we present a small simulation study of the frequentist coverage
of the credible intervals for 
the range and the marginal variance, and the behaviour of the joint posterior 
when using the PC prior, the Jeffreys' rule prior, and the Jeffreys prior
for variance combined with a bounded uniform prior on range and a bounded uniform
prior on the logarithm of range.

\subsection{Study setup}
\begin{figure}
        \centering
        \includegraphics[width=7cm]{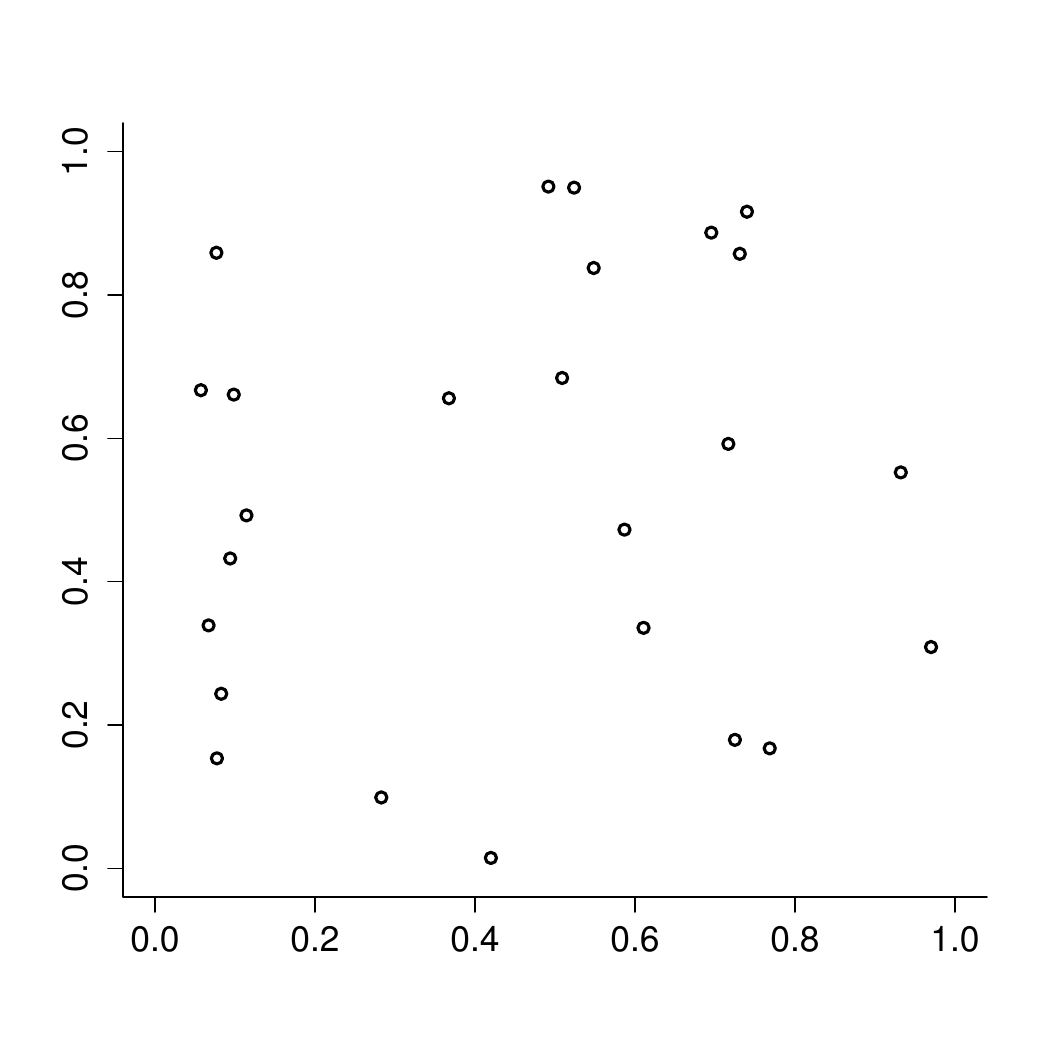}
        \caption{Spatial design for the simulation study.}
        \label{fig:spatialDesign}
\end{figure}

We choose the observation domain $[0,1]^2 \subset \mathbb{R}^2$ and select
the 25 observation locations, $\boldsymbol{s}_1, \boldsymbol{s}_2, \ldots, \boldsymbol{s}_{25}$,
shown in Figure \ref{fig:spatialDesign} at random. Then we simulate observations, 
$\boldsymbol{u} = (u(\boldsymbol{s}_1), u(\boldsymbol{s_2}), \ldots, u(\boldsymbol{s}_{25}))$, 
for these observation locations for a GRF with an exponential covariance
function \(c(r) = \exp(-2r/R_0)\) for $R_0 = 0.1$ and 
$R_0 = 1$. We generate 1000 realizations of the observations for $R_0 = 0.1$ and $R_0 = 1$
and collect them
in datasets Data1 and Data2 , respectively. Additionally, for each of the 1000 realizations in Data1 a third dataset
(Data3) is generated by simulating $y_i | p_i \sim \mathrm{Binomial}(20, p_i)$, where
$\mathrm{probit}(p_i) = u_i$, for $i = 1, 2, \ldots, 25$.

Two models are used to fit the data: a spatial regression model (Model1) and a 
spatial logistic regression model (Model2).
In Model1 observations are modelled as
\(y_i = u(\boldsymbol{s}_i)\), for $i = 1, 2, \ldots, 25$,
where $u$ is an exponential GRF with the covariance function, \(c(r) = \sigma^2\exp(-2r/\rho)\),
where $\sigma^2$ is the marginal variance and $\rho$ is the range, and
in Model 2 the observations are modelled as $y_i | p_i \sim \mathrm{Binomial}(20, p_i)$,
where $\mathrm{probit}(p_i) = u(\boldsymbol{s}_i)$, 
for $i = 1, 2, \ldots, 25$, where $u$ is an exponential GRF with 
covariance function, \(c(r) = \sigma^2\exp(-2r/\rho)\),
where $\sigma^2$ is the marginal variance and $\rho$ is the range.

Four different priors are used for the parameters: the PC prior (PriorPC), the 
Jeffreys' rule prior (PriorJe), 
a uniform prior on range on a bounded interval combined with the Jeffreys' prior
for variance (PriorUn1) and a uniform prior on the log-range on
a bounded interval combined with the Jeffreys' prior for
variance (PriorUn2). The full expressions for the priors are given in Table~\ref{tab:freqPriors}.

\begin{table}
        \centering
        \caption{The four different priors used in the simulation study. The
                         Jeffreys' rule prior uses the spatial design of the problem through
                         \(U = (\frac{\partial}{\partial \rho}\Sigma)\Sigma^{-1}\), where
                         \(\Sigma\) is the correlation matrix of the observations 
                         (See \citet{Berger2001}).}
        \begin{tabular}{lll}
                \textbf{Prior} & \textbf{Expression} & \textbf{Parameters} \\
                \toprule
                PriorPC & \(\displaystyle{\pi_1(\rho, \sigma) = \lambda_1\lambda_2\rho^{-2}\exp\left(-\lambda_1\rho^{-1}-\lambda_2\sigma\right)}\) & \parbox{5cm}{\(\rho, \sigma > 0\) \\ Hyperparameters:\\ \(\alpha_\rho\), \(\rho_0\), \(\alpha_\sigma\), \(\sigma_0\)} \\[1.5cm]
                PriorJe & \(\displaystyle{\pi_2(\rho, \sigma) = \sigma^{-1}\left(\mathrm{tr}(U^2)-\frac{1}{n}\mathrm{tr}(U)^2\right)^{1/2}}\) & \parbox{5cm}{\(\rho, \sigma > 0\)\\ Hyperparameters:\\ None} \\[1.5cm]
                PriorUn1 & \(\displaystyle{\pi_3(\rho, \sigma) \propto \sigma^{-1}}\) & \parbox{5cm}{\(\rho\in[A, B]\), \(\sigma>0\) \\ Hyperparameters:\\ \(A\), \(B\)} \\[1.5cm]
                PriorUn2 & \(\displaystyle{\pi_4(\rho, \sigma) \propto \sigma^{-1}\cdot\rho^{-1}}\) & \parbox{5cm}{\(\rho\in[A, B]\), \(\sigma>0\)\\ Hyperparameters:\\ \(A\), \(B\)}
        \end{tabular}
        \label{tab:freqPriors}
\end{table}

\subsection{Frequentist coverage}
\label{sec:FreqCov}
The series of papers on reference priors for GRFs starting with \citet{Berger2001} 
evaluated the priors by studying frequentist properties of the resulting 
Bayesian inference. A prior intended for use as a default prior should lead to
good frequentist properties such as 
frequentist coverage of the equal-tailed \(100(1-\alpha)\%\) Bayesian
credible intervals that is close to the nominal \(100(1-\alpha)\%\). In this
paper, the study is replicated with one key difference: no covariates are included.
This choice is made because the PC prior is derived for a zero-mean GRF, and
if a mean were desired, it would be handled by extending the hierarchical model
with another latent component that had its own, separate prior.
Without covariates the reference prior approach results in the Jeffreys' rule
prior as there are no 
nuisance parameters to integrate out when constructing the spatial reference prior.
Furthermore, we compute the $100(1-\alpha)\%$ highest posterior density (HPD) credible
intervals \citep{HPDinterval}
since the resulting posteriors will be highly skewed and the HPD
intervals may differ substantially from the quantile-based intervals.

In this section Model1 is combined with PriorJe, PriorPC, PriorUn1 and
PriorUn2. Data1 and Data2 each contains 1000 realizations and the frequentist coverage
is estimated for the equal-tailed $95\%$ credible intervals and the HPD $95\%$ credible
intervals for the range and
the marginal variance by counting how many times the true parameter value is included in
the credible intervals. The equal-tailed intervals are calculated based on the quantiles of
the samples from an MCMC chain and the HPD intervals are calculated using the BOA package
\citep{smith2007boa}. We split the presentation of the results for the quantile-based credible
intervals and the HPD credible intervals: in this section we discuss the quantile-based intervals and
their associated results, and in the next section we discuss the results for the 
HPD credible intervals and
differences from the results for the quantile-based intervals.

PriorJe has no hyperparameters, but PriorPC, PriorUn1 and PriorUn2 each has
hyperparameters that need to be set before using the priors. For PriorUn1 and
PriorUn2 it is hard to give guidelines about which values should be selected
since the main purpose of limiting the prior distributions to a bounded interval
is to avoid an improper posterior and the choice tends to be \emph{ad-hoc}. For
PriorPC, on the other hand, there is an interpretable statement for selecting
the hyperparameters, which helps give an idea about which prior assumptions
the chosen hyperparameters are expressing.

For PriorPC we need to make an \emph{a priori} decision about the scales of the range and the 
marginal variance. The prior is set through four hyperparameters that describe
our prior beliefs about the spatial field. We use
\(
        \mathrm{P}(\rho<\rho_0) = 0.05
\)
for \(\rho_0 = 0.025\rho_\mathrm{T}\), \(\rho_0 = 0.1\rho_\mathrm{T}\), 
\(\rho_0 = 0.4\rho_\mathrm{T}\) and
\(\rho_0 = 1.6\rho_\mathrm{T}\), where \(\rho_\mathrm{T}\) is the true range. 
This covers a prior where \(\rho_0\) is much smaller than the 
true range, two priors where \(\rho_0\) is smaller than the true range, but not far
away, and one prior where \(\rho_0\) is higher than the true range.
For the marginal variance we use
\(
        \mathrm{P}(\sigma^2>\sigma_0^2) = 0.05,
\)
for \(\sigma_0 = 0.625\), \(\sigma_0 = 2.5\),  \(\sigma_0 = 10\) and \(\sigma_0 = 40\).
We follow the same logic as for range and cover too small and too large \(\sigma_0\)
and two reasonable values. For PriorUn1 and PriorUn2, we set the lower and upper limits 
for the nominal range according to the values 
\(A = 0.05\), \(A = 0.005\) and \(A = 0.0005\), and \(B = 2\), \(B = 20\) and \(B = 200\).
Some of the values are intentionally extreme to see the effect of misspecification.

The results for PriorPC are given in Tables \ref{tab:PriorPC_01} and \ref{tab:PriorPC_10}
for the true ranges \(\rho_\mathrm{T} = 0.1\) and \(\rho_\mathrm{T} = 1\), respectively, and the tables for PriorUn1 and PriorUn2
are given in Section \ref{app:coverageTables}. PriorJe resulted in 98.3\% coverage with 
average length of the credible intervals of 0.78 for range and 96.7\% coverage and average length of the credible
intervals of 2.6 for marginal variance for \(\rho_\mathrm{T}= 0.1\), and 95.6\% coverage with average
length of the credible intervals of 376 for range and 95.6\% coverage with average length of
the credible intervals of 295 for variance for \(\rho_\mathrm{T} = 1\). The tables show that for PriorPC,
PriorUn1 and PriorUn2 the 
coverage and the length of the credible intervals are sensitive to the 
choice of hyperparameters. 
The lengths of the credible intervals are, in general, more well-behaved for \(\rho_\mathrm{T} = 0.1\)
than for \(\rho_\mathrm{T} = 1\) because there is more information about the range available in the domain
when the range is shorter.

\begin{table}
        \centering
        \caption{Frequentist coverage of the 95\% credible intervals for the range 
                 and the marginal variance when the true range is \(\rho_\mathrm{T} = 0.1\) 
                 using PriorPC. The average lengths of the credible 
                 intervals are shown in brackets.
                        }
        
                \subfloat[Range]{
                        \begin{tabular}{lllll}
                                \(\rho_0\)\textbackslash\(\sigma_0\) & 40 & 10 & 2.5 & 0.625 \\
                                \toprule
                                0.0025 & 0.755 [0.24] & 0.776 [0.22] & 0.760 [0.20] & 0.709 [0.18] \\
                                0.01   & 0.969 [0.33] & 0.970 [0.32] & 0.958 [0.28] & 0.924 [0.21] \\
                                0.04   & 0.988 [0.46] & 0.990 [0.41] & 0.991 [0.33] & 0.990 [0.25] \\
                                0.16   & 0.723 [0.99] & 0.685 [0.82] & 0.733 [0.55] & 0.798 [0.34]
                        \end{tabular}
                        \label{tab:PriorPC_01_range}
                }\\
                \subfloat[Marginal variance]{
                        \begin{tabular}{lllll}
                                \(\rho_0\)\textbackslash\(\sigma_0\) & 40 & 10 & 2.5 & 0.625 \\
                                \toprule
                                0.0025 & 0.960 [1.5] & 0.943 [1.4] & 0.946 [1.3] & 0.898 [0.97] \\
                                0.01   & 0.934 [1.6] & 0.966 [1.6] & 0.960 [1.4] & 0.923 [0.99] \\
                                0.04   & 0.949 [2.0] & 0.945 [1.8] & 0.953 [1.5] & 0.941 [1.1]  \\
                                0.16   & 0.895 [3.8] & 0.905 [3.2] & 0.947 [2.2] & 0.977 [1.3]
                        \end{tabular}
                        \label{tab:PriorPC_01_var}
                }
        
        \label{tab:PriorPC_01}
\end{table}

\begin{table}
        \centering
        \caption{Frequentist coverage of the 95\% credible intervals for the range and 
                 the marginal variance when the true range is \(\rho_\mathrm{T} = 1\) using 
                 PriorPC. The average lengths of the credible intervals 
                 are shown in brackets.
                        }
        
                \subfloat[Range]{
                        \begin{tabular}{lllll}
                                \(\rho_0\)\textbackslash\(\sigma_0\) & 40 & 10 & 2.5 & 0.625 \\
                                \toprule
                                0.025 & 0.957 [12] & 0.947 [7.3] & 0.921 [3.3] & 0.782 [1.4] \\
                                0.1   & 0.977 [14] & 0.967 [8.5] & 0.962 [3.5] & 0.861 [1.5] \\
                                0.4   & 0.963 [25] & 0.970 [13]  & 0.988 [5.2] & 0.980 [1.9] \\
                                1.6   & 0.63 [73] & 0.301 [32]  & 0.711 [11]  & 0.945 [3.3]
                        \end{tabular}
                        \label{tab:PriorPC_10_range}
                }\\
                \subfloat[Marginal variance]{
                        \begin{tabular}{lllll}
                                \(\rho_0\)\textbackslash\(\sigma_0\) & 40 & 10 & 2.5 & 0.625 \\
                                \toprule
                                0.025 & 0.956 [11] & 0.949 [6.5] & 0.927 [2.8] & 0.771 [1.1] \\
                                0.1   & 0.964 [13] & 0.966 [7.5] & 0.950 [3.1] & 0.848 [1.2] \\
                                0.4   & 0.953 [22] & 0.964 [12]  & 0.980 [4.5] & 0.965 [1.5] \\
                                1.6   & 0.435 [69] & 0.549 [29]  & 0.804 [9.1] & 0.988 [2.5]
                        \end{tabular}
                        \label{tab:PriorPC_10_var}
                }
        
        \label{tab:PriorPC_10}
\end{table}

The results verifies the observations by \citet{Berger2001} that the inference is
overly sensitive to the hyperparameters for PriorUn1. The coverage and the length
of the credible intervals are strongly dependent on the upper limit of the prior.
For PriorUn2 the coverage is good in both the short range and long range case, but
the length of the credible intervals are sensitive to the upper limit of the prior.
For PriorJe the coverage is good, but the credible intervals are excessively long
and the prior is computationally expensive and only computationally feasible for
a low number of observation locations. The average length of the credible intervals
for \(\rho_\mathrm{T} = 1\) for 
marginal variance is 295, which imply unreasonably high standard deviations. 
The high standard deviations do not seem consistent with observations 
drawn with true marginal variance equal to 1.

Further, the results show that the coverage for PriorPC is stable when a too low lower limit
for range or a too high upper limit for marginal variance is specified, but that specifying 
a too high lower limit for the range or a too low
upper limit for variance produces large changes in the coverage. This is
not unreasonable as the prior is then explicitly stating that the true value
for range or variance is unlikely. The average length of the credible intervals
are more sensitive to the hyperparameters than the coverages, but we see less extreme sizes for
the credible
intervals than for PriorJe. 

With respect to computation time and ease of use versus coverage and length of
credible intervals PriorUn2 and PriorPC appear to be the best choices. If coverage
is the only concern, PriorUn2 performs the best, but if one also wants to control
the length of the credible intervals by disallowing unreasonably high variances,
PriorPC offers the most interpretable alternative. In a realistic situation it is
highly likely that the researcher has prior knowledge, for example, that the
spatial effect should not be greater than, say 4, and by encoding this information in PriorPC one can
limit the upper limits of the credible intervals both for range and
marginal variance.

\subsection{Differences in results between equal-tailed and HPD credible intervals}
For each case discussed in the previous section, we also calculated the 95\% credible
intervals using HPD intervals
and the results and tables are found in Section \ref{app:hpdCoverageTables}.
In general, the average length of the credible intervals are significantly 
shorter for HPD credible intervals than for
quantile-based  credbile intervals, and the most substantial
decrease is seen for true range equal to $1.0$ using PriorJe, where the average
length of the credible interval decreases from 376 to 95 for range and
from 295 to 75 for marginal variance. However, the conclusions in the previous
section on differences
in average lengths of credible intervals between priors
and between true range equal to $0.1$ and $1.0$ remain valid since the relative
differences remain similar. In particular, the average length
of the HPD credible intervals for marginal variance for PriorJe with
true range equal to $1.0$ is around 95, which is still unreasonably high
when prior knowledge about the marginal standard deviation is available.

The coverage of the credible intervals constructed using HPD intervals
differ from the quantile-based intervals. For PriorPC, the coverage of the 
HPD intervals is more sensitive to the hyperparameters and if 
$\rho_0 = 0.4 \rho_\mathrm{T}$ the coverage of the HPD intervals for range
are almost 100\%. Further, the coverage of the HPD intervals for marginal 
variance are excessively high when $\sigma_0 = 40$ or $\sigma_0 = 10$, and
there is no recommendation for hyperparameters that perform consistently well
in both true range equal to $0.1$ and $1.0$ and for both range and marginal
variance. Similarly, the coverage of the credible interval for 
range is almost 100\%  when
the true range is $0.1$ with PriorJe. This contrasts the quantile-based 
credible intervals where PriorJe performs well with respect to coverage
for both true range equal to $0.1$ and $1.0$. For PriorUn1 and PriorUn2
the coverage of the HPD credible intervals are less sensitive to hyperparameters
than the quantile-based credible intervals, but the the HPD intervals tend to
have higher coverage than the nominal level.

We use the equal-tailed 95\% credible intervals in what follows since 
the 95\% HPD credible intervals are further away from nominal level and more 
sensitive to hyperparameters than equal-tailed 95\% credible intervals for the PC prior.

\subsection{Behaviour of the joint posterior}
In the previous section we only studied the marginal properties of the posterior, but
these do not tell the entire story because there is strong dependence between 
the range and the marginal variance in the posterior distribution. 
We study this dependence using one realization from Data2 where the
true range is 1 and the observed values lie in the range $-1$ to $3$.
An MCMC sampler is used to draw samples from the posterior of
the marginal standard deviation and the range. 
Model1 is combined with PriorJe, and PriorPC with
hyperparameters \(\alpha_\rho = 0.05\), \(\rho_0 = 0.1\),  
\(\alpha_\sigma = 0.05\) and \(\sigma_0 = 10\).

\begin{figure}
        \centering
        \includegraphics[width=7cm]{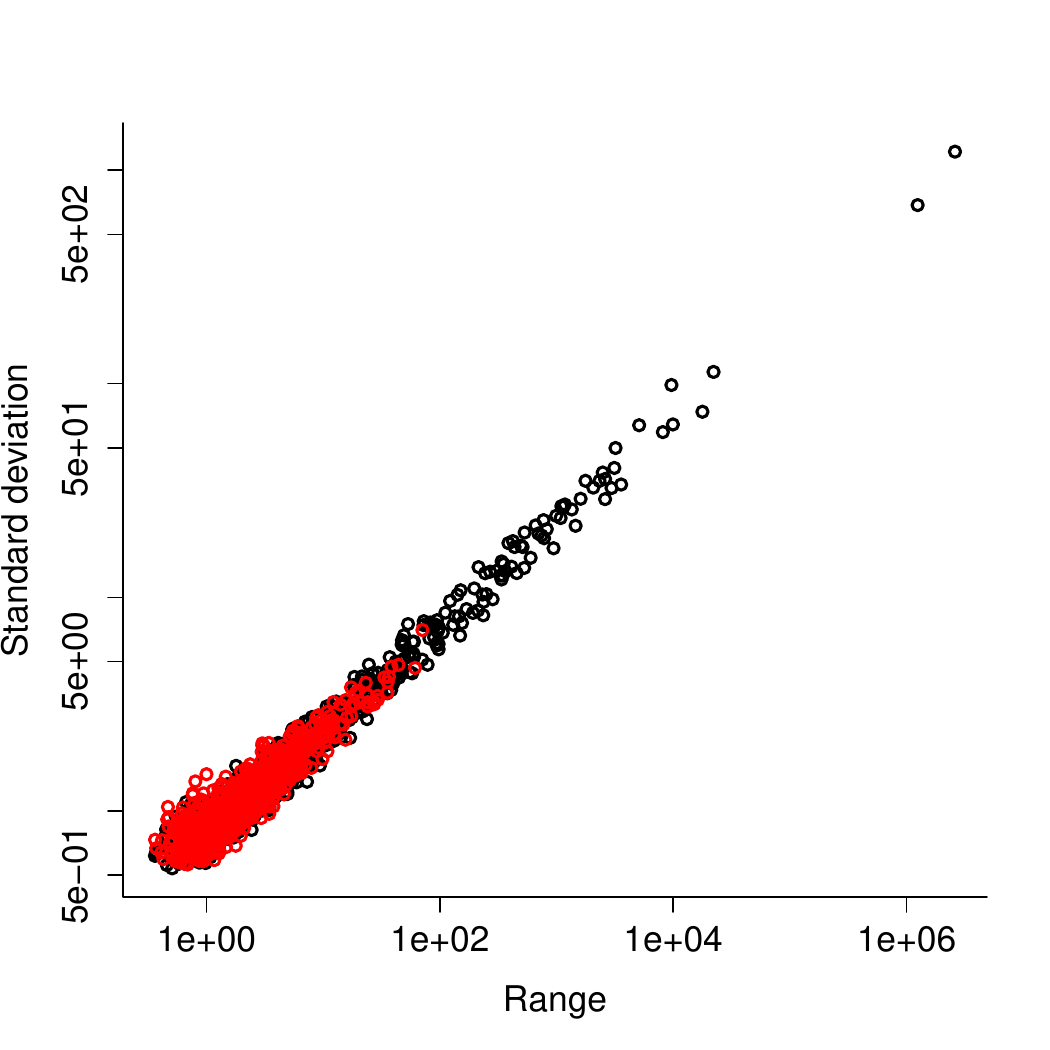}
        \caption{Samples from the joint posterior of range and marginal 
                 standard deviation. The red circles are samples using the 
                 PC-prior and the black circles are samples using the 
                 Jeffreys' rule prior.}
        \label{fig:jointPost}
\end{figure}

Figure~\ref{fig:jointPost} shows the strong posterior dependence between
the marginal standard deviation and the range in the tail of the distribution.
The long tails are not a major
concern for predictions since the asymptotic predictions are the same
along the ridge, but they pose a concern for the interpretability
of the range and the marginal variance. Since the values of the observations lie within the range
\(-1\) to \(3\), it is unlikely that the true standard deviation should be on the
order of 20. 

\begin{figure}
        \centering
        \subfloat[Posterior for the logarithm of range\label{fig:sub:nomRange}]{
                \includegraphics[width=5cm]{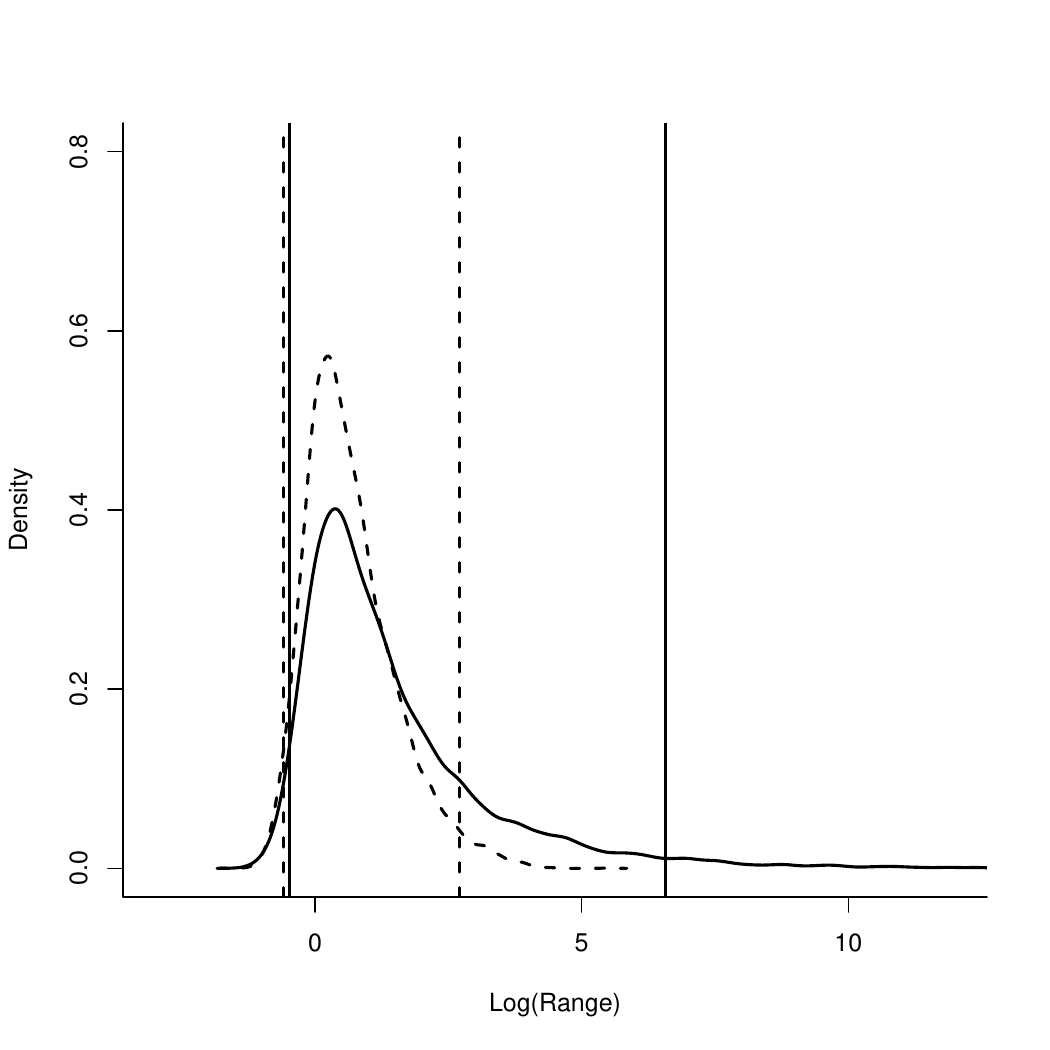}
        }
        \subfloat[Posterior for the logarithm of marginal standard deviation\label{fig:sub:stdDev}]{
                \includegraphics[width=5cm]{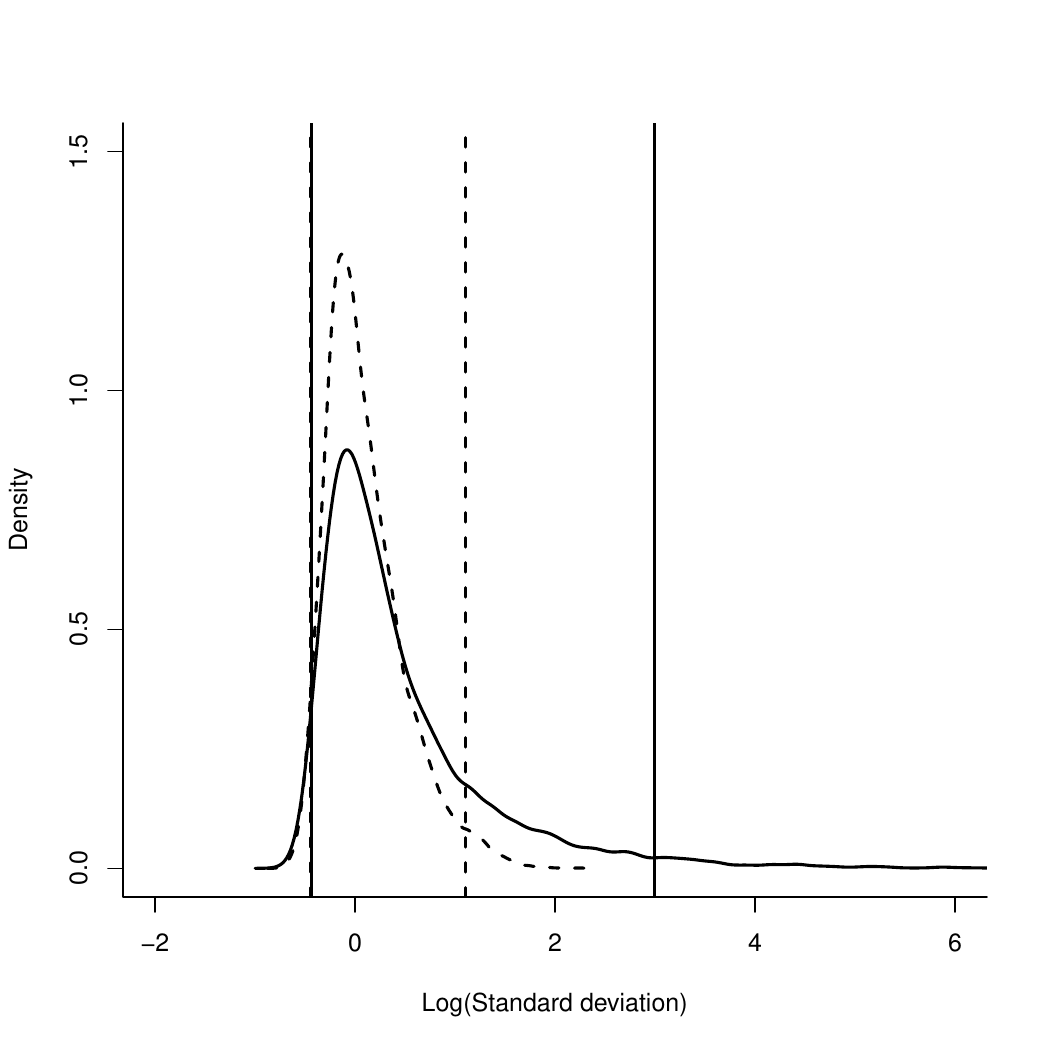}
        }
        \caption{Marginal posteriors of the logarithms of range and marginal standard deviation. 
                         The dashed lines corresponds to the PC prior and the solid
                         corresponds to the Jeffreys' rule prior. Equal-tailed 95\% credible
                         intervals are shown as vertical lines.}
        \label{fig:postComp}
\end{figure}

As seen in Figure \ref{fig:postComp}, 
the heavier upper tail for the joint posterior when using 
PriorJe compared to using PriorPC results
in heavier tails also for the marginal posteriors. The lower endpoints of the equal-tailed credible
intervals are similar using both priors, but there is a large difference in the
upper endpoints. The PC prior for range has a heavy upper tail and the upper tail
of the posterior for the range is controlled through the prior on the marginal variance.
The light upper tail of the prior on marginal variance restricts the joint posterior
from moving far along the ridge in the likelihood.

Intrinsic models have a place in statistics, but the results show that PriorJe
favors intrinsic GRFs with
large marginal standard deviations and ranges even though they might not be physically
reasonable for the application. PriorPc offers a way to
introduce prior belief about the size of the marginal standard deviations, and thus a way
to reduce the preference for the intrinsic GRFs and limit the size of the
credible intervals according to knowledge about the process.

\subsection{Example: Spatial logistic regression}
A clear weakness of the reference priors is that they must be
re-derived when components are added to the model or the observation
process is changed. The PC prior can be used in any model since its derivation
is observation-process agnostic and results in a prior for the model
component itself and not the whole model. The frequentist
coverage resulting from using Model2 with PriorPC for 500 of the realizations
in Data3 is estimated similarly as in Section \ref{sec:FreqCov}.

The experiment is repeated for 64 different settings of the prior: the
hyperparameter \(\rho_0\) 
varies over \(\rho_0 = 0.0025, 0.01, 0.04, 0.16\) and 
the hyperparameter \(\sigma_0\) varies 
over \(\sigma_0 = 40, 10, 2.5, 0.625\). This covers a broad range of values 
from too small to too large. The values in Table \ref{tab:SLR} are similar
to the values in Table~\ref{tab:PriorPC_01} except that the equal-tailed credible
intervals are slightly longer. The longer credible 
intervals are reasonable since the 
binomial likelihood gives less information about the spatial field than 
direct observations. The coverage for the marginal variance
is good even for grossly miscalibrated priors, but the coverage for range 
is sensitive to bad calibration for range and the coverage is somewhat higher
than nominal for the well-calibrated priors. This is a feature also seen in the 
directly observed case in Section~\ref{sec:FreqCov}. For completeness, 
the corresponding 95\% HPD credible intervals are shown in Table \ref{tab:hpd_SLR}.
The table shows that the coverage for range is too low for $\rho_0 = 0.0025$ and
$\rho_0 = 0.01$ and that the coverage is too high for $\rho_0 = 0.04$ and 
$\rho_0 = 0.16$. This was also the case for Gaussian observations, and 
the HPD intervals are more sensitivity to the hyperparameters
than the equal-tailed credible intervals for PriorPC.

\begin{table}
        \centering
        \caption{Frequentist coverage of the 95\% credible intervals for range
                 and marginal variance when the true range is 0.1 and true 
                 marginal variance is 1, where the average length of the 
                 credible intervals are given in brackets, for the spatial 
                 logistic regression example.
                         }
        
                \subfloat[Range\label{tab:SLR_range}]{
                        \begin{tabular}{lllll}
                                        \(\rho_0\)\textbackslash\(\sigma_0\) & 40 & 10 & 2.5 & 0.625 \\
                                        \toprule
                                        0.0025 & 0.790 [0.32] & 0.775 [0.25] & 0.760 [0.22] & 0.720 [0.19] \\
                                        0.01   & 0.982 [0.42] & 0.981 [0.37] & 0.974 [0.30] & 0.960 [0.25] \\
                                        0.04   & 0.990 [0.65] & 0.987 [0.53] & 0.995 [0.40] & 0.985 [0.30] \\
                                        0.16   & 0.621 [1.6]  & 0.638 [1.2]  & 0.682 [0.71] & 0.779 [0.43]
                        \end{tabular}
                }\\
                
                \subfloat[Marginal variance\label{tab:SLR_var}]{
                        \begin{tabular}{lllll}
                                \(\rho_0\)\textbackslash\(\sigma_0\) & 40 & 10 & 2.5 & 0.625 \\
                                \toprule
                                0.0025 & 0.953 [2.1] & 0.936 [1.9] & 0.941 [1.7] & 0.913 [1.2] \\
                                0.01   & 0.952 [2.2] & 0.949 [2.1] & 0.954 [1.7] & 0.931 [1.2] \\
                                0.04   & 0.949 [2.7] & 0.942 [2.5] & 0.960 [1.9] & 0.923 [1.3] \\
                                0.16   & 0.906 [5.5] & 0.923 [4.2] & 0.961 [2.7] & 0.972 [1.5]
                        \end{tabular}
                }
        \label{tab:SLR}
\end{table}

\section{Additional tables for simulation study using quantile-based credible intervals}
\label{app:coverageTables}
The simulation study in Section \ref{sec:SimStud} was run with four different priors: the PC prior
(PriorPC), the Jeffreys' rule prior (PriorJe), a uniform prior on range on a bounded interval
combined with the Jeffreys' prior for variance (PriorUn1) and a uniform prior on the log-range
on a bounded interval combined with the Jeffreys' prior for variance (PriorUn2). For each prior 
a selection of hyperparameters were tested on datasets generated from true ranges \(\rho_\mathrm{T} = 0.1\) and \(\rho_\mathrm{T} = 1.0\), and the 
frequentist coverages of the 95\% credible intervals and the lengths of the credible intervals
were estimated. For \(\rho_\mathrm{T} = 0.1\), PriorJe gave 98.3\% coverage with average length 0.78 for range and 
96.7\% coverage with average length 2.6 for marginal variance, and for \(\rho_\mathrm{T} = 1.0\),
PriorJe gave 95.6\% coverage with average length 376 for range and 95.6\% coverage
with average length of 295 for marginal variance. The results for PriorPC is given in Section
\ref{sec:SimStud} and the results for the two other priors are collected in the tables:
\begin{center}
        \begin{tabular}{lll}
                \textbf{Prior} & \(\mathbf{\rho_\mathrm{T} = 0.1}\) & \(\mathbf{\rho_\mathrm{T} = 1.0}\) \\
                \hline
                PriorUn1 & Table \ref{tab:PriorUn1_01} & Table \ref{tab:PriorUn1_10} \\
                PriorUn2 & Table \ref{tab:PriorUn2_01} & Table \ref{tab:PriorUn2_10}
        \end{tabular}
\end{center}

\begin{table}
        \centering
        \caption{Frequentist coverage of 95\% credible intervals for range and 
                 marginal variance when the true range \(\rho_\mathrm{T} = 0.1\) using 
                 PriorUn1, where the average lengths of the credible intervals 
                 are shown in brackets.}
        
                \subfloat[Range]{
                        \begin{tabular}{lllll}
                                \(A\)\textbackslash \(B\) & 2 & 20 & 200 \\
                                \toprule
                                \(5\cdot 10^{-2}\) & 0.920 [0.93] & 0.886 [8.5] & 0.840 [119] \\
                                \(5\cdot 10^{-3}\) & 0.937 [0.94] & 0.910 [8.1] & 0.866 [104] \\
                                \(5\cdot 10^{-4}\) & 0.937 [0.91] & 0.925 [8.0] & 0.864 [108]
                        \end{tabular}
                        \label{tab:PriorUn1_01_range}
                }\\

                \subfloat[Marginal variance]{
                        \begin{tabular}{lllll}
                                \(A\)\textbackslash \(B\) & 2 & 20 & 200 \\
                                \toprule
                                \(5\cdot 10^{-2}\) & 0.941 [3.5] & 0.937 [30] & 0.900 [443] \\
                                \(5\cdot 10^{-3}\) & 0.934 [3.4] & 0.924 [27] & 0.924 [383] \\
                                \(5\cdot 10^{-4}\) & 0.934 [3.3] & 0.945 [27] & 0.922 [388]
                        \end{tabular}
                        \label{tab:PriorUn1_01_var}
                }
        \label{tab:PriorUn1_01}
\end{table}

\begin{table}
        \centering
        \caption{Frequentist coverage of 95\% credible intervals for range and 
                 marginal variance when the true range \(\rho_\mathrm{T} = 0.1\) using 
                 PriorUn2, where the average lengths of the credible intervals 
                 are shown in brackets.}
        
                \subfloat[Range]{
                        \begin{tabular}{lllll}
                                \(A\)\textbackslash \(B\) & 2 & 20 & 200\\
                                \toprule
                                \(5\cdot 10^{-2}\) & 0.987 [0.44] & 0.983 [0.72] & 0.985 [1.1] \\
                                \(5\cdot 10^{-3}\) & 0.959 [0.43] & 0.972 [0.74] & 0.965 [1.3] \\
                                \(5\cdot 10^{-4}\) & 0.923 [0.39] & 0.944 [0.68] & 0.933 [1.1]
                        \end{tabular}
                        \label{tab:PriorUn2_01_range}
                } \\

                \subfloat[Marginal variance]{
                        \begin{tabular}{lllll}
                                \(A\)\textbackslash \(B\) & 2 & 20 & 200 \\
                                \toprule
                                \(5\cdot 10^{-2}\) & 0.954 [1.9] & 0.954 [2.7] & 0.961 [3.5] \\
                                \(5\cdot 10^{-3}\) & 0.957 [1.7] & 0.957 [2.4] & 0.950 [3.8] \\
                                \(5\cdot 10^{-4}\) & 0.947 [1.6] & 0.954 [2.3] & 0.939 [3.2]
                        \end{tabular}
                        \label{tab:PriorUn2_01_var}
                }       
        \label{tab:PriorUn2_01}
\end{table}

\begin{table}
        \centering
        \caption{Frequentist coverage of 95\% credible intervals for range and 
                 marginal variance when the true range \(\rho_\mathrm{T} = 1\) using 
                 PriorUn1, where the average lengths of the credible intervals 
                 are shown in brackets.}
        
                \subfloat[Range]{
                        \begin{tabular}{lllll}
                                \(A\)\textbackslash \(B\) & 2 & 20 & 200 \\
                                \toprule
                                \(5\cdot 10^{-2}\) & 0.995 [1.5] & 0.840 [18] & 0.562 [188] \\
                                \(5\cdot 10^{-3}\) & 0.997 [1.5] & 0.831 [18] & 0.560 [188] \\
                                \(5\cdot 10^{-4}\) & 0.993 [1.5] & 0.823 [18] & 0.550 [188]
                        \end{tabular}
                        \label{tab:PriorUn1_10_range}
                }\\

                \subfloat[Marginal variance]{
                        \begin{tabular}{lllll}
                                \(A\)\textbackslash \(B\) & 2 & 20 & 200 \\
                                \toprule
                                \(5\cdot 10^{-2}\) & 0.975 [2.0] & 0.848 [20] & 0.574 [205] \\
                                \(5\cdot 10^{-3}\) & 0.978 [2.0] & 0.822 [21] & 0.600 [203] \\
                                \(5\cdot 10^{-4}\) & 0.983 [2.0] & 0.837 [20] & 0.564 [206]
                        \end{tabular}
                        \label{tab:PriorUn1_10_var}
                }
        \label{tab:PriorUn1_10}
\end{table}

\begin{table}
        \centering
        \caption{Frequentist coverage of 95\% credible intervals for range and 
                 marginal variance when the true range \(\rho_\mathrm{T} = 1\) using 
                 PriorUn2, where the average lengths of the credible intervals 
                 are shown in brackets.}
        
                \subfloat[Range]{
                        \begin{tabular}{lllll}
                                \(A\)\textbackslash \(B\) & 2 & 20 & 200\\
                                \toprule
                                \(5\cdot 10^{-2}\) & 0.978 [1.5] & 0.965 [13] & 0.963 [69] \\
                                \(5\cdot 10^{-3}\) & 0.969 [1.5] & 0.951 [12] & 0.944 [67] \\
                                \(5\cdot 10^{-4}\) & 0.978 [1.5] & 0.957 [13] & 0.947 [68]
                        \end{tabular}
                        \label{tab:PriorUn2_10_range}
                } \\

                \subfloat[Marginal variance]{
                        \begin{tabular}{lllll}
                                \(A\)\textbackslash \(B\) & 2 & 20 & 200 \\
                                \toprule
                                \(5\cdot 10^{-2}\) & 0.964 [1.8] & 0.961 [12] & 0.949 [61] \\
                                \(5\cdot 10^{-3}\) & 0.957 [1.8] & 0.953 [11] & 0.934 [60] \\
                                \(5\cdot 10^{-4}\) & 0.958 [1.8] & 0.945 [12] & 0.941 [59]
                        \end{tabular}
                        \label{tab:PriorUn2_10_var}
                }       
        \label{tab:PriorUn2_10}
\end{table}

\section{Results of simulation study using HPD credible intervals}
\label{app:hpdCoverageTables}
The simulation study in Section \ref{sec:SimStud} was run with four different priors: the PC prior
(PriorPC), the Jeffreys' rule prior (PriorJe), a uniform prior on range on a bounded interval
combined with the Jeffreys' prior for variance (PriorUn1) and a uniform prior on the log-range
on a bounded interval combined with the Jeffreys' prior for variance (PriorUn2). For each prior 
a selection of hyperparameters were tested on datasets generated from true ranges \(\rho_\mathrm{T} = 0.1\) and \(\rho_\mathrm{T} = 1.0\), and the 
frequentist coverages of the 95\% highest posterior density (HPD) credible intervals and the average
lengths of the HPD credible intervals
were estimated. For \(\rho_\mathrm{T} = 0.1\), PriorJe gave 99.9\% coverage with average length 0.46 for range and 
98.2\% coverage with average length 1.8 for marginal variance, and for \(\rho_\mathrm{T} = 1.0\),
PriorJe gave 95.7\% coverage with average length 95 for range and 96.5\% coverage
with average length of 75 for marginal variance. The results for PriorPC, PriorUn1 and PriorUn2 are given in the tables:
\begin{center}
        \begin{tabular}{lll}
                \textbf{Prior} & \(\mathbf{\rho_\mathrm{T} = 0.1}\) & \(\mathbf{\rho_\mathrm{T} = 1.0}\) \\
                \hline
                PriorPC  & Table \ref{tab:hpd_PriorPC_01} & Table \ref{tab:hpd_PriorPC_10} \\
                PriorUn1 & Table \ref{tab:hpd_PriorUn1_01} & Table \ref{tab:hpd_PriorUn1_10} \\
                PriorUn2 & Table \ref{tab:hpd_PriorUn2_01} & Table \ref{tab:hpd_PriorUn2_10} \\
                PriorPC and logistic regression & Table \ref{tab:hpd_SLR} & N/A
        \end{tabular}
\end{center}

\begin{table}
        \centering
        \caption{Frequentist coverage of the 95\% HPD credible intervals for the range 
                 and the marginal variance when the true range is \(\rho_\mathrm{T} = 0.1\) 
                 using PriorPC. The average lengths of the credible 
                 intervals are shown in brackets.
                        }
        
                \subfloat[Range]{
                        \begin{tabular}{lllll}
                                \(\rho_0\)\textbackslash\(\sigma_0\) & 40 & 10 & 2.5 & 0.625 \\
                                \toprule
                                0.0025 & 0.571 [0.17] & 0.586 [0.17] & 0.584 [0.16] & 0.535 [0.14] \\
                                0.01   & 0.903 [0.25] & 0.912 [0.25] & 0.900 [0.23] & 0.841 [0.18] \\
                                0.04   & 1.000 [0.35] & 0.999 [0.33] & 0.999 [0.28] & 0.998 [0.22] \\
                                0.16   & 0.990 [0.67] & 0.992 [0.60] & 0.980 [0.45] & 0.957 [0.31]
                        \end{tabular}
                        \label{tab:hpd_PriorPC_01_range}
                }\\
                \subfloat[Marginal variance]{
                        \begin{tabular}{lllll}
                                \(\rho_0\)\textbackslash\(\sigma_0\) & 40 & 10 & 2.5 & 0.625 \\
                                \toprule
                                0.0025 & 0.961 [1.3] & 0.947 [1.3] & 0.947 [1.2] & 0.857 [0.92] \\
                                0.01   & 0.959 [1.4] & 0.969 [1.4] & 0.958 [1.2] & 0.882 [0.93] \\
                                0.04   & 0.980 [1.7] & 0.967 [1.6] & 0.961 [1.3] & 0.908 [1.0]  \\
                                0.16   & 0.991 [2.8] & 0.988 [2.5] & 0.990 [1.9] & 0.962 [1.2]
                        \end{tabular}
                        \label{tab:hpd_PriorPC_01_var}
                }
        
        \label{tab:hpd_PriorPC_01}
\end{table}

\begin{table}
        \centering
        \caption{Frequentist coverage of 95\% HPD credible intervals for range and 
                 marginal variance when the true range \(\rho_\mathrm{T} = 0.1\) using 
                 PriorUn1, where the average lengths of the credible intervals 
                 are shown in brackets.}
        
                \subfloat[Range]{
                        \begin{tabular}{lllll}
                                \(A\)\textbackslash \(B\) & 2 & 20 & 200 \\
                                \toprule
                                \(5\cdot 10^{-2}\) & 0.977 [0.71] & 0.992 [5.7] & 0.989 [92] \\
                                \(5\cdot 10^{-3}\) & 0.977 [0.74] & 0.994 [5.6] & 0.990 [78] \\
                                \(5\cdot 10^{-4}\) & 0.970 [0.71] & 0.988 [5.4] & 0.993 [82]
                        \end{tabular}
                        \label{tab:hpd_PriorUn1_01_range}
                }\\

                \subfloat[Marginal variance]{
                        \begin{tabular}{lllll}
                                \(A\)\textbackslash \(B\) & 2 & 20 & 200 \\
                                \toprule
                                \(5\cdot 10^{-2}\) & 0.991 [2.7] & 0.993 [19] & 1.00 [312] \\
                                \(5\cdot 10^{-3}\) & 0.985 [2.7] & 0.993 [18] & 0.993 [263] \\
                                \(5\cdot 10^{-4}\) & 0.981 [2.6] & 0.989 [18] & 0.993 [270]
                        \end{tabular}
                        \label{tab:hpd_PriorUn1_01_var}
                }
        \label{tab:hpd_PriorUn1_01}
\end{table}

\begin{table}
        \centering
        \caption{Frequentist coverage of 95\% HPD credible intervals for range and 
                 marginal variance when the true range \(\rho_\mathrm{T} = 0.1\) using 
                 PriorUn2, where the average lengths of the credible intervals 
                 are shown in brackets.}
        
                \subfloat[Range]{
                        \begin{tabular}{lllll}
                                \(A\)\textbackslash \(B\) & 2 & 20 & 200\\
                                \toprule
                                \(5\cdot 10^{-2}\) & 0.998 [0.34] & 0.999 [0.45] & 1.000 [0.54] \\
                                \(5\cdot 10^{-3}\) & 0.922 [0.33] & 0.936 [0.46] & 0.922 [0.62] \\
                                \(5\cdot 10^{-4}\) & 0.831 [0.30] & 0.866 [0.42] & 0.864 [0.54]
                        \end{tabular}
                        \label{tab:hpd_PriorUn2_01_range}
                } \\

                \subfloat[Marginal variance]{
                        \begin{tabular}{lllll}
                                \(A\)\textbackslash \(B\) & 2 & 20 & 200 \\
                                \toprule
                                \(5\cdot 10^{-2}\) & 0.978 [1.6] & 0.977 [2.0] & 0.976 [2.1] \\
                                \(5\cdot 10^{-3}\) & 0.957 [1.5] & 0.974 [1.8] & 0.960 [2.2] \\
                                \(5\cdot 10^{-4}\) & 0.949 [1.4] & 0.966 [1.7] & 0.958 [2.0]
                        \end{tabular}
                        \label{tab:hpd_PriorUn2_01_var}
                }       
        \label{tab:hpd_PriorUn2_01}
\end{table}

\begin{table}
        \centering
        \caption{Frequentist coverage of the 95\% HPD credible intervals for the range and 
                 the marginal variance when the true range is \(\rho_\mathrm{T} = 1\) using 
                 PriorPC. The average lengths of the credible intervals 
                 are shown in brackets.
                        }
        
                \subfloat[Range]{
                        \begin{tabular}{lllll}
                                \(\rho_0\)\textbackslash\(\sigma_0\) & 40 & 10 & 2.5 & 0.625 \\
                                \toprule
                                0.025 & 0.927 [7.2] & 0.915 [4.8] & 0.869 [2.5] & 0.690 [1.2] \\
                                0.1   & 0.973 [8.2] & 0.961 [5.6] & 0.924 [2.7] & 0.783 [1.3] \\
                                0.4   & 1.000 [14] & 1.000 [8.6]  & 0.997 [4.0] & 0.949 [1.6] \\
                                1.6   & 0.993 [44] & 0.994 [22]  & 0.992 [8.3]  & 0.990 [2.9]
                        \end{tabular}
                        \label{tab:hpd_PriorPC_10_range}
                }\\
                \subfloat[Marginal variance]{
                        \begin{tabular}{lllll}
                                \(\rho_0\)\textbackslash\(\sigma_0\) & 40 & 10 & 2.5 & 0.625 \\
                                \toprule
                                0.025 & 0.946 [6.5] & 0.936 [4.3] & 0.889 [2.2] & 0.666 [0.95] \\
                                0.1   & 0.975 [7.5] & 0.980 [4.9] & 0.940 [2.4] & 0.754 [1.1] \\
                                0.4   & 1.000 [13] & 1.000 [7.8]  & 0.998 [3.4] & 0.912 [1.3] \\
                                1.6   & 0.996 [41] & 0.986 [20]  & 0.984 [7.3] & 0.999 [2.2]
                        \end{tabular}
                        \label{tab:hpd_PriorPC_10_var}
                }
        
        \label{tab:hpd_PriorPC_10}
\end{table}

\begin{table}
        \centering
        \caption{Frequentist coverage of 95\% HPD credible intervals for range and 
                 marginal variance when the true range \(\rho_\mathrm{T} = 1\) using 
                 PriorUn1, where the average lengths of the credible intervals 
                 are shown in brackets.}
        
                \subfloat[Range]{
                        \begin{tabular}{lllll}
                                \(A\)\textbackslash \(B\) & 2 & 20 & 200 \\
                                \toprule
                                \(5\cdot 10^{-2}\) & 0.980 [1.5] & 0.946 [17] & 0.979 [179] \\
                                \(5\cdot 10^{-3}\) & 0.989 [1.5] & 0.938 [17] & 0.967 [178] \\
                                \(5\cdot 10^{-4}\) & 0.979 [1.5] & 0.931 [17] & 0.967 [178]
                        \end{tabular}
                        \label{tab:hpd_PriorUn1_10_range}
                }\\

                \subfloat[Marginal variance]{
                        \begin{tabular}{lllll}
                                \(A\)\textbackslash \(B\) & 2 & 20 & 200 \\
                                \toprule
                                \(5\cdot 10^{-2}\) & 0.977 [1.8] & 0.989 [17] & 0.996 [176] \\
                                \(5\cdot 10^{-3}\) & 0.979 [1.8] & 0.985 [18] & 0.991 [175] \\
                                \(5\cdot 10^{-4}\) & 0.979 [1.8] & 0.983 [17] & 0.987 [177]
                        \end{tabular}
                        \label{tab:hpd_PriorUn1_10_var}
                }
        \label{tab:hpd_PriorUn1_10}
\end{table}

\begin{table}
        \centering
        \caption{Frequentist coverage of 95\% HPD credible intervals for range and 
                 marginal variance when the true range \(\rho_\mathrm{T} = 1\) using 
                 PriorUn2, where the average lengths of the credible intervals 
                 are shown in brackets.}
        
                \subfloat[Range]{
                        \begin{tabular}{lllll}
                                \(A\)\textbackslash \(B\) & 2 & 20 & 200\\
                                \toprule
                                \(5\cdot 10^{-2}\) & 0.945 [1.4] & 0.974 [9.8] & 0.985 [40] \\
                                \(5\cdot 10^{-3}\) & 0.936 [1.4] & 0.959 [9.6] & 0.973 [39] \\
                                \(5\cdot 10^{-4}\) & 0.954 [1.4] & 0.961 [9.5] & 0.966 [39]
                        \end{tabular}
                        \label{tab:hpd_PriorUn2_10_range}
                } \\

                \subfloat[Marginal variance]{
                        \begin{tabular}{lllll}
                                \(A\)\textbackslash \(B\) & 2 & 20 & 200 \\
                                \toprule
                                \(5\cdot 10^{-2}\) & 0.936 [1.6] & 0.983 [8.7] & 0.987 [35] \\
                                \(5\cdot 10^{-3}\) & 0.933 [1.6] & 0.971 [8.6] & 0.980 [34] \\
                                \(5\cdot 10^{-4}\) & 0.937 [1.6] & 0.969 [8.7] & 0.968 [33]
                        \end{tabular}
                        \label{tab:hpd_PriorUn2_10_var}
                }       
        \label{tab:hpd_PriorUn2_10}
\end{table}

\begin{table}
        \centering
        \caption{Frequentist coverage of the 95\% HPD credible intervals for range
                 and marginal variance when the true range is 0.1 and true 
                 marginal variance is 1, where the average length of the 
                 credible intervals are given in brackets, for the spatial 
                 logistic regression example.
                         }
        
                \subfloat[Range\label{tab:hpd_SLR_range}]{
                        \begin{tabular}{lllll}
                                        \(\rho_0\)\textbackslash\(\sigma_0\) & 40 & 10 & 2.5 & 0.625 \\
                                        \toprule
                                        0.0025 & 0.582 [0.22] & 0.575 [0.18] & 0.577 [0.17] & 0.539 [0.15] \\
                                        0.01   & 0.922 [0.30] & 0.925 [0.28] & 0.906 [0.24] & 0.883 [0.21] \\
                                        0.04   & 0.999 [0.46] & 0.999 [0.40] & 1.000 [0.33] & 0.998 [0.27] \\
                                        0.16   & 0.994 [1.0]  & 0.995 [0.8]  & 0.983 [0.57] & 0.972 [0.38]
                        \end{tabular}
                }\\
                
                \subfloat[Marginal variance\label{tab:hpd_SLR_var}]{
                        \begin{tabular}{lllll}
                                \(\rho_0\)\textbackslash\(\sigma_0\) & 40 & 10 & 2.5 & 0.625 \\
                                \toprule
                                0.0025 & 0.968 [1.8] & 0.945 [1.7] & 0.944 [1.6] & 0.867 [1.1] \\
                                0.01   & 0.973 [1.9] & 0.961 [1.8] & 0.954 [1.6] & 0.885 [1.1] \\
                                0.04   & 0.982 [2.2] & 0.978 [2.1] & 0.961 [1.7] & 0.893 [1.2] \\
                                0.16   & 0.993 [3.9] & 0.991 [3.3] & 0.991 [2.3] & 0.950 [1.4]
                        \end{tabular}
                }
        \label{tab:hpd_SLR}
\end{table}

\section{Prior for extra flexibility in the covariance structure}
\label{sec:NonStationarity}
\citet{Lindgren2011} represented Mat\'ern GRFs as the stationary solutions
to the stochastic partial differential equation (SPDE)
\begin{equation}
        [\kappa^2 - \Delta]^{\alpha/2}(\tau u(\boldsymbol{s})) = \mathcal{W}(\boldsymbol{s}), \quad \boldsymbol{s} \in \mathbb{R}^d,
        \label{eq:SPDEstat}
\end{equation}
where $\kappa > 0$ and $\tau > 0$ are parameters, $\alpha$ is connected to the smoothness
$\nu$ through $\alpha = \nu + d/2$, $\Delta$ is the Laplacian, and $\mathcal{W}$ is standard
Gaussian white noise. 
\citet{ingebrigtsen2014spatial} allowed the 
parameters of the SPDE to be spatially
varying functions, \(\log(\kappa(\cdot))\) and 
\(\log(\tau(\cdot))\), through low-dimensional bases using a small number of covariates, and used
independent Gaussian priors for the extra parameters. However, they experienced numerical
problems and prior sensitivity, and \citet{ingebrigtsen2014estimation} 
developed an improved scheme for selecting the hyperparameters of the priors
based on the properties of the resulting spatially varying local ranges and 
marginal variances. However, the inherent problem of their specification is that
\(\kappa(\cdot)\) affects both the correlation structure and the marginal variances of the 
spatial field. This makes it challenging to set priors on \(\kappa(\cdot)\) and 
\(\tau(\cdot)\), and we aim to improve their procedure by first improving the parametrization
of the non-stationarity, and then developing a prior  using
the improved parametrization. 

\subsection{Parametrizing the extra flexibility}

Instead of adding spatial variation to the coefficients of the SPDE in
Equation \eqref{eq:SPDEstat}, \(\kappa\) and
\(\tau\), one can vary the geometry of the space in a similar way as the deformation
method \citep{Sampson1992}. If \(E\) is the Euclidean space \(\mathbb{R}^2\), the simple SPDE
\begin{equation}
        (1-\Delta_E) u(\boldsymbol{s}) = \sqrt{4\pi} \mathcal{W}_E(\boldsymbol{s}), \quad \boldsymbol{s}\in E,
        \label{eq:SPDEtensor}
\end{equation}
generates a stationary Mat\'ern GRF with range
\(\rho = \sqrt{8}\), marginal variance 
\(\sigma^2 = 1\), and smoothness \(\nu = 1\). We introduce spatially varying distances
in the space by giving the space geometric structure according to the metric tensor 
\(\mathbf{g}(\boldsymbol{s}) = R(\boldsymbol{s})^{-2} \mathbf{I}_2\), where 
\(R(\cdot)\) is a strictly positive scalar function. This locally scales
distances by a factor \(R(\boldsymbol{s})^{-1}\),
\begin{equation}
        \mathrm{d}\sigma^2 = \begin{bmatrix}\mathrm{d}s_1 & \mathrm{d}s_2\end{bmatrix} \mathbf{g}(\boldsymbol{s}) \begin{bmatrix} \mathrm{d}s_1 \\ \mathrm{d}s_2 \end{bmatrix}= R(\boldsymbol{s})^{-2} (\mathrm{d}s_1^2+\mathrm{d}s_2^2),
        \label{eq:SPDEman}
\end{equation}
where \(\mathrm{d}\sigma\) is the line element, and \(s_1\) and \(s_2\) are the two coordinates
of \(E = \mathbb{R}^2\). 

The non-stationarity in the correlation structure is then described through the
spatially varying geometry in Equation~\eqref{eq:SPDEman}, which results
in a curved two-dimensional manifold that must be embedded in a space with dimension higher than 2 to
exist in Euclidean space. The resulting spatial field does not have exactly constant marginal 
variance because the curvature of the space is non-constant unless \(R(\cdot)\) does not
vary in space, but there will be less interaction between \(R(\cdot)\) and 
the marginal variance than between \(\kappa(\cdot)\) and the marginal variance. 
And when \(R(\cdot)\) varies slowly, the variation in marginal variances
is small.

We can relate
the Laplace-Beltrami operator in \(E\) to the usual Laplacian in \(\mathbb{R}^2\)
through
\[
        \Delta_{E} = \frac{1}{\sqrt{\det(g)}} \nabla_{\mathbb{R}^2}\cdot (\sqrt{\det(g)} g^{-1} \nabla_{\mathbb{R}^2}) = R(\boldsymbol{s})^2\Delta_{\mathbb{R}^2},
\]
and the Gaussian standard white noise in \(E\) to the Gaussian standard white
noise in \(\mathbb{R}^2\) through
\[
        \mathcal{W}_E(\boldsymbol{s}) = \det(g)^{1/4}\mathcal{W}_{\mathbb{R}^2}(\boldsymbol{s})=R(s)^{-1} \mathcal{W}_{\mathbb{R}^2}(\boldsymbol{s}).
\]
Thus the equivalent SPDE in \(\mathbb{R}^2\) can be written as
\[
        R(\boldsymbol{s})^{-2}\left[1 - R(\boldsymbol{s})^2\Delta_{\mathbb{R}^2}\right]u(\boldsymbol{s}) = R(s)^{-1} \sqrt{4\pi}\mathcal{W}_{\mathbb{R}^2}(\boldsymbol{s}), \quad \boldsymbol{s}\in\mathbb{R}^2,
\]
where the first factor is needed because the volume element 
\(\mathrm{d}V_E = \sqrt{\det(g)}\mathrm{d}V_{\mathrm{R}^2}\).
The SPDE can be written as
\begin{equation}
        (R(s)^{-2} - \Delta_{\mathbb{R}^2})u(\boldsymbol{s}) = \sqrt{4\pi}R(s)^{-1}\mathcal{W}_{\mathbb{R}^2}, \quad \boldsymbol{s}\in\mathbb{R}^2,
        \label{eq:deformed}
\end{equation}
in Euclidean space, but we can interpret the non-stationarity through the implied metric
tensor. The procedure is similar to  the simple reparametrization \(\kappa(\cdot) = R(\cdot)^{-1}\), 
but the extra factor on the right-hand side of the equation reduces the variability of 
the marginal variances due to changes in \(\kappa(\cdot)\).

\begin{figure}
        \centering
        \includegraphics[width=9cm]{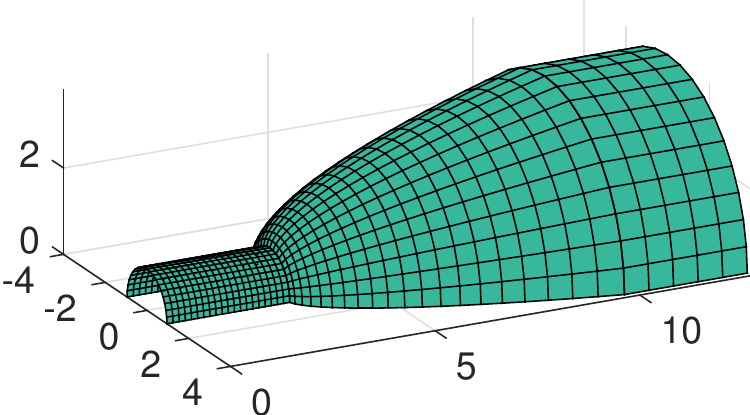}
        \caption{Half cylinder deformed according to the spatially varying
                 metric tensor. The lines formed a regular grid on the half
                 cylinder before deformation.  }
        \label{fig:deform}
\end{figure}

For example, the space \([0,9]\times[0,3]\) with the Euclidean
distance metric can be visualized as a rectangle,
which exists in \(\mathbb{R}^2\), or as a
half cylinder with radius \(3/\pi\) and height 9, which exists in
\(\mathbb{R}^3\), but if the space is given the spatially
varying metric tensor defined by the local range function 
\begin{equation}
        R(s_1,s_2) = \begin{cases} 1 & 0 \leq s_1 < 3, 0 \leq s_2 \leq \pi, \\
                               (s_1-2) & 3 \leq s_1 < 6, 0 \leq s_2 \leq \pi,\\
                                4 & 6 \leq s_1  \leq 9, 0 \leq s_2 \leq \pi,
                \end{cases}
\end{equation}
the space cannot be embedded in \(\mathbb{R}^2\). With this
metric tensor, the space is no longer flat, and must be embedded
in \(\mathbb{R}^3\) as, for example, the deformed cylinder shown in 
Figure~\ref{fig:deform}. Thus, solving Equation~\eqref{eq:deformed} 
with the spatially varying coefficient is the same as solving 
Equation~\eqref{eq:SPDEtensor} on the deformed 
space. This means that unlike the deformation method, a spatially
varying \(R(\cdot)\) does not correspond to a deformation
of \(\mathbb{R}^2\) to \(\mathbb{R}^2\), but rather from
\(\mathbb{R}^2\) to a higher-dimensional space.

Since the variation in the marginal variances 
due to variations in the local ranges is small if \(R(\cdot)\) does not vary too much, 
we introduce a separate function \(S(\cdot)\) that controls the marginal standard deviations
of the process and limit the SPDE to a region of interest, \(\mathcal{D}\), 
with Neumann boundary conditions,
\[
        (R(s)^{-2} - \Delta_{\mathbb{R}^2})\left(\frac{u(\boldsymbol{s})}{S(\boldsymbol{s})}\right) = \sqrt{4\pi}R(s)^{-1}\mathcal{W}_{\mathbb{R}^2}(\boldsymbol{s}), \quad \boldsymbol{s}\in\mathcal{D}.
\]
This introduces boundary effects as was discussed in the paper by
\citet{Lindgren2011}, but we will not discuss the effects of the
boundary in this paper.

This SPDE allows for greater separation of the parameters
that affect correlation structure and the parameters that affect marginal 
standard deviations than the previous approach, and demonstrates the usefulness of
careful consideration of how the spatially varying behaviour is
introduced and parametrized. The SPDE derived based on the metric tensor allows 
for separate priors for correlation structure and marginal standard deviations through
expansions of \(\log(R(\cdot))\) and \(\log(S(\cdot))\) into bases.

\subsection{Setting priors on the parameters}
There are two sources of non-stationarity in the flexible
extension from stationarity: a function 
\(R(\cdot)\) that controls local range and a function \(S(\cdot)\) that controls the
marginal standard deviation. The degree of flexibility in each of these
sources of non-stationarity must be controlled to limit the risk of overfitting.
Due to the issues of singular and equivalent Gaussian measures discussed in the
main paper, we will not follow the
PC prior framework, but instead use a construction motivated by the principles of the PC priors
to make the non-stationary model contract towards a base model of stationarity.
Denote by \(\boldsymbol{\theta}\) the extra parameters added to the GRF that
move the model away from the base model of stationarity, \(\boldsymbol{\theta} = \boldsymbol{0}\). 
The prior on \(\boldsymbol{\theta}\)
will be constructed conditionally on the parameters of the stationary GRF, \(\rho\) and
\(\sigma^2\), and for each choice of these parameters, \(\boldsymbol{\theta}\) should
shrink towards \(\boldsymbol{0}\).

We parametrize the local distance, \(R(\cdot)\), and the approximate
marginal standard deviations, \(S(\cdot)\), through
\begin{equation}
\begin{aligned}
        \log(R(\boldsymbol{s})) &= \log\left(\frac{\rho}{\sqrt{8}}\right)+\sum_{i=1}^{n_1} \theta_{1,i} f_{1,i}(\boldsymbol{s}), \quad \boldsymbol{s}\in\mathcal{D},\\
        \log(S(\boldsymbol{s})) &= \log(\sigma) + \sum_{i=1}^{n_2} \theta_{2,i} f_{2,i}(\boldsymbol{s}), \quad \boldsymbol{s}\in\mathcal{D},
\end{aligned}\label{eq:BasisExpansion}
\end{equation}
where \(\{f_{1,i}\}\) is a set of basis functions for the local range centred such
that \(\langle f_{1,i}, 1\rangle_\mathcal{D} = 0\), for \(i=1,2,\ldots,n_1\), and
\(\{f_{2,i}\}\) is a set of basis functions for the marginal standard deviations centred
such that \(\langle f_{2,i}, 1\rangle_\mathcal{D} = 0\) for \(i = 1,2,\ldots, n_2\).
We collect the parameters in vectors 
\(\boldsymbol{\theta}_1 = (\theta_{1,1}, \ldots, \theta_{1,n_1})\) and
\(\boldsymbol{\theta}_2 = (\theta_{2,1},\ldots, \theta_{2,n_2})\) such that
\(\boldsymbol{\theta}_1\) controls the non-stationarity in the correlation
structure and \(\boldsymbol{\theta}_2\) controls the non-stationarity in the
marginal standard deviations.

We want the prior for each source of non-stationarity to be invariant to scaling of the covariates and
to handle linear dependencies between the covariates in a reasonable way, and we follow the basic
idea of the g-priors~\citep{zellner1986assessing} (with \(g = 1\)),
\[
        \boldsymbol{\theta}_1 | \tau_1 \sim \mathcal{N}(\boldsymbol{0}, \tau_1^{-1} \mathbf{S}_1^{-1}) \quad \text{and} \quad \boldsymbol{\theta}_2 | \tau_2 \sim \mathcal{N}(\boldsymbol{0}, \tau_2^{-1} \mathbf{S}_2^{-1}),
\]
where \(S_1\) is the Gramian,
\[
        S_{1,i,j} = \frac{\left\langle f_{1,i} , f_{1,j} \right\rangle_\mathcal{D}}{\left\langle 1 , 1 \right\rangle_\mathcal{D}},\quad \text{for \(i,j = 1,2,\ldots, n_1\)},
\]
and \(S_2\) is similarly the Gramian based on \(\{f_{2,i}\}\).
In this set-up the Gramians account for the structure in the basis functions
and the strictness of the priors are controlled by two precisions parameters
\(\tau_1\) and \(\tau_2\). If the precision parameters are
fixed hyperparameters, the resulting priors are Gaussian. However, the Gaussian
probability density is flat at zero due to the infinite differentiability of the density
function, and we prefer a prior that has a spike at zero.

This can be achieved by selecting the hyperpriors to be the PC prior for the precision parameter in 
a Gaussian distribution~\citep{Martins2014}, which
is designed to shrink towards a base model of zero variance. We combine the selection for the hyperpriors
with an \emph{a priori} ansatz that the independence between the correlation
structure and the marginal variance in the prior for the stationary model also can be applied to
the non-stationarity,
\[
        \pi(\tau_1) = \frac{\lambda_1}{2}\tau_1^{-3/2}\exp\left(-\lambda\tau_1^{-1/2}\right) \,\,\, \text{and} \,\,\, \pi(\tau_2) = \frac{\lambda_2}{2}\tau_2^{-3/2}\exp\left(-\lambda_2\tau_2^{-1/2}\right).
\]
These hyperpriors for the precision parameters have so heavy tails that integrating them out
will introduce infinite spikes in the
marginal priors for \(\boldsymbol{\theta}_1\) and \(\boldsymbol{\theta}_2\) at zero.

The hyperparameters \(\lambda_1\) and \(\lambda_2\) control the spread of
the priors and can be selected either based on expert knowledge
or on frequentist properties. The parameters \(\boldsymbol{\theta}_1\) and
\(\boldsymbol{\theta}_2\) give multiplicative effects to local range
and marginal standard deviations, respectively, and one possibility is to control the size of the multiplicative effect through
\[
         \text{Prob}\left(\max_{\boldsymbol{s}\in\mathcal{D}} \left| \log\left(\frac{R(\boldsymbol{s})}{\rho/\sqrt{8}}\right)\right| > C_1 \middle\vert \rho, \sigma^2\right) = \text{Prob}\left(\max_{\boldsymbol{s}\in\mathcal{D}} \left| \log\left(\frac{R(\boldsymbol{s})}{\rho/\sqrt{8}}\right)\right| > C_1\right) = \alpha_1,
\]
\[      
        \text{Prob}\left(\max_{\boldsymbol{s}\in\mathcal{D}} \left| \log\left(\frac{S(\boldsymbol{s})}{\sigma^2}\right)\right| > C_2 \middle\vert \rho, \sigma^2\right) = \text{Prob}\left(\max_{\boldsymbol{s}\in\mathcal{D}} \left| \log\left(\frac{S(\boldsymbol{s})}{\sigma^2}\right)\right| > C_2 \right) = \alpha_2.
\]
One can see from Equation~\eqref{eq:BasisExpansion} that the relative differences do not depend on 
the parameters of the stationary model, and the full prior factors as
\(\pi(\rho, \sigma^2, \boldsymbol{\theta}) =\pi(\rho)\pi(\sigma^2)\pi(\boldsymbol{\theta}_1)\pi(\boldsymbol{\theta}_2)\).

In practice, it is difficult to have an informed,
\emph{a priori} opinion on the non-stationary part of the model,
but the hyperparameters \(\lambda_1\) and \(\lambda_2\) can be chosen in such a way
that they give a conservative prior. Since stationarity is our base model and the non-stationarity is
 provided as extra flexibility, we will require that the hyperparameters are set such
that the inference behaves
well when the true data-generating distribution is stationary. We propose to set the 
hyperparameter by first fitting the
stationary model, using the maximum aposteriori estimate of the parameters
to make multiple simulated datasets from the stationary
GRF and nugget effect, fit the non-stationary GRF with a nugget effect to
each dataset, and calculate the frequentist coverage of the non-stationarity parameters.
The hyperparameters can then be set such that the coverage of the credible 
intervals of the non-stationary parameters is close to nominal coverage. This ensures
that the prior provides enough regularization that each posterior 
marginal for the non-stationarity parameters do not suggest non-stationarity when a 
stationary data-generating function is used. 

\subsection{Supplementary details for example on precipitation}
The example is described in the main paper and this section provides 
extra figures and details that supplements the presentation in the main paper.

\begin{figure} 
    \centering
    \includegraphics[width = 8cm]{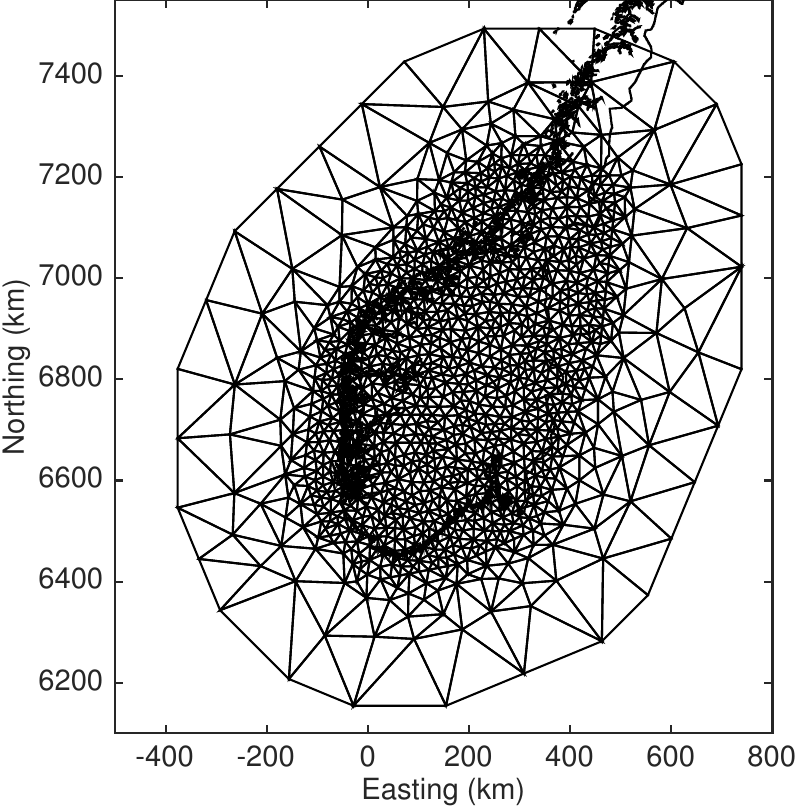}
    \caption{Mesh used for the SPDE approach.}
    \label{fig:mesh}
\end{figure}

\begin{figure}
    \centering
    \subfloat[Elevation (km)]{
        \includegraphics[width = 7.7cm]{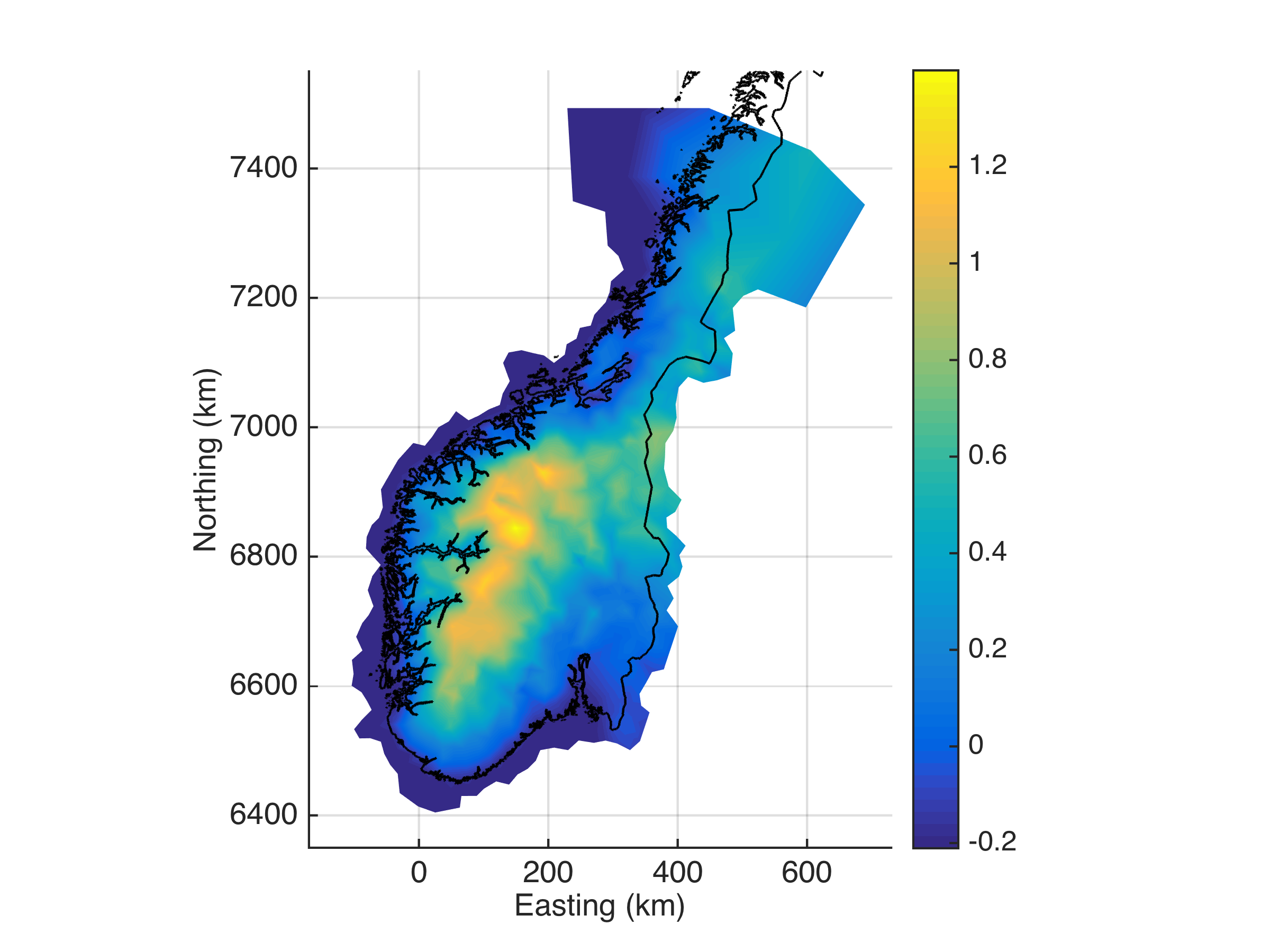}
        \label{fig:covCov:elev}
    }
    \subfloat[Magnitude of gradient (100m/km)]{
        \includegraphics[width = 7.7cm]{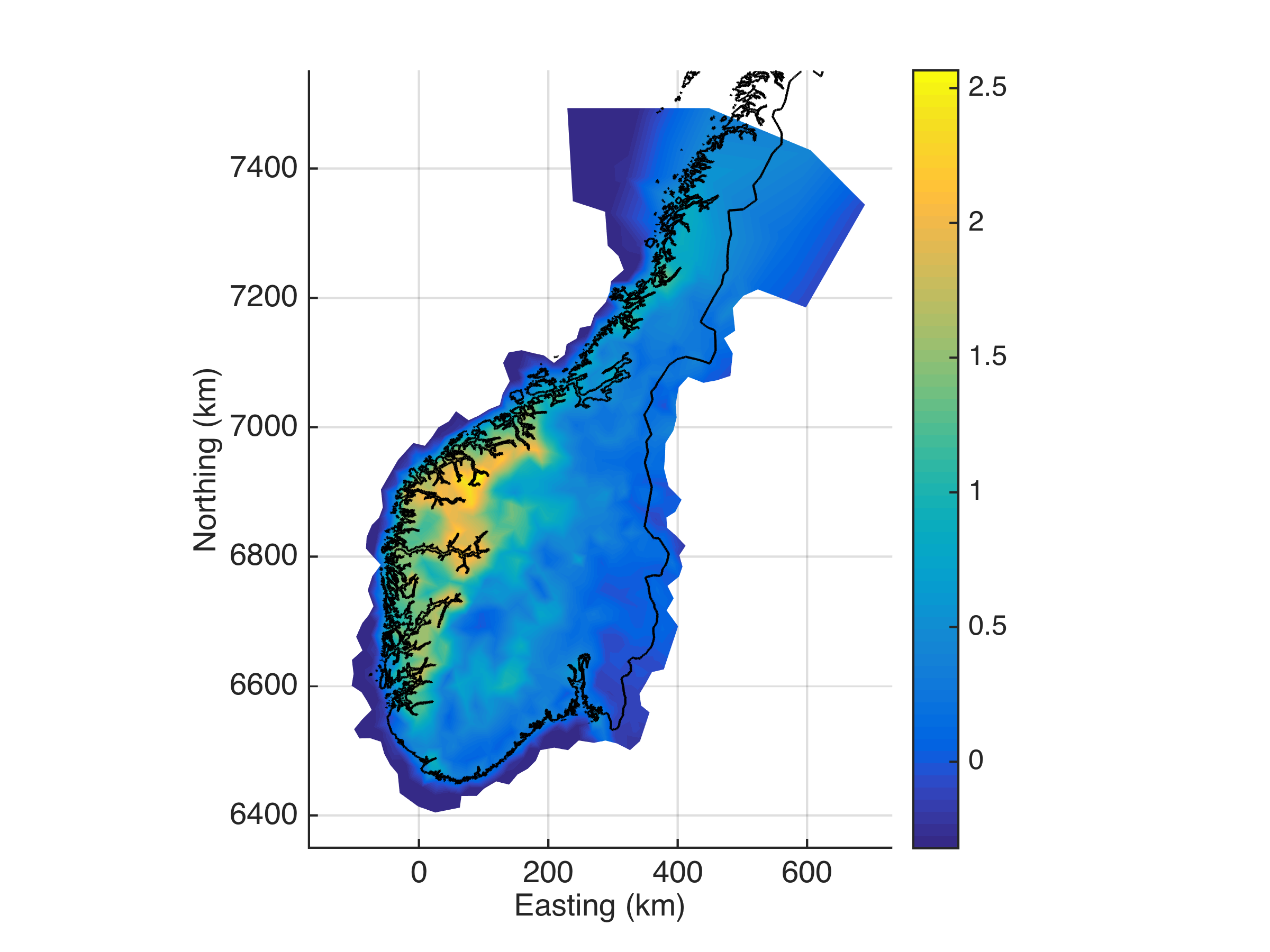}
        \label{fig:covCov:grad}
    }
    \caption{The covariates \protect\subref{fig:covCov:elev} elevation and
             \protect\subref{fig:covCov:grad} magnitude of the gradient used 
             for the covariance structure.}
    \label{fig:covCov}
\end{figure}

In the SPDE approach the spatial field is defined on a triangular mesh and the values of the 
spatial field within
the triangles are defined through linear interpolation based on the values on the nodes of the mesh. 
This means that the elevation covariate in the first-order structure is only needed at each
observation location, but that the elevation and gradient covariates in the
second-order structure are needed at every location within the triangulation. We use
the mesh shown in Figure~\ref{fig:mesh} and project elevation and 
gradient values from the high resolution digital elevation map
GLOBE~\citep{Globe1999} onto the mesh. The projection is piece-wice linear
on each triangle of the mesh and minimizes the integrated square deviation over the domain
covered by the mesh. This results in the piece-wise linear covariates shown in  
Figure~\ref{fig:covCov}.

The coefficients, \(\boldsymbol{\theta}_1\), of the two linear covariates in 
\(\log(R(\cdot))\) are given the prior
\begin{align*}
    \boldsymbol{\theta}_1 | \tau_1 &\sim \mathcal{N}(\boldsymbol{0}, S_1/\sqrt{\tau_1}) \\
    \tau_1 &\sim \frac{\lambda_1}{2}\tau_1^{-3/2}\mathrm{e}^{-\lambda_1/\sqrt{\tau_1}} 
    \end{align*}
as described in the previous section,
and the coefficients, \(\boldsymbol{\theta}_2\), of the two linear covariates in 
\(\log(S(\cdot))\) are given a similar prior, but with hyperparameter
\(\lambda_2\). The non-stationary model is more difficult to fit in the INLA framework than
the stationary model because the 
priors for \(\boldsymbol{\theta}_1\) and \(\boldsymbol{\theta}_2\) have 
infinite spikes in \(\boldsymbol{0}\) that makes the posteriors non-Gaussian in the area
around the origin. The optimization can be improved by reparametrizing as
\(\boldsymbol{\theta}_1' = \boldsymbol{\theta}_1\sqrt{\tau}_1\) and
\(\boldsymbol{\theta}_2' = \boldsymbol{\theta}_2\sqrt{\tau}_2\), but
the marginal posteriors will not be sufficiently peaked at the origin and
will miss the multimodality that should be present when there is a mode
close to zero. However, we still use INLA as a fast approximation for
the repeated fitting of the datasets needed for selecting the hyperparameters
of the prior for non-stationarity. 

\begin{figure}
    \centering
    \subfloat[Range]{
        \includegraphics[width = 7.7cm]{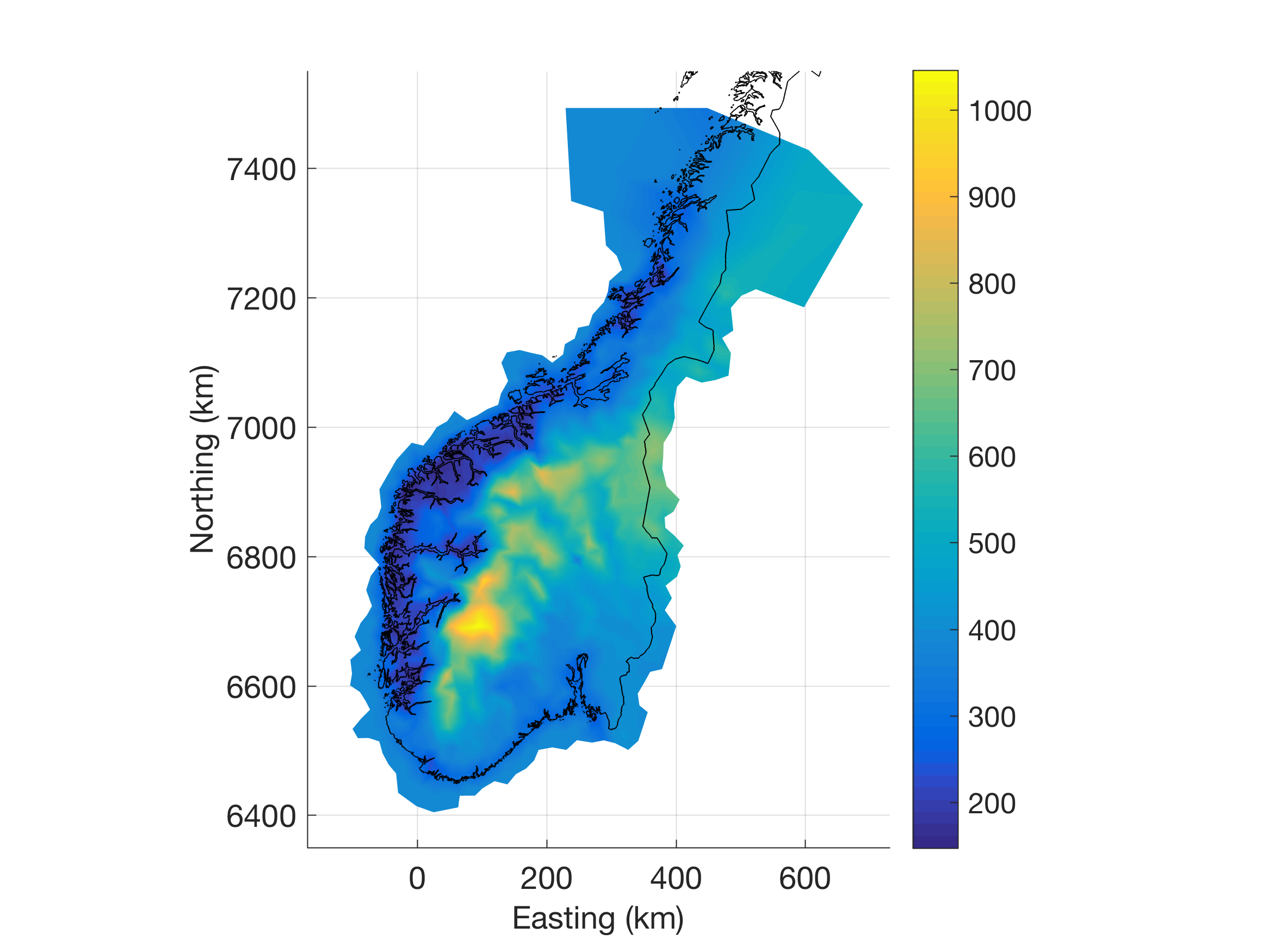}
        \label{fig:nonStat:postEffect:range}
    }
    \subfloat[Standard deviation]{
        \includegraphics[width = 7.7cm]{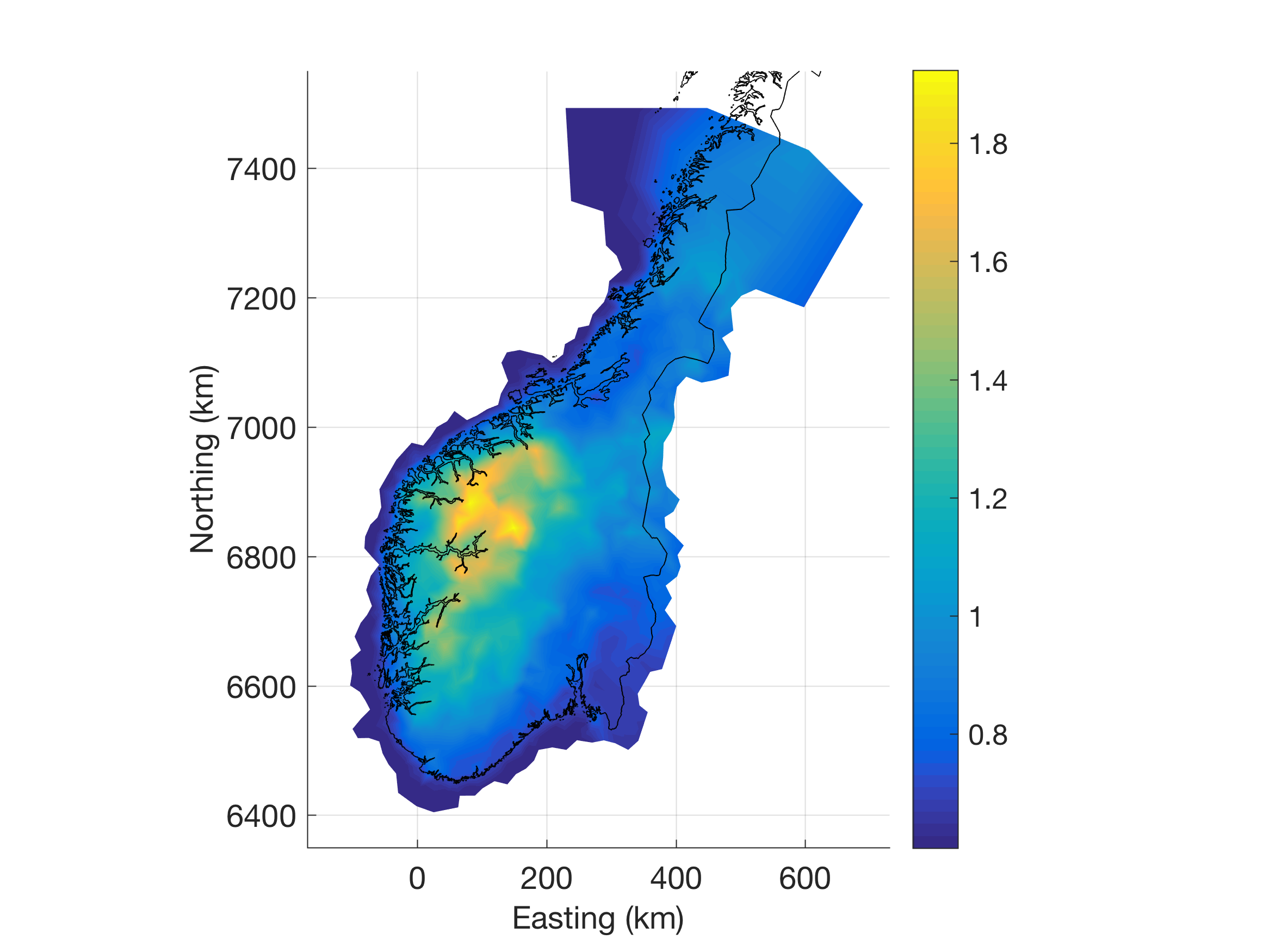}
        \label{fig:nonStat:postEffect:stddev}
    }
    \caption{Posterior mean of 
             \protect\subref{fig:nonStat:postEffect:range} range and
             \protect\subref{fig:nonStat:postEffect:stddev} standard
             deviation.}
    \label{fig:nonStat:posterior}
\end{figure}

\begin{figure}
    \centering
    \subfloat[Level curves of correlations]{ 
        \includegraphics[width = 6cm]{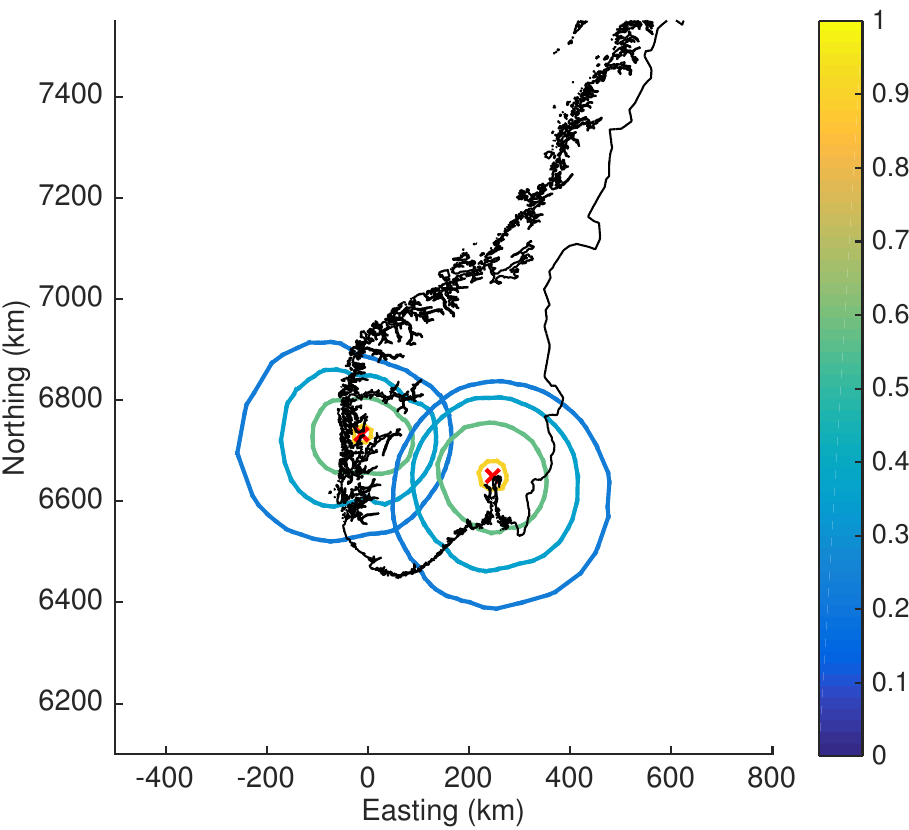} 
        \label{fig:nonStat:corStructure:cor}
    } 
    \subfloat[Marginal standard deviations]{
        \includegraphics[width = 7.7cm]{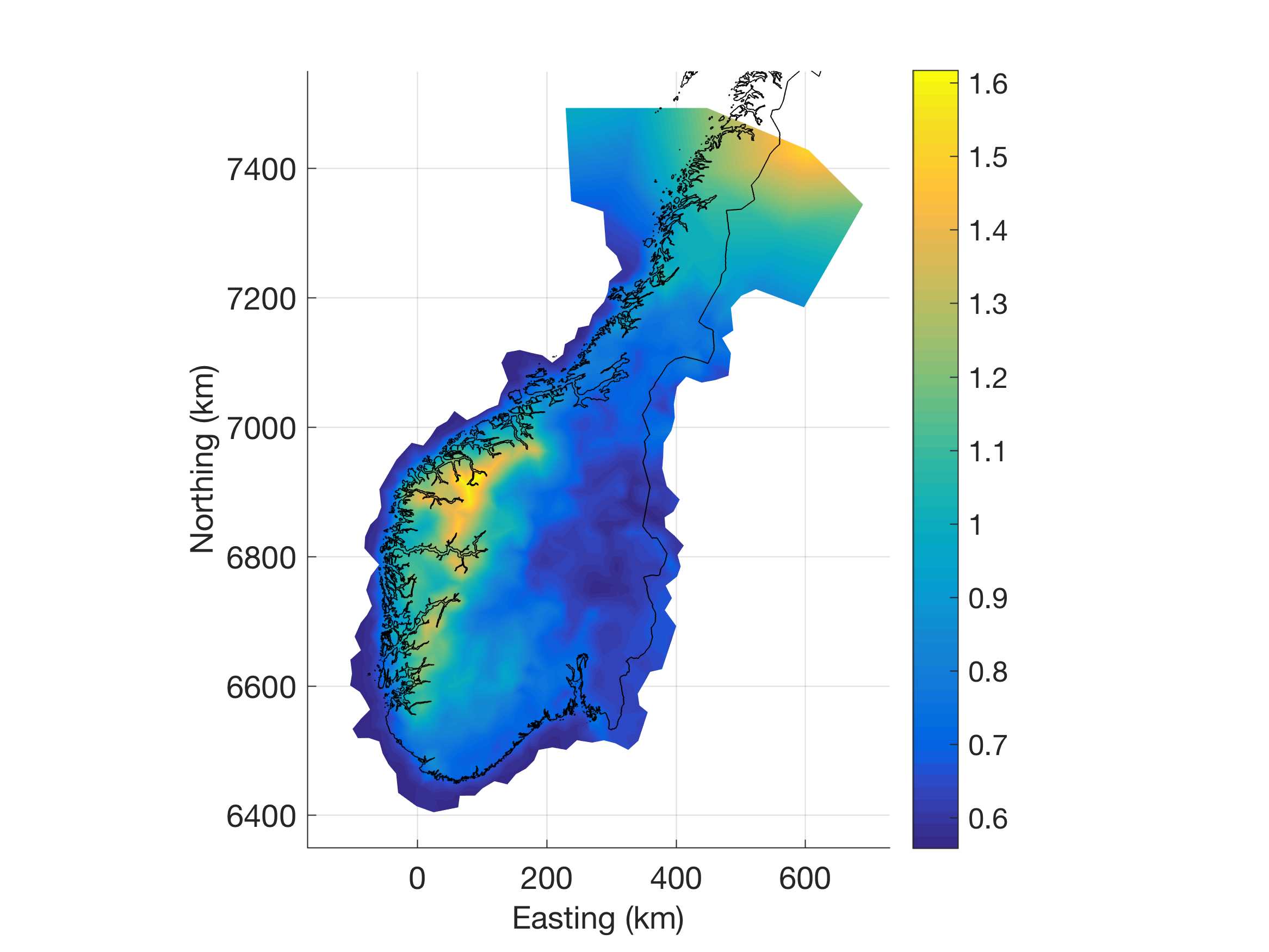}
        \label{fig:nonStat:corStructure:stddev}
    }
    \caption{Covariance structure described through 
             \protect\subref{fig:nonStat:corStructure:cor} 0.90, 0.57, 0.36 and 0.22
             level curves of correlation with respect to the two locations marked
             with red crosses and
             \protect\subref{fig:nonStat:corStructure:stddev} marginal
             standard deviations.}
    \label{fig:nonStat:corStructure}
\end{figure}

\begin{figure}
    \centering
    \subfloat[Multiplicative effect on range]{
        \includegraphics[width = 6cm]{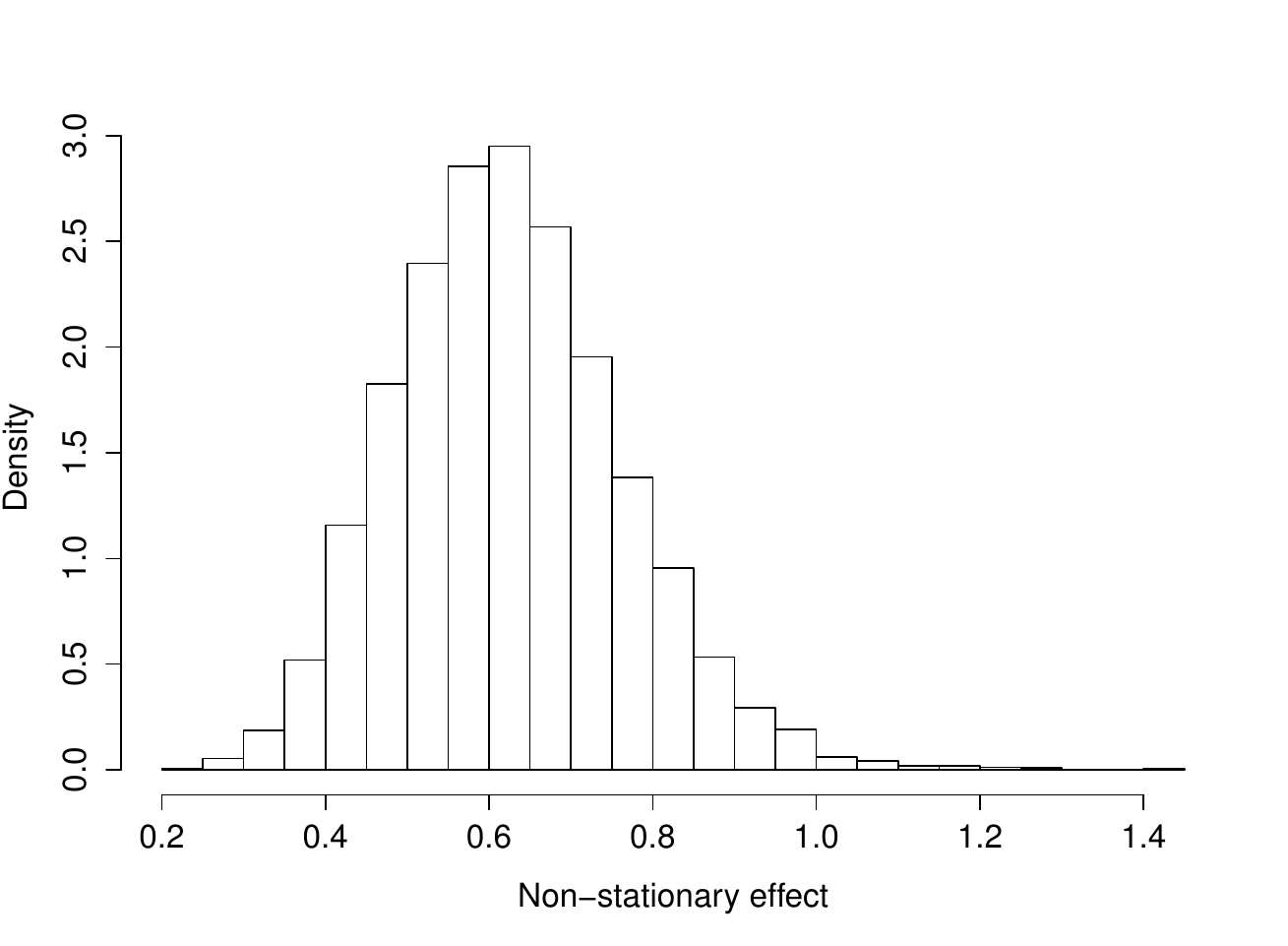}
        \label{fig:nonStat:locEffect:range}
    }
    \subfloat[Multiplicative effect on standard deviation]{
        \includegraphics[width = 6cm]{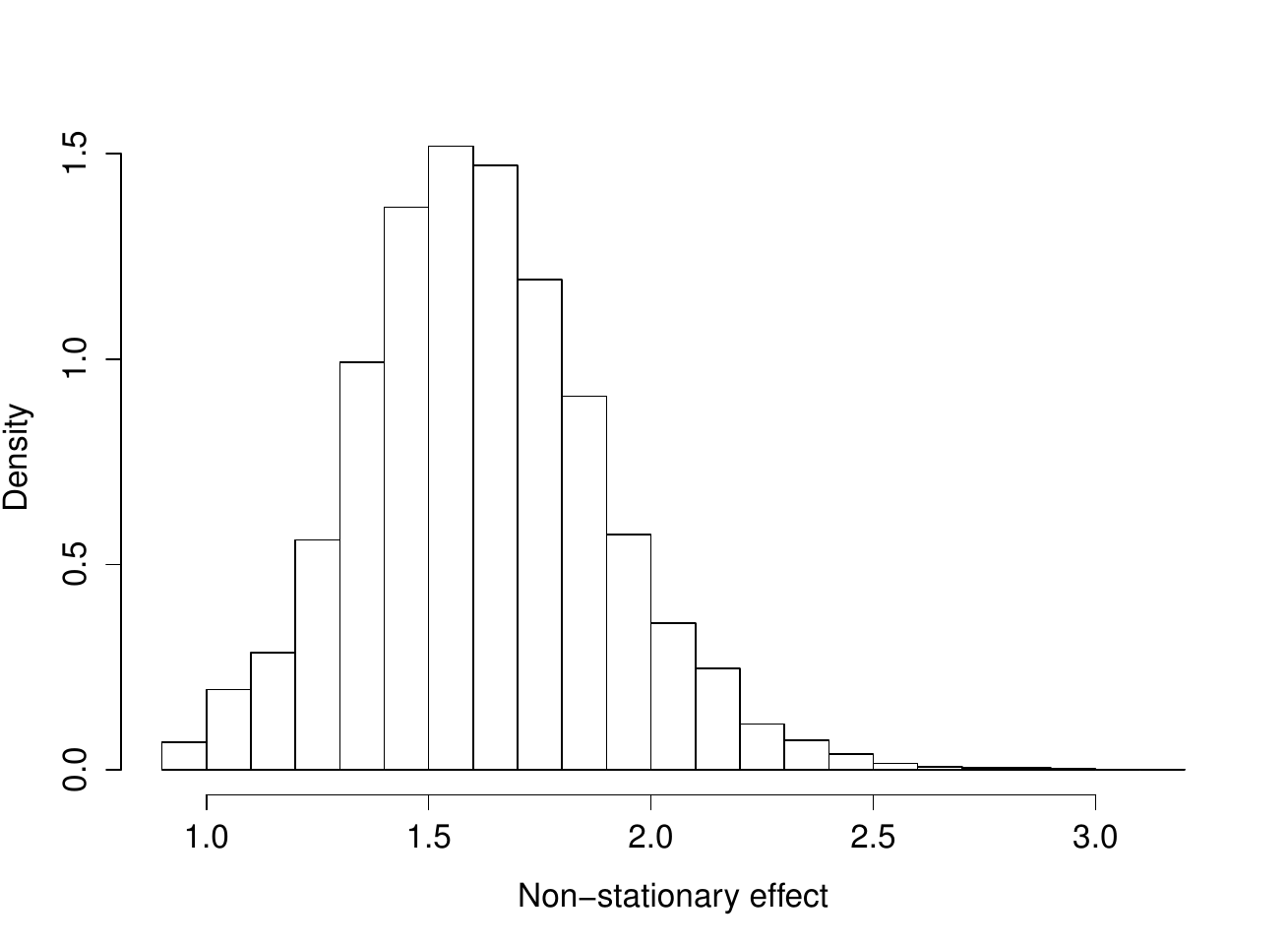}
        \label{fig:nonStat:locEffect:stddev}
    }
    \caption{Posteriors of the multiplicative effect on the stationary
             \protect\subref{fig:nonStat:locEffect:range} range and
             \protect\subref{fig:nonStat:locEffect:stddev} standard
             deviation at the western location in Figure~\ref{fig:nonStat:corStructure:cor}.}
    \label{fig:nonStat:locEffect}
\end{figure}

The non-stationary model was fitted using an MCMC sampler and 
the resulting posterior means of the range and the standard deviation are shown in 
Figure~\ref{fig:nonStat:posterior}. From 
Figure~\ref{fig:nonStat:corStructure} one can see that the spatially
varying range and standard deviation leads to non-stationarity in
the correlation structure and the marginal standard deviations of
the spatial effect. However, the effect in standard deviations 
appear to be stronger than the effect of the spatially varying
range. The posteriors for the multiplicative effects on the stationary
range and standard deviation for the western location in Figure \ref{fig:nonStat:corStructure:cor}
shown in Figure \ref{fig:nonStat:locEffect} shows that the effects
are significant in that location. The posterior
probabilities for the effects to be less than 1 and greater than 1
are \(99\%\) and \(99\%\), respectively. This shows that the the more
flexible non-stationary model is preferring to move away from the
stationary model even under a conservatively selected prior.

\section{Additional theorem}
    In the proof of the main result in the paper it is necessary to 
    show that the integral used to calculate the KLD
    is finite. The following theorem shows that this holds in dimensions
    $d = 1$, $d = 2$ and $d = 3$.
        \begin{thm}
        \label{thm:finInt}
        The definite integral
        \[
                I_d = \int_{\mathbb{R}^d} \left[\left(\frac{||\boldsymbol{w}||^2}{(1 + ||\boldsymbol{w}||^2)}\right)^\alpha - 1 - \log \left(\frac{||\boldsymbol{w}||^2}{(1 + ||\boldsymbol{w}||^2)}\right)^\alpha\right] \mathrm{d}\boldsymbol{w},
        \]
        where $\alpha > 0$, is finite for $d \leq 3$.
        \end{thm}
        \begin{proof}
                The definite integral can be expressed an an integral in \(d\)-dimensional spherical coordinates,
                \begin{equation}
                        I_d = C_d \int_{0}^\infty\left[\left(\frac{r^2}{1+r^2}\right)^\alpha-1-\log\left(\frac{r^2}{1+r^2}\right)^\alpha\right]r^{d-1} \mathrm{d}r,
                        \label{eq:KLDintegral}
                \end{equation}
                where \(C_d\) is a finite constant that varies with dimension. There are two issues: the
                behaviour for small \(r\) and the behaviour for large \(r\). For \(d=1\),
                \[
                        0 \leq  I_d \leq -C_1\alpha \int_0^\infty \log\frac{r^2}{1+r^2} \mathrm{d}r = \pi \alpha C_1 < \infty,
                \]
                and the definite integral is finite for $d = 1$. Furthermore, the factor $r^{d-1}$ makes the value of the integrand
                smaller close to $0$ for larger $d$ and we can conclude that the behaviour around \(0\) is not a problem for any \(d\geq 1\). 

                The behaviour of the integrand
                for large \(r\) can be studied through an expansion of the integrand in \((1+r^2)^{-1}\). The part between the square
                brackets in Equation~\eqref{eq:KLDintegral} behaves as
                \[
                        \frac{\alpha^2}{2} \frac{1}{(1+r^2)^2} + \mathcal{O}\left(\frac{1}{(1+r^2)^3}
                        \right).
                \]
                This means that there exists a \(0 < r_0 < \infty\) such that
                \begin{align*}
                        \int_{0}^\infty&\left[\left(\frac{r^2}{1+r^2}\right)^\alpha-1-\log\left(\frac{r^2}{1+r^2}\right)^\alpha\right]r^{d-1} \mathrm{d}r \\
                        & \phantom{2222222222}\leq C_1 + \int_{r_0}^\infty \left[\frac{\alpha^2}{2} \frac{1}{(1+r^2)^2} + \frac{C_2}{(1+r^2)^3}
                        \right] r^{d-1}\mathrm{d}r,
                \end{align*}
                where \(|C_1| \leq \infty\) due to the finiteness for $d = 1$, and $C_2$ is a constant. For \(d\leq 3\) both terms on the right hand side 
                are finite. Thus $I_d$ is finite for $d \leq 3$.
        \end{proof}

\renewcommand{\bibfont}{\small}
\bibliographystyle{apalike}
\bibliography{references}

\end{document}